%% file: paper.tex
\documentclass[12pt]{article}
\usepackage{jheppub}
\pdfoutput=1

\usepackage{amsmath,bbm,array,amsfonts,graphicx,wrapfig,lscape,float,mathtools,multirow,longtable}

\input{pref}

\definecolor{darkspringgreen}{rgb}{0.09, 0.45, 0.27}
\definecolor{forestgreen}{rgb}{0.13, 0.55, 0.13}

\title{Brane Brick Models, Toric Calabi-Yau 4-Folds \\ and 2d (0,2) Quivers}

\author[a,b]{Sebasti\'an Franco,}
\author[c,d,e,f]{Sangmin Lee,}
\author[g]{Rak-Kyeong Seong}

\affiliation[a]{
Physics Department, The City College of the CUNY \\
160 Convent Avenue, New York, NY 10031, USA}

\affiliation[b]{The Graduate School and University Center, The City University of New York  \\
365 Fifth Avenue, New York NY 10016, USA }

\affiliation[c]{
Center for Theoretical Physics, Seoul National University, Seoul 08826, Korea
}

\affiliation[d]{
Department of Physics and Astronomy, Seoul National University, Seoul 08826, Korea
}

\affiliation[e]{
College of Liberal Studies, Seoul National University, Seoul 08826, Korea
}

\affiliation[f]{
School of Natural Sciences, Institute for Advanced Study, Princeton, NJ 08540, USA
}

\affiliation[g]{
School of Physics, Korea Institute for Advanced Study, Seoul 02455, Korea
}

\emailAdd{sfranco@ccny.cuny.edu}
\emailAdd{sangmin@snu.ac.kr}
\emailAdd{rkseong@kias.re.kr}

\preprint{
\begin{flushright}
CCNY-HEP-15-07 \\
SNUTP15-009\\
KIAS-P15039
\end{flushright}
}

\abstract{
We introduce brane brick models, a novel type of Type IIA brane configurations consisting of D4-branes ending on an NS5-brane. Brane brick models are T-dual to D1-branes over singular toric Calabi-Yau 4-folds. They fully encode the infinite class of $2d$ (generically) $\mathcal{N} = (0,2)$ gauge theories on the worldvolume of the D1-branes and streamline their connection to the probed geometries. For this purpose, we also introduce new combinatorial procedures for deriving the Calabi-Yau associated to a given gauge theory and vice versa.
}

\begin{document}

\maketitle

\section{Introduction}

In recent years, there have been intense efforts in mapping the landscape of quantum field theories and uncovering their dynamics. As part of this enterprise, quantum field theories in various dimensions and with diverse amounts of supersymmetry have been investigated and connections between many of them have been explored.

This article is devoted to $2d$ $\mathcal{N}=(0,2)$ theories, which are particularly interesting for various reasons. Despite the reduced amount of SUSY, chirality and holomorphy provide substantial control of their dynamics. The recent discovery of a new IR equivalence between different theories known as triality \cite{Gadde:2013lxa} is an interesting example of this. In addition, these theories arise on the worldsheet of heterotic strings. Another exciting development is the geometric realization of a wide class of these theories via compactification of the $6d$ $(2,0)$ theory on 4-manifolds \cite{Gadde:2013sca}.

It is extremely desirable to understand how to engineer $2d$ $(0,2)$ gauge theories in terms of branes. Early steps in this direction were taken in \cite{GarciaCompean:1998kh,Ahn:1999iu,Sarkar:2000iz}. This question was revived in our previous work \cite{Franco:2015tna}, which initiated an ambitious program aimed at understanding in detail the infinite class of gauge theories arising on D1-branes probing arbitrary singular toric Calabi-Yau (CY) 4-folds and developing T-dual brane setups.\footnote{Other interesting approaches to the D-brane engineering of $2d$ $(0,2)$ theories can be found in \cite{Kutasov:2013ffl,Kutasov:2014hha,Tatar:2015sga}.}

The part of the story regarding D1-branes on toric CY$_4$ singularities was completely developed in \cite{Franco:2015tna}. In particular, an algorithm for connecting gauge theories to the probed geometry, which arises as the their classical mesonic moduli space, was developed. We referred to it as the {\it forward algorithm}. In addition, a systematic procedure for obtaining the gauge theory for an arbitrary toric CY$_4$ singularity by means of partial resolution was introduced. 
 
In this work, we will fully develop the second part of the story: the T-dual brane setups. These brane configurations are called {\it brane brick models} and their basic features were already anticipated in \cite{Franco:2015tna}. They substantially simplify the connection between geometry and gauge theory.

This article is organized as follows. Section \sref{section_CY4_and_gauge theory} reviews general features of the gauge theories on D1-branes over toric CY$_4$ folds, their description in terms of periodic quivers and the brane box configurations for abelian orbifolds of $\mathbb{C}^4$. Section \sref{section_brane_brick_models} introduces brane brick models and presents the dictionary connecting them to gauge theories. Section \sref{sgeomtobrick} introduces the {\it fast inverse algorithm}, for going from the toric diagram of a CY$_4$ to the corresponding brane brick model. Section \sref{sbricktogeom} presents the {\it fast forward algorithm}, which goes in the opposite direction and determines the CY$_4$ associated to a brane brick model. A key ingredient of this approach is a correspondence between GLSM fields and a new class of combinatorial objects, denoted {\it brick matchings}. This combinatorial computation of the geometry represents a tremendous simplification over the standard forward algorithm of \cite{Franco:2015tna}. Section \sref{section_partial_resolution} discusses partial resolutions in terms of brane brick models, extending the comprehensive study presented in \cite{Franco:2015tna}. Section \sref{section_CY3_x_C} is devoted to Calabi-Yau 4-folds of the form CY$_3 \times \mathbb{C}$. The corresponding $2d$ gauge theories have $(2,2)$ SUSY and can be obtained from the $4d$ $\mathcal{N}=1$ theories associated to the CY$_3$ by dimensional reduction. A lifting algorithm for generating the brane brick model for the $2d$ theory from the brane tiling for the $4d$ one is introduced. Section \sref{section_general_CY4} goes beyond orbifolds and dimensionally reduced theories and studies the brane brick models for generic toric CY$_4$ singularities. Section \sref{section_conclusions} presents our conclusions and some directions for future research.

\section{$2d$ (0,2) Theories from D1-Branes over Toric CY$_4$ Cones \label{2d_(0,2)_generalities}}

For a thorough discussion on the structure of general $2d$ $(0,2)$ theories, including their supermultiplet structure and the construction of their Lagrangians in terms of $(0,2)$ superspace, we refer to \cite{Witten:1993yc,GarciaCompean:1998kh,Gadde:2013lxa,Kutasov:2013ffl}. This paper focuses on $2d$ theories on the worldvolume of D1-branes probing toric Calabi-Yau (CY) 4-folds. As explained in \cite{Franco:2015tna}, these theories have a special structure which is the reason for their beautiful connection to toric geometry and to certain combinatorial objects that are going to be introduced in this paper.

\paragraph{Symmetries and Quivers.} 

D1-branes probing a generic toric CY$_4$ singularity preserve $(0,2)$ SUSY. When the CY$_4$ is of the form CY$_3 \times \mathbb{C}$, CY$_2 \times \mathbb{C}^2$ and $\mathbb{C}^4$, there is a non-chiral enhancement of SUSY to $(2,2)$, $(4,4)$ and $(8,8)$, respectively. Such theories can be constructed by dimensional reduction from $4d$ $\mathcal{N}=1$, $2$ and $4$. Chiral SUSY enhancement to $(0,4)$ occurs for CY$_2 \times$CY$_2$. Finally, enhancement to $(0,6)$ and $(0,8)$ is possible for certain orbifolds \cite{GarciaCompean:1998kh}.

The gauge symmetry and matter content of these theories can be encoded in terms of generalized quiver diagrams involving two types of matter fields in bifundamental or adjoint representations: chiral and Fermi multiplets. The gauge group for these theories is a product of $U(N_i)$ factors. As usual, each of these factors is represented by a node in the quiver. The total number of gauge nodes in the quiver is given by the volume of the toric diagram normalized with respect to a minimal tetrahedron. 

All matter multiplets are in adjoint or bifundamental representation of the gauge group. Chiral multiplets are represented by oriented arrows in the quiver diagram. We typically label the chiral fields as $X_{ij}$ with $i$ and $j$ gauge node indices. Fermi multiplets are labeled similarly, $\Lambda_{ij}$, but they are represented by red unoriented lines in the quiver diagram. The reason for this is that $2d$ $(0,2)$ theories are invariant under the exchange of any $\Lambda_{ij}$ with its conjugate $\bar{\Lambda}_{ij}$, i.e. Fermi fields are intrinsically unoriented.

\paragraph{Anomalies.} 

Cancellation of $SU(N_i)^2$ gauge anomalies imposes severe constraints on $2d$ $(0,2)$ theories.\footnote{In theories on D1-branes at singularities, abelian gauge anomalies are cancelled by a generalized Green-Schwarz mechanism through interactions with bulk RR fields \cite{Mohri:1997ef}.} Throughout this paper we will restrict to the case of $N$ regular D1-branes, for which all nodes are $U(N)$. More general rank assignments are possible in the presence of fractional D1-branes. For the case when all ranks are equal, $N_i=N$, cancellation of $SU(N_i)^2$ anomalies at node $i$ require
\beal{esb1a0}
n_i^{\chi} - n_i^F = 2 ~,~
\eea
where $n_i^\chi$ and $n_i^F$ are the total number of chiral and Fermi fields that are attached to node $i$, respectively. Adjoint chiral or Fermi fields contribute 2 to $n_i^{\chi}$ or $n_i^{F}$, respectively.

\paragraph{Toric $J$- and $E$-Terms.} 

In a general $2d$ $(0,2)$ theory, every Fermi field $\Lambda_{ij}$ is associated with a pair of holomorphic functions of chiral fields: $E_{ij}(X)$ (with the same gauge quantum numbers of $\Lambda_{ij}$) and $J_{ji}(X)$ (with conjugate gauge quantum numbers) \cite{Witten:1993yc,GarciaCompean:1998kh,Gadde:2013lxa,Kutasov:2013ffl}. In the theories dual to toric Calabi-Yau 4-folds, these functions take a very special form. This restriction was called the \textit{toric condition} in \cite{Franco:2015tna}, and implies that $J$- and $E$-terms take the following general form
\beal{esb1a1}
J_{ji} = J_{ji}^{+} - J_{ji}^{-} ~,~ E_{ij} = E_{ij}^{+} - E_{ij}^{-} ~,~
\eea
where $J_{ji}^{\pm}$ and $E_{ij}^{\pm}$ are holomorphic monomials in chiral fields.

\paragraph{CY$_4$ Geometry from Gauge Theory.} 

\label{section_CY4_and_gauge theory}

The CY$_4$ geometry probed by the D1-branes is recovered as the {\it classical mesonic moduli space} of the gauge theory that lives on the worldvolume of the D1-branes. Given that the mesonic moduli space of the worldvolume theory of a stack of $N$ D1-branes is the $N$-th symmetric product of the worldvolume theory on a single D1-brane, we focus in this paper on the moduli space of abelian theories.

The mesonic moduli space is obtained by demanding vanishing $J$-, $E$- and $D$-terms. The so-called {\it forward algorithm} for systematically computing these mesonic moduli spaces for arbitrary toric quiver gauge theories has been developed in \cite{Franco:2015tna}. The algorithm solves for vanishing $J$- and $E$-terms expressing chiral fields as products of GLSM fields.  Let $n^\chi$ be the total number of chiral fields. $J$- and $E$-terms impose $n^F-3$ independent constraints, with $n^F$ being the total number of Fermi fields. Demanding invariance under complexified gauge charges gives rise to $G-1$ further constraints, where $G$ is the number of gauge nodes in the quiver.\footnote{Since all fields in the class of theories under study are bifundamental or adjoint, they are neutral under the diagonal combination of all nodes.} Finally, the sum of anomaly cancellation conditions \eref{esb1a0} over all nodes implies that $n^\chi - n^F = G$. Combining all the relations, we find that the mesonic moduli space has complex dimension $n^\chi - (n^F-3)-(G-1) = 4$, as expected. 

Arbitrary toric singularities can be obtained from abelian orbifolds by a series of partial resolutions, which translate to higgsing in the gauge theory. This approach can be exploited for deriving the gauge theories associated to generic toric singularities. A systematic implementation of this method has been developed in \cite{Franco:2015tna}.

\subsection{Unification of Quiver and Toric $J$- and $E$-Terms: Periodic Quivers \label{section_periodic_quiver}}

$2d$ $(0,2)$ theories are specified by the quiver, namely the gauge symmetry and matter content, and the $J$- and $E$-terms for all Fermi fields. Remarkably, for theories corresponding to toric $\text{CY}_4$, this information can be encapsulated in a single graphical object: the {\it periodic quiver}. Periodic quivers were originally introduced in the context of abelian orbifolds of $\mathbb{C}^4$ in \cite{GarciaCompean:1998kh} and were later extended to generic toric singularities in \cite{Franco:2015tna}. 

\begin{figure}[ht!!]
\begin{center}
\resizebox{1\hsize}{!}{
\includegraphics[trim=0cm 0cm 0cm 0cm,totalheight=10 cm]{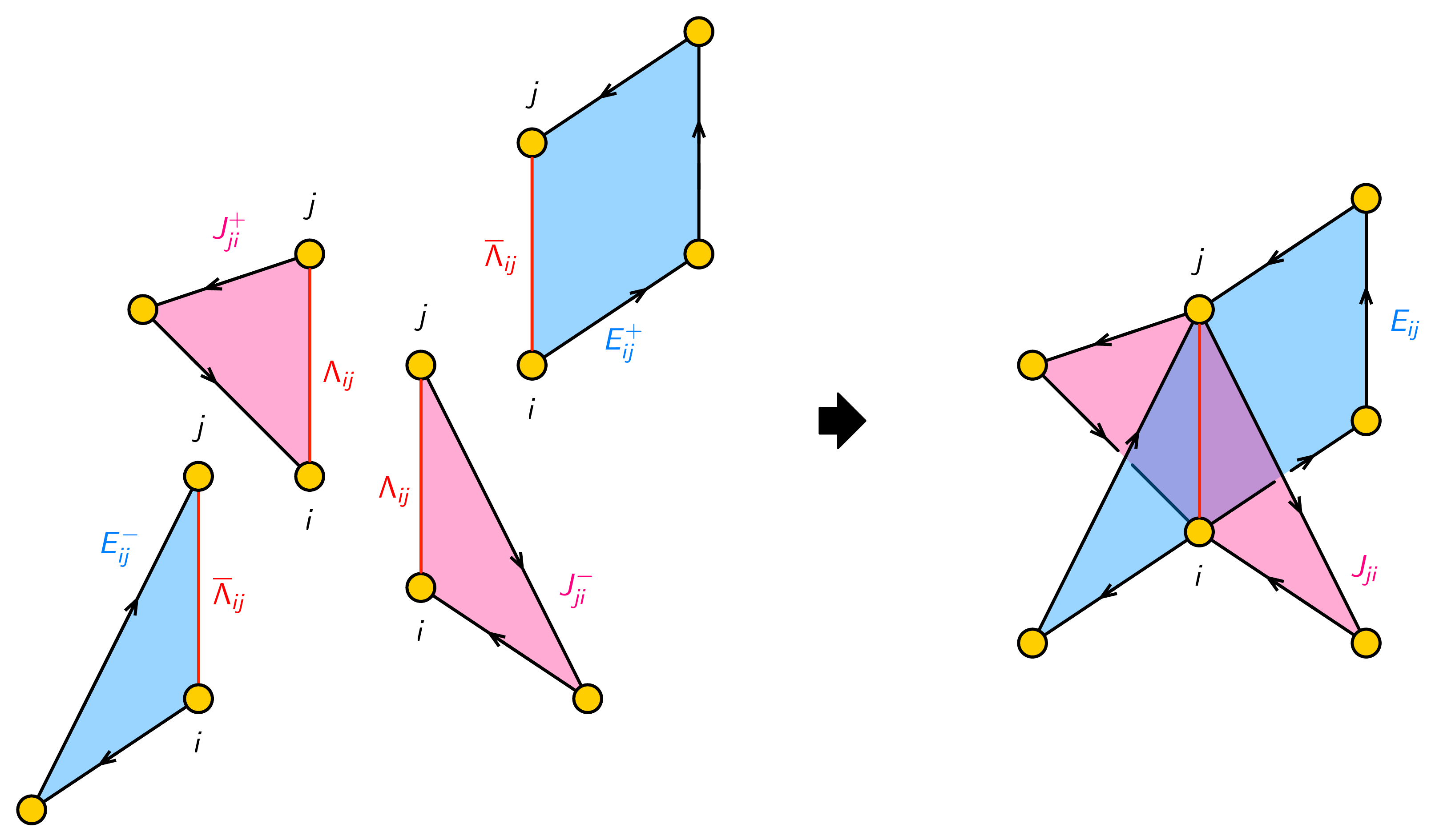}
}  
\caption{
The four plaquettes $(\Lambda_{ij}, J_{ji}^{\pm})$ and  $(\overline{\Lambda}_{ij}, E_{ij}^{\pm})$ corresponding to a Fermi field $\Lambda_{ij}$. 
\label{fplaquettes}}
 \end{center}
 \end{figure}

A periodic quiver lives on a 3-torus $T^3$ and is such that the individual contributions to $J$- and $E$-terms are encoded in terms of certain {\it minimal plaquettes}, as schematically shown in \fref{fplaquettes}.\footnote{What we precisely mean by ``minimal" will be clarified in later sections, once we consider the dual of the periodic quiver.} A plaquette is defined as a closed loop in the quiver consisting of an arbitrary number of chiral fields and a single Fermi field. The chiral fields in a plaquette form an oriented path with two endpoints connected by the Fermi field, which closes the loop. The toric condition \eref{esb1a1} implies that for every Fermi field $\Lambda_{ij}$ there are four plaquettes  $(\overline{\Lambda}_{ij}, E_{ij}^{\pm})$ and $(\Lambda_{ij}, J_{ji}^{\pm})$, which share the undirected edge associated to $\Lambda_{ij}$. 

\begin{figure}[h]
	\centering
	\includegraphics[height=5.5cm]{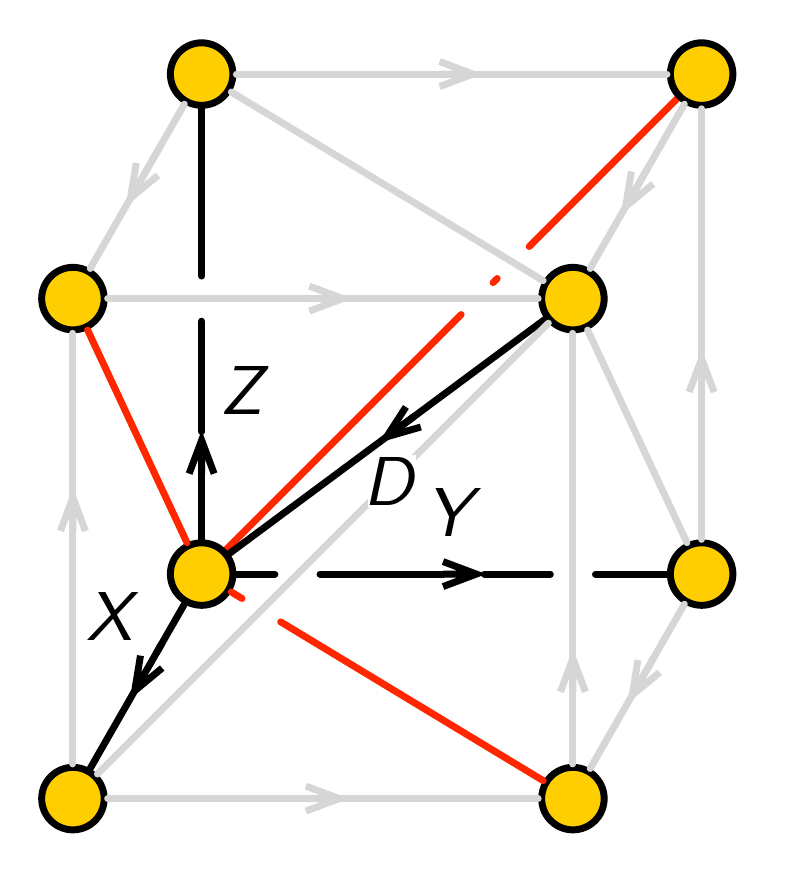}
\caption{A unit cell of the periodic quiver of $\mathbb{C}^4$.}
	\label{fig:bcc}
\end{figure}

It is sometimes useful to visualize the periodic quiver as a tessellation of $\mathbb{R}^3$ by a unit cell. The simplest example of a periodic quiver corresponds to D1-branes over $\mathbb{C}^4$, for which the unit cell is shown in \fref{fig:bcc} \cite{GarciaCompean:1998kh}. All abelian orbifolds of $\mathbb{C}^4$ can be constructed by combining copies of the $\mathbb{C}^4$ unit cell with periodicity conditions determined by the action of the generators of the orbifold group \cite{GarciaCompean:1998kh,Franco:2015tna}.

\begin{figure}[ht!!]
\begin{center}
\resizebox{0.9\hsize}{!}{
\includegraphics[trim=0cm 0cm 0cm 0cm,totalheight=10 cm]{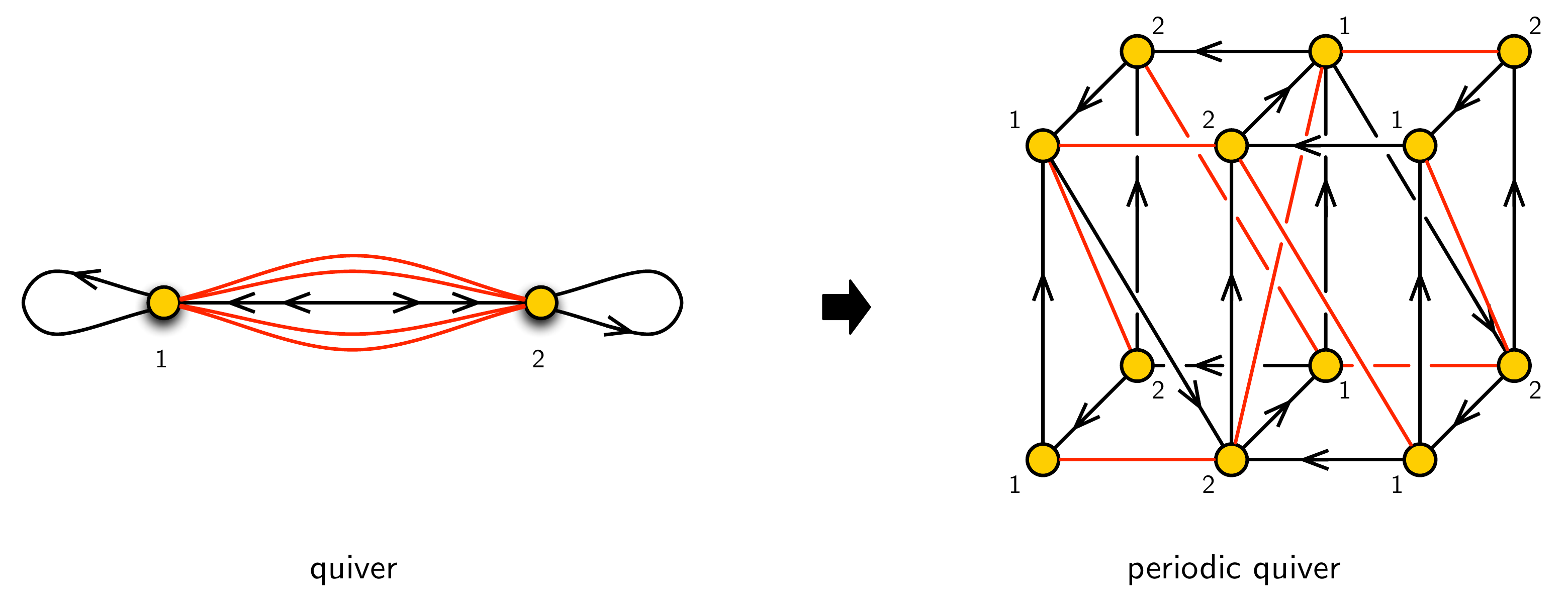}
}  
\caption{
The standard and periodic quivers for $\mathcal{C}\times\mathbb{C}$.
\label{fquivercomp}}
 \end{center}
 \end{figure}

\fref{fquivercomp} shows the periodic quiver for $\mathcal{C}\times\mathbb{C}$, where $\mathcal{C}$ indicates the conifold. Several additional examples of periodic quivers can be found in \cite{Franco:2015tna} and in the following sections.

\subsection{Earlier Constructions: Brane Box Models for Orbifolds}

The construction of brane setups realizing $2d$ $(0,2)$ theories was pioneered in \cite{GarciaCompean:1998kh} with the introduction of \textit{brane box models}, which are reviewed in this section. Brane box models are Type IIA configurations consisting of three types of orthogonal NS5-branes: $\text{NS}$, $\text{NS}^\prime$ and $\text{NS}^{\prime\prime}$-branes, which extend along the $(012345)$, $(012367)$ and $(014567)$ directions, respectively. In addition, there are D4-branes spanning $(01246)$. The $(246)$ directions are compactified on a $T^3$. The $2d$ gauge theories live on the two directions $(01)$ common to all the branes. Each type of NS5-branes breaks SUSY by one half. The D4-branes break SUSY by an additional half leading, generically, to $2d$ $(0,2)$. The NS5-branes divide $T^3$ into cubic ``boxes" within each of which there is a stack of $N_i$ D4-branes, giving rise to a $U(N_i)$ gauge group in the $2d$ theory. All branes sit at the same position in the $(89)$ directions. The $U(1)$ R-symmetry is given by the rotational symmetry in the $(89)$ plane. Table \ref{tbrane} summarizes the brane configuration. 

\begin{table}[ht!]
\centering
\begin{tabular}{l|cccccccccc}
\; & 0 & 1 & 2 & 3 & 4 & 5 & 6 & 7 & 8 & 9 
\\
\hline
$\text{NS}$ & $\times$ & $\times$ & $\times$ & $\times$ & $\times$ & $\times$ & $\cdot$ & $\cdot$ & $\cdot$ & $\cdot$
\\
$\text{NS}^\prime$ & $\times$ & $\times$ & $\times$ & $\times$ &  $\cdot$ & $\cdot$ & $\times$ & $\times$ & $\cdot$ & $\cdot$
\\
$\text{NS}^{\prime\prime}$ & $\times$ & $\times$ & $\cdot$ & $\cdot$ &  $\times$ & $\times$& $\times$ & $\times$  & $\cdot$ & $\cdot$ 
\\
D4 & $\times$ & $\times$ & $\times$ & $\cdot$ &  $\times$ & $\cdot$& $\times$ & $\cdot$  & $\cdot$ & $\cdot$ 
\end{tabular}
\caption{
Brane configuration for brane box models. The $(246)$ directions are compactified on a $T^3$.
\label{tbrane}
}
\end{table}

Brane box models are related to systems of D1-branes over abelian orbifolds of $\mathbb{C}^4$ by T-duality along $(246)$ \cite{GarciaCompean:1998kh}. These configurations are natural generalizations of brane boxes with a single type of NS5-branes (also known as elliptic models), which correspond to the $6d$ theories associated to orbifolds of $\mathbb{C}^2$ \cite{Hanany:1997gh}, and brane boxes with two types of NS5-branes, which describe the $4d$ theories for orbifolds of $\mathbb{C}^3$ \cite{Hanany:1997tb,Hanany:1998it}. 

 \begin{figure}[ht!!]
\begin{center}
\resizebox{0.4\hsize}{!}{
\includegraphics[trim=0cm 0cm 0cm 0cm,totalheight=10 cm]{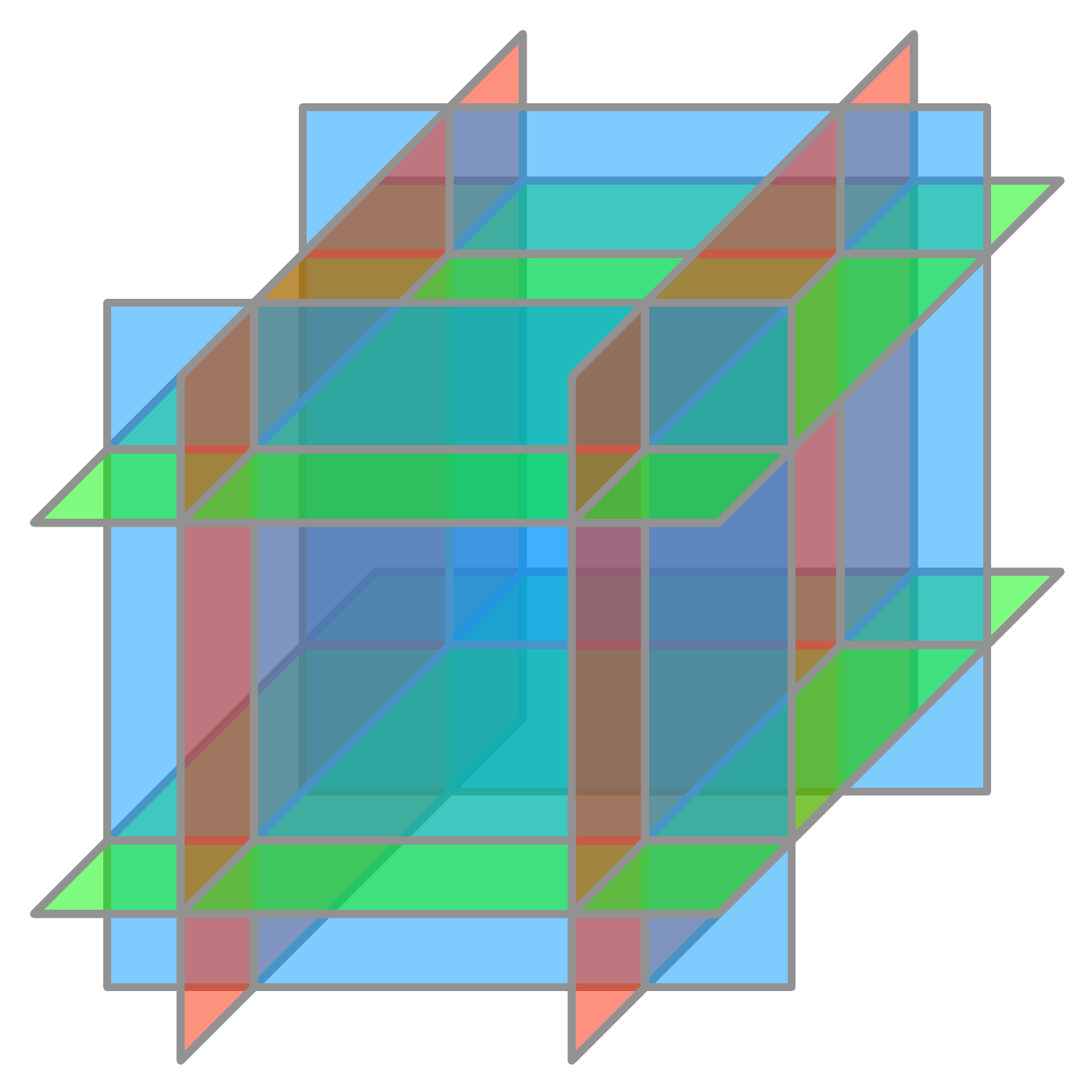}
}  
\caption{
\textit{Brane box models.} Schematic representation of the internal $(2,4,6)$ directions. The blue, red and green planes correspond to $\text{NS}$, $\text{NS}^\prime$ and $\text{NS}^{\prime\prime}$-branes that extend along $(24)$, $(26)$ and $(46)$ directions. D4-branes span the $(246)$ directions, filling the boxes. The geometric action of the dual abelian orbifold of $\mathbb{C}^4$ is translated into the periodicity conditions on $T^3$.
\label{fbranebox}}
 \end{center}
 \end{figure}
 
It is possible to place $k$ $\text{NS}$, $k^\prime$ $\text{NS}^\prime$ and $k^{\prime\prime}$ $\text{NS}^{\prime\prime}$-branes such that they divide the $T^3$ into $k k^\prime k^{\prime\prime}$ boxes. Such a configuration is T-dual to a $\mathbb{C}^4/(\mathbb{Z}_k \times \mathbb{Z}_{k^{\prime}} \times \mathbb{Z}_{k^{\prime\prime}})$ orbifold. The geometric action of the orbifold is encoded in how the brane boxes are periodically identified. \fref{fbranebox} illustrates the basic features of a brane box model along the internal $T^3$.

 \begin{figure}[ht!!]
\begin{center}
\resizebox{0.8\hsize}{!}{
\includegraphics[trim=0cm 0cm 0cm 0cm,totalheight=10 cm]{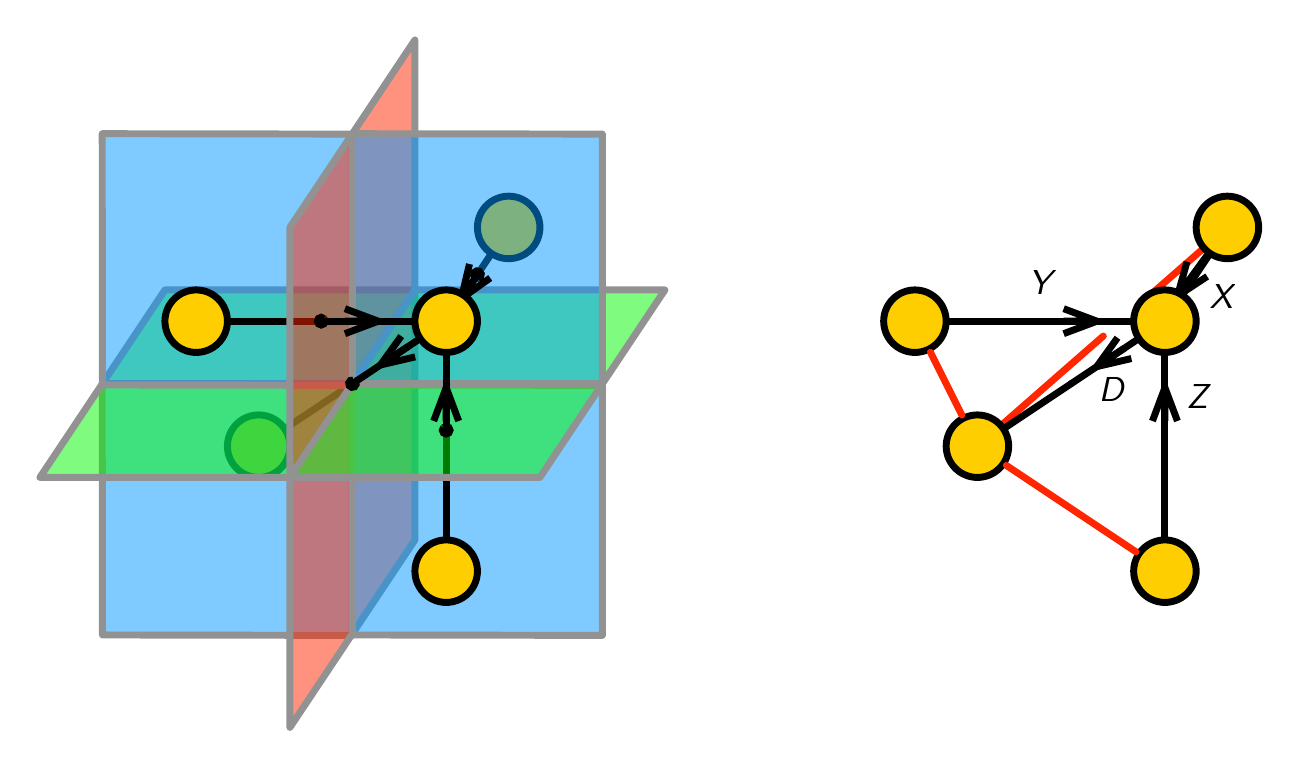}
}  
\caption{
\textit{Brane box models and periodic quivers.} 
This figure presents the brane box model for $\mathbb{C}^4$, which has a single NS5-brane of each type. It also shows how the brane configuration gives rise to the corresponding periodic quiver.
Orbifold models are obtained from this configuration by enlarging the unit cell. 
\label{fd333}}
 \end{center}
 \end{figure}

There is a straightforward translation between brane box models and the corresponding periodic quivers mentioned in section \sref{section_periodic_quiver}. The simplest $\mathbb{C}^4$ theory corresponds to $k=k^{\prime}=k^{\prime\prime}=1$ and hence has a single box. The theory has four chiral fields. Three of them define unit vectors in the $T^3$: $(1,0,0)$, $(0,1,0)$, $(0,0,1)$ and are hence transverse to the NS5-branes as illustrated in \fref{fd333}. We call them $X$, $Y$, $Z$, respectively. The fourth chiral field points in the $(-1,-1,-1)$ direction. In the same basis, the three Fermi fields lie along the $(0,1,1)$, $(1,0,1)$, $(1,1,0)$ directions, i.e. along the diagonals of the square faces of the box.  

Brane box models have several positive features. First, they can be used to deduce the gauge theories associated to arbitrary abelian orbifolds of $\mathbb{C}^4$. In addition, they introduce the basic ingredients for brane configurations that are T-dual to D1-branes at toric singularities as well as some of their key characteristics such as their compactification on $T^3$. 

Despite all these successes, brane box models have several shortcomings. Overcoming them is one of the main goals of this paper. First of all, they do not provide the gauge theories for D1-branes on Calabi-Yau 4-folds beyond orbifolds. Furthermore, there is no one-to-one map between objects in the gauge theory and elements in the brane box model. Most notably, while $X$, $Y$ and $Z$-type chiral fields map to box faces, this is not true for $D$-type chiral fields. Similar arguments apply to the plaquettes involving these fields. In turn, this implies that basic symmetries of the gauge theories are not manifest in brane box models. Finally, brane box models do not relate to combinatorial objects that streamline their connection between CY$_4$ geometry and gauge theory.

In the next section we introduce new constructions that overcome all these limitations. In fact, brane box models can be regarded as degenerate limits of these more general setups.

\section{Brane Brick Models \label{section_brane_brick_models}}

In this section, we introduce {\it brane brick models}, a novel class of brane configurations that provide a direct connection between toric Calabi-Yau 4-folds and the corresponding $2d$ $(0,2)$ gauge theories. Brane brick models play a role analogous to the one of brane tilings, which correspond to $4d$ $\mathcal{N}=1$ gauge theories on D3-branes probing toric Calabi-Yau 3-folds. Not surprisingly, given the peculiarities of $2d$ $(0,2)$ theories, brane brick models exhibit several original features as explained in the following sections.

\subsection{Brane Brick Models as Type IIA Brane Configurations}
\label{section_brane_brick_model_generalities}

Brane brick models are Type IIA brane configurations that share basic features with brane box models \cite{GarciaCompean:1998kh}. They generalize the brane box construction to generic toric Calabi-Yau 4-folds that are not necessarily of $\mathbb{C}^4$.

A brane brick model consists of an NS5-brane and D4-branes. The NS5-brane extends along the $(01)$ directions and wraps a holomorphic surface (i.e. four real dimensions) embedded into the $(234567)$ directions. The directions $(246)$ are periodically identified to form a $T^3$. It is therefore natural to pairwise combine $(23)$, $(45)$ and $(67)$ into three complex variables $x$, $y$ and $z$ of which $(246)$ are the arguments. The surface $\Sigma$ wrapped by the NS5-brane is the zero locus of the Newton polynomial associated to the CY$_4$,
\beq
\sum_{(a,b,c)\in V} c_{(a,b,c)} x^a y^b z^c = 0 \,,
\label{eq_Sigma}
\eeq
where $(x, y, z)$ take values in $(\mathbb{C}^*)^3$ and $V$ is the set of points in the toric diagram on $\mathbb{Z}^3$. Stacks of D4-branes extend along $(01)$ and are suspended within each of the voids cut out by the NS5-brane surface within the $(246)$ 3-torus. Generically, the holomorphically embedded NS5-brane breaks 1/8 of the SUSY, while the D4-branes break an additional 1/2, resulting in a $2d$ $(0,2)$ theory in the two common dimensions $(01)$. As for brane boxes, the NS5-brane and the D4-branes sit at the same point in the transverse $(89)$ dimensions and the $U(1)$ R-symmetry of the gauge theory is geometrically realized as rotations on this plane.

From now on, we will primarily focus on a simpler object, which is obtained by replacing $\Sigma$ by its skeleton or tropical limit. For simplicity, we will also refer to this object as the brane brick model. In this limit, $\Sigma$ is replaced by a collection of $2d$ faces that separate $T^3$ into a collection of $3d$ polytopes filled by D4-branes. We call the $3d$ polytopes ${\it bricks}$.

\subsection{The Brane Brick Model - Gauge Theory Dictionary}

\label{section_dictionary}

\begin{figure}[ht!!]
\begin{center}
\resizebox{0.6\hsize}{!}{
\includegraphics[trim=0cm 0cm 0cm 0cm,totalheight=10 cm]{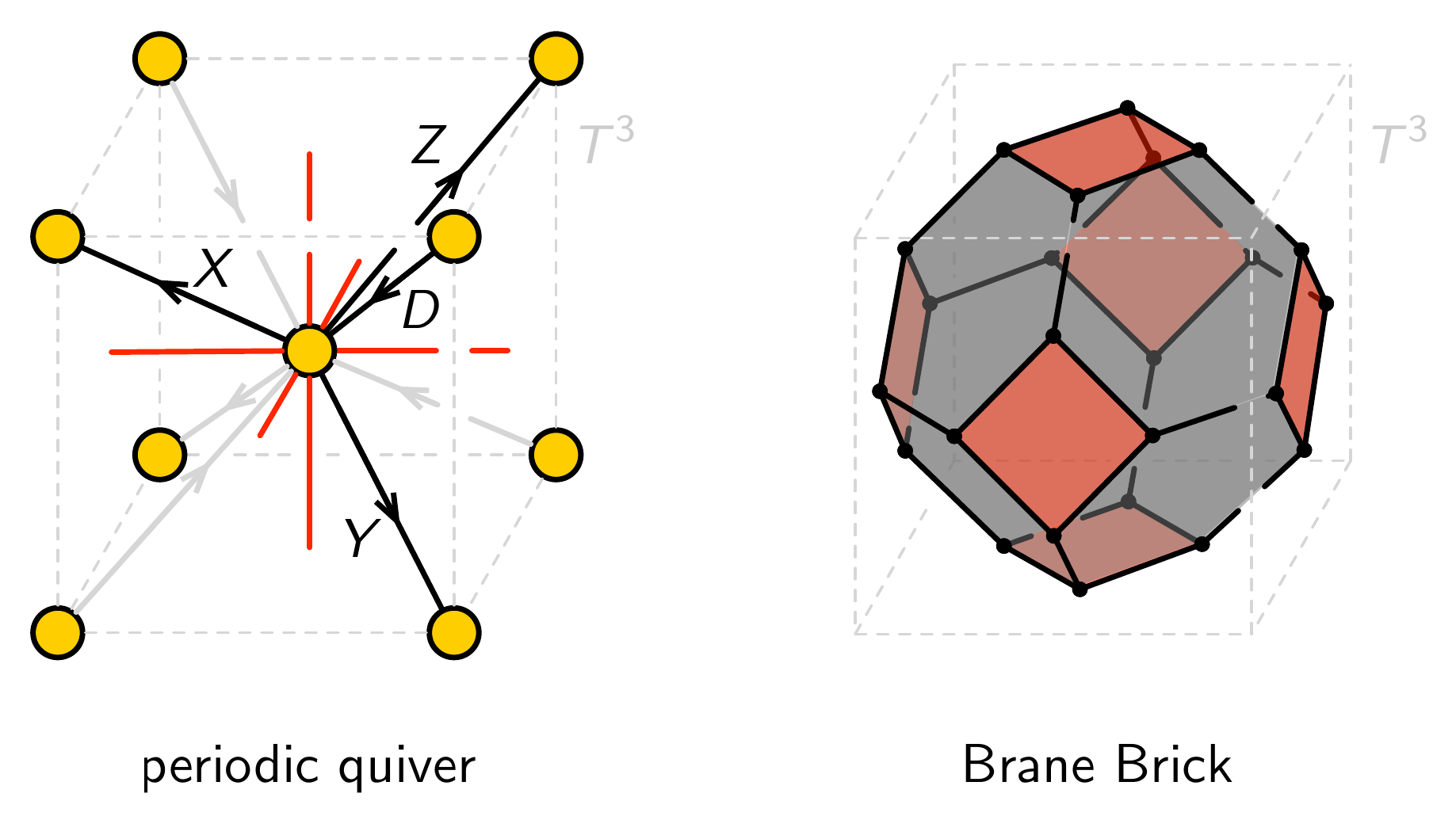}
}  
\caption{
The periodic quiver and dual brane brick model $T^3$ for the $\mathbb{C}^4$ theory.
\label{ft3diagrams}}
 \end{center}
 \end{figure}

In section \sref{section_periodic_quiver}, we explained how the periodic quiver combines the quiver and $J$- and $E$-terms of a $2d$ $(0,2)$ gauge theory into a single object. In analogy to the connection between brane tilings and periodic quivers for $4d$ $\mathcal{N}=1$ toric gauge theories \cite{Franco:2005rj}, brane brick models can be constructed from the periodic quiver by graph dualization. Both constructions therefore contain precisely the same information. The dualization procedure for $\mathbb{C}^4$ is illustrated in \fref{ft3diagrams}. 

The brane brick model for $\mathbb{C}^4$ contains a single brick, which corresponds to the only gauge group of the theory. This brick takes the form of a truncated octahedron consisting of eight hexagonal and four square faces, which correspond to chiral and Fermi fields, respectively.\footnote{Truncated octahedra have appeared in a similar context in \cite{2010arXiv1001.4858F}.} More generally, orbifolds of $\mathbb{C}^4$ are obtained by tessellating $T^3$ with additional copies of the same type of brick. From now on, motivated by the convention for quivers, we will use black faces to indicate chiral fields and red ones to indicate Fermi fields. For $\mathbb{C}^4$, the faces of the brick are pairwise identified in $T^3$ resulting, as expected, in four chiral fields and three Fermi fields in the adjoint representation of the gauge group. We can regard brane box models as degenerate limits of brane brick models for $\mathbb{C}^4$ and its orbifolds in which some faces shrink to zero size. 

Plaquettes in the periodic quiver correspond to edges in a brane brick model. The toric condition thus implies that Fermi fields correspond to squares. The converse is however not true and certain geometries lead to brane box models with square chiral faces. The four edges of a Fermi face split into two pairs, each of them contributing to a $J$- or $E$-term. For the $\mathbb{C}^4$ orbifold examples, one Fermi and two chiral faces meet at every edge. This arrangement captures the structure of plaquettes in these theories and gives rise to the corresponding $J$- and  $E$-terms, all of which involve a pair of quadratic terms in chiral fields \cite{GarciaCompean:1998kh,Franco:2015tna}.

The basic dictionary between brane brick models and the corresponding gauge theories is summarized in \tref{tbrick}. 

\bigskip

\begin{table}[h]
\centering
\begin{tabular}{|l|l|}
\hline
\ \ \ \ \ \ {\bf Brane Brick Model} \ \ \ \ \ & \ \ \ \ \ \ \ \ \ \ \ \ \ \ \ \ \ \ \ \ {\bf Gauge Theory} \ \ \ \ \ \ \ \ \ \ \ \ 
\\
\hline\hline
Brick  & Gauge group \\
\hline
Oriented face between bricks & Chiral field in the bifundamental representation \\
$i$ and $j$ & of nodes $i$ and $j$ (adjoint for $i=j$) \\
\hline
Unoriented square face between & Fermi field in the bifundamental representation \\
bricks $i$ and $j$ & of nodes $i$ and $j$ (adjoint for $i=j$) \\
\hline
Edge  & Plaquette encoding a monomial in a \\ 
& $J$- or $E$-term \\
\hline
\end{tabular}
\caption{
Dictionary between brane brick models and gauge theories.
\label{tbrick}
}
\end{table}

This article focuses only on topological properties of brane brick models. Whether there are preferred shapes for them and their significance is an interesting question that we postpone for further studies. A comparable example for brane tilings is given by {\it isoradial embeddings} and, in particular, the one encoding superconformal R-charges \cite{Kennaway:2007tq}. We now elaborate on some entries of the dictionary in further detail.

\paragraph{Chirality.} 


Let us explain how brane brick models incorporate chirality, namely how to assign orientations to their faces. As usual, the orientation of a face can be translated into the orientation of its edges. By convention, if all edges in the perimeter of a face are oriented clockwise/counterclockwise, as seen from the interior of a brick, we say that it corresponds to a chiral field in the dual periodic quiver pointing towards the exterior/interior of the brick. 

\begin{figure}[ht!!]å
\begin{center}
\resizebox{0.8\hsize}{!}{
\includegraphics[trim=0cm 0cm 0cm 0cm,totalheight=10 cm]{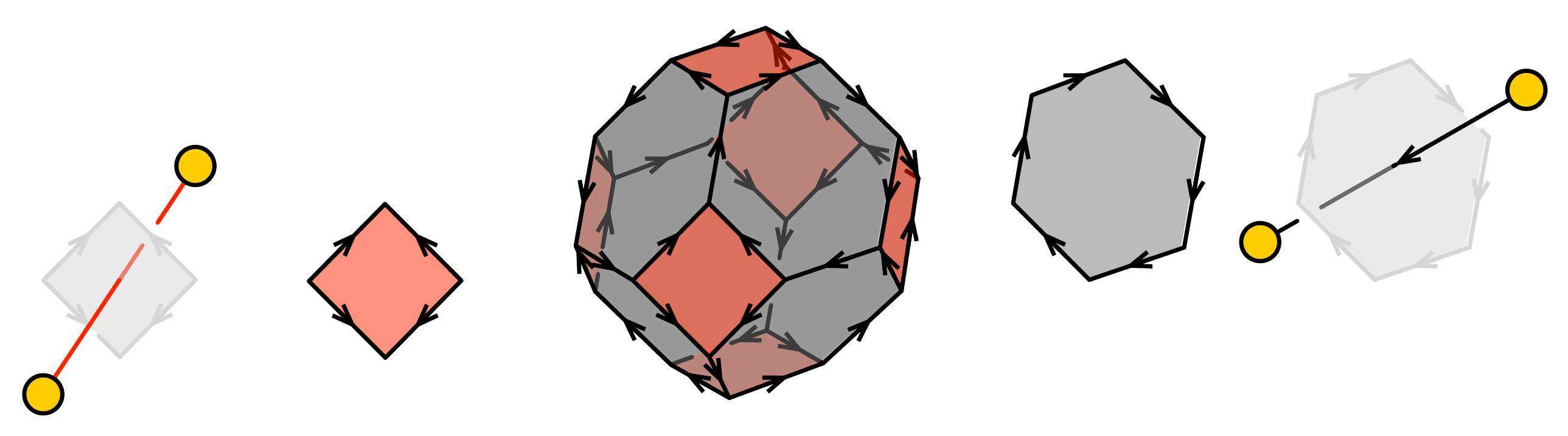}
}  
\caption{Faces in a brane brick model can be systematically oriented starting from one associated to a chiral field. Faces corresponding to chiral fields are oriented while faces corresponding to Fermi fields are not.
\label{fbrickorientation}}
 \end{center}
 \end{figure}

The entire brane brick model can be systematically oriented as follows. We start from a face associated to a chiral field and assign to it the corresponding orientation. We then continue consistently orienting adjacent faces, whenever possible, until covering all edges. At the end of this process, some square faces will turn out not to have a definite orientation. These unoriented faces are precisely those that correspond to Fermi fields, which lack a notion of chirality. \fref{fbrickorientation} illustrates this procedure for $\mathbb{C}^4$. A more intrinsic algorithm for identifying unoriented faces corresponding to Fermi fields, which does not depend on the initial choice of a face associated to a chiral field, is presented in section \sref{sgeomtobrick}.

\paragraph{Edges, Plaquettes and Fermi Fields.} 

\begin{figure}[ht!!]
\begin{center}
\resizebox{0.8\hsize}{!}{
\includegraphics[trim=0cm 0cm 0cm 0cm,totalheight=10 cm]{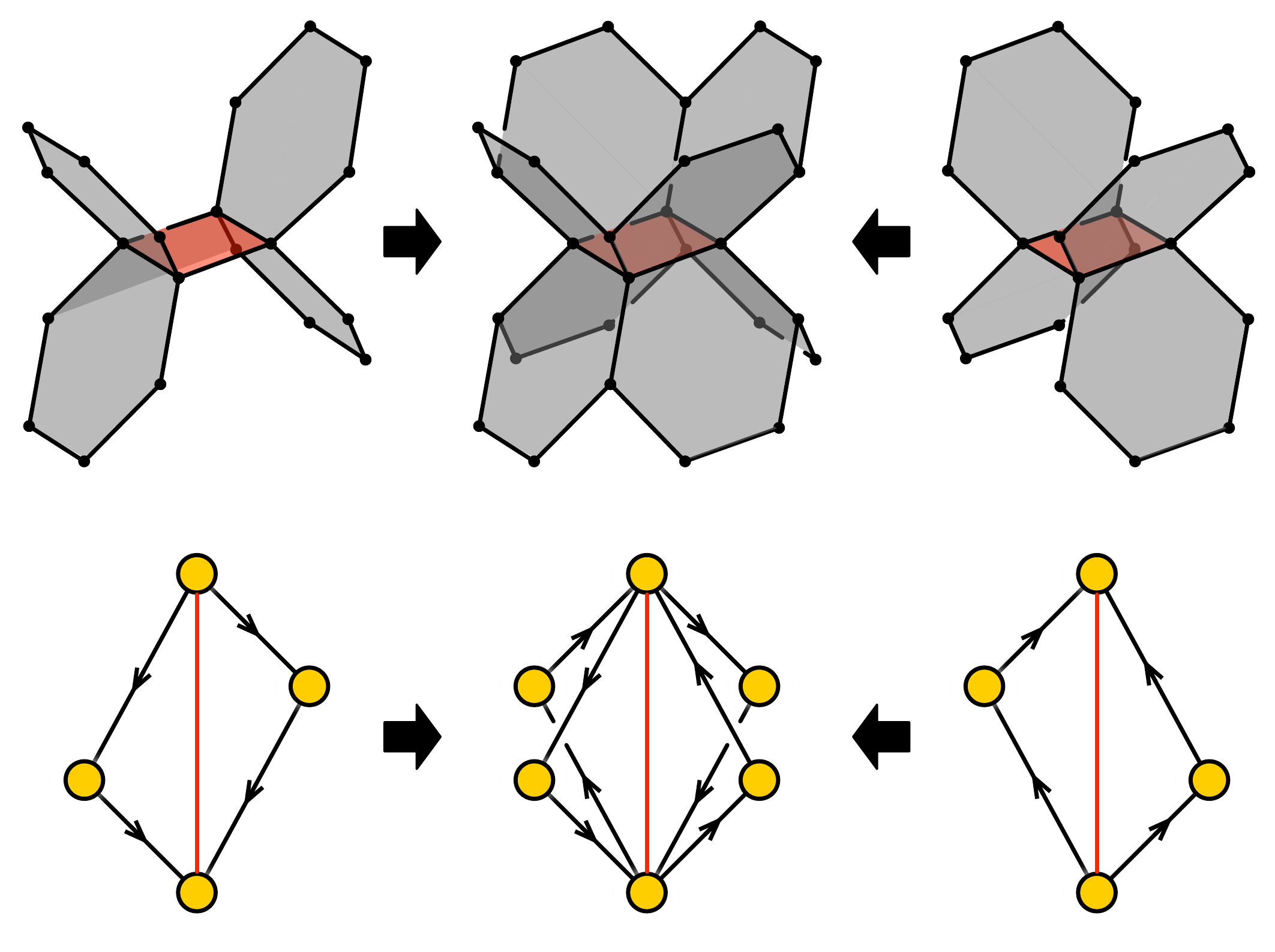}
}  
\caption{
The four plaquettes corresponding to a Fermi field face in brane brick models for $\mathbb{C}^4$ and its orbifolds. 
\label{fplaquette}}
 \end{center}
 \end{figure}

As previously mentioned, every edge in a brane brick model corresponds to a plaquette. Since every plaquette is associated to a Fermi field, we conclude that every edge is the boundary of at least one Fermi face. This is illustrated in \fref{fplaquette}, which shows the neighborhood of a Fermi face in the brane brick models for $\mathbb{C}^4$ and its orbifolds. While the faces associated to the initial and final chiral field in a plaquette must share the corresponding edge with the Fermi, intermediate chiral fields may not, as we now explain.\footnote{In the case of linear contributions to $J$- or $E$-terms associated to chiral-Fermi massive pairs, the initial and final chiral field in the corresponding plaquette coincide.} 

\begin{figure}[ht!!]
\begin{center}
\resizebox{1\hsize}{!}{
\includegraphics[trim=0cm 0cm 0cm 0cm,totalheight=10 cm]{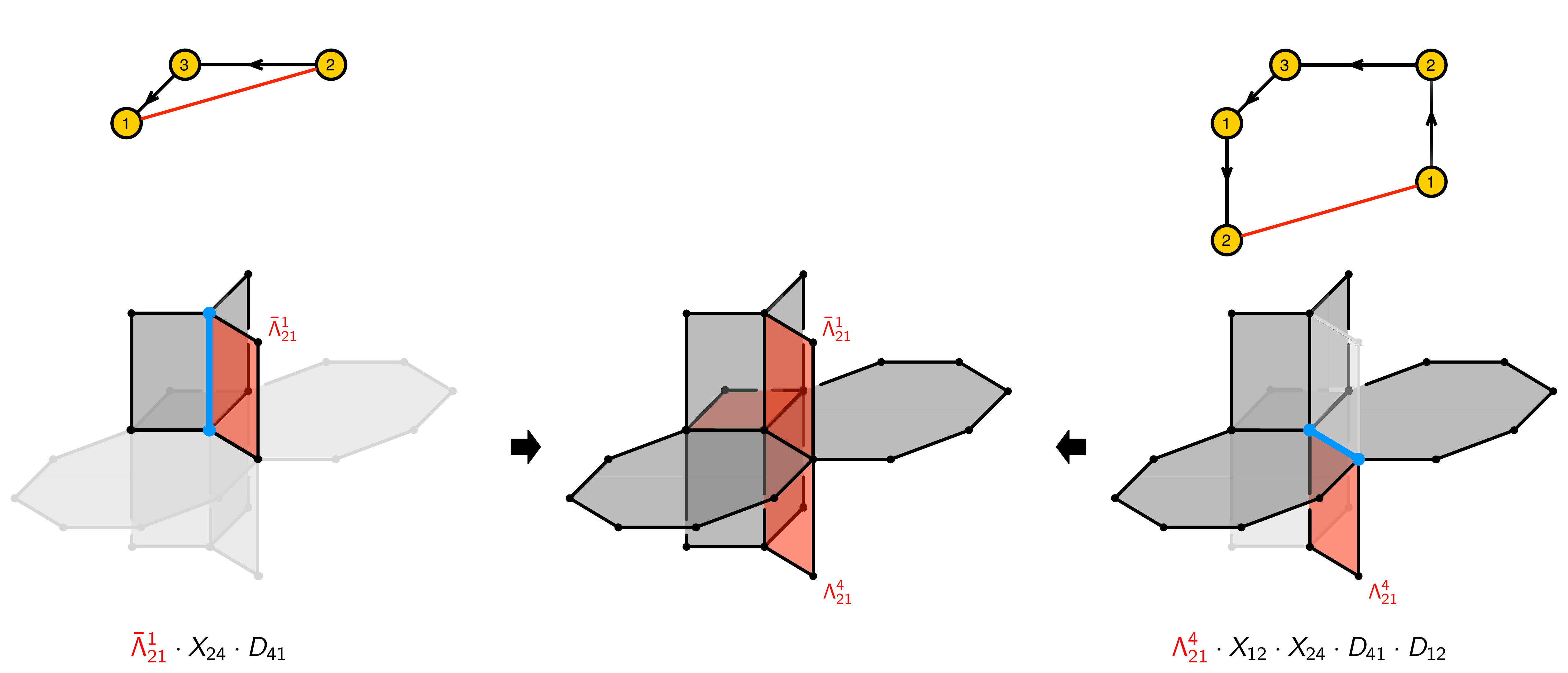}
}  
\caption{
Two adjacent Fermi fields in the brane brick model for $Q^{1,1,1}$ and two of the corresponding plaquettes. The chiral fields in the small plaquette are contained inside the second one.
\label{fq111embedded}}
 \end{center}
 \end{figure}

It is possible for more than one Fermi face to be adjacent to the same edge. This is the case when the chiral fields in a plaquette are a subset of those in a larger plaquette. This phenomenon occurs in $Q^{1,1,1}$ for which a brane brick model will be studied in detail in section \sref{sq111}. The $J$- and $E$-terms of this theory contain the following contributions
\beal{es00a1}
J_{21}^{4-} = X_{12} \cdot X_{24} \cdot D_{41} \cdot D_{12} ~,~
E_{21}^{1-} = X_{24} \cdot D_{41} ~.~
\eea
We see that $E_{21}^{1-} \subset J_{21}^{4-}$. The corresponding plaquettes are shown in \fref{fq111embedded} and are associated to adjacent Fermi fields $\Lambda_{21}^4$ and $\bar\Lambda_{21}^{1}$.

\bigskip

To conclude this section, let us note that there are several ways of constructing brane brick models. First, they can be obtained by dualizing the periodic quiver, as explained in this section. In addition, they can be systematically obtained by partial resolution/higgsing from known ones, such as the ones for $\mathbb{C}^4$ orbifolds. Finally, it is possible to construct them directly from the probed geometry by means of a procedure which we call the \textit{fast inverse algorithm}, as explained in section \sref{sgeomtobrick}.

\subsection{Mass Terms and Higgsing}

\begin{figure}[!ht]
\begin{center}
\includegraphics[width=10cm]{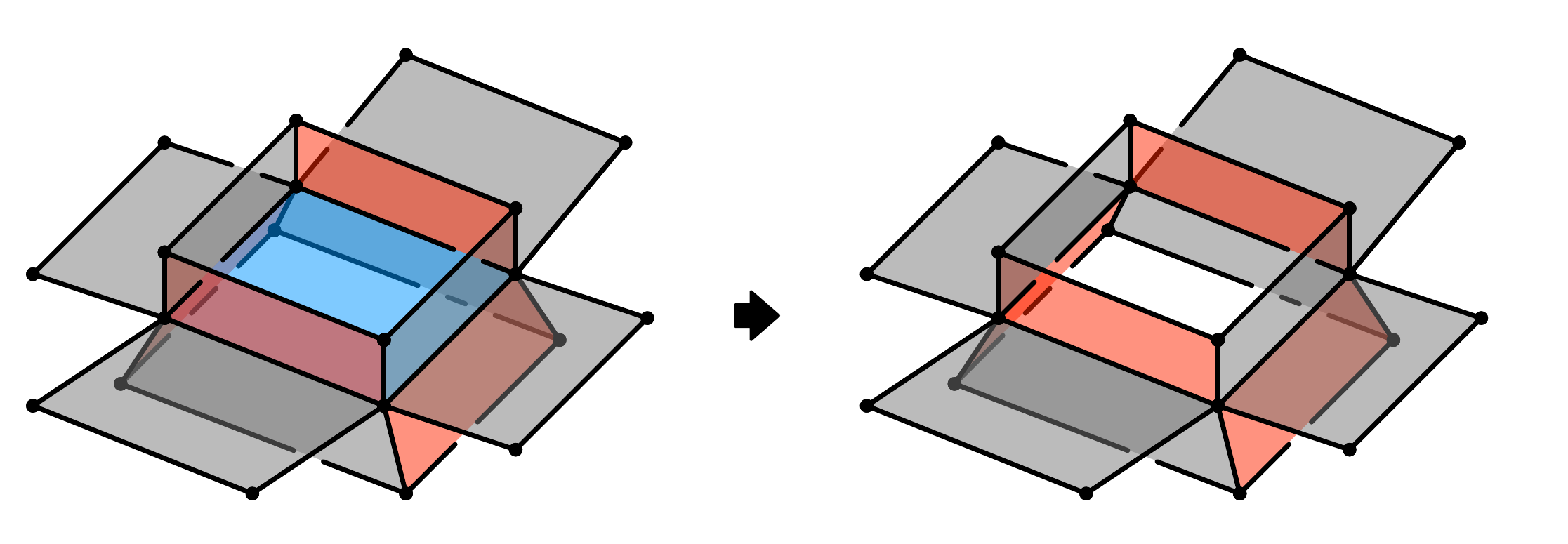}
\caption{Giving a non-zero VEV to a chiral field maps to deleting the corresponding face, here shown in blue, in the brane brick model. This results in the combination of two adjacent bricks into a single one.}
\label{brickhiggs}
\end{center}
\end{figure}

\paragraph{Higgsing.} When a non-zero VEV is turned on for the scalar component of a bifundamental chiral field $X_{ij}$, the gauge groups associated to quiver nodes $i$ and $j$ are higgsed to the diagonal subgroup. From the perspective of the brane brick model, this amounts to removing the face associated to $X_{ij}$, which results in the combination of the bricks for nodes $i$ and $j$ into a single one, as schematically shown in \fref{brickhiggs}. Deleting the face for $X_{ij}$ also has the desired effect of replacing it by its VEV, which without loss of generality is taken to be equal to 1, in all the plaquettes containing it. 

\paragraph{Massive Fields.} Massive fields correspond to Fermi-chiral pairs extending between the same pair of gauge groups, such that either the $J$- or $E$-term for the Fermi field contains a term that is linear in the chiral field. In the brane brick model, these linear terms are represented by edges that are connected to a single Fermi face and a single chiral face. We refer to such edges as {\it massive edges}.

\begin{figure}[ht!]
\begin{center}
\resizebox{0.95\hsize}{!}{
\includegraphics[trim=0cm 0cm 0cm 0cm,totalheight=10 cm]{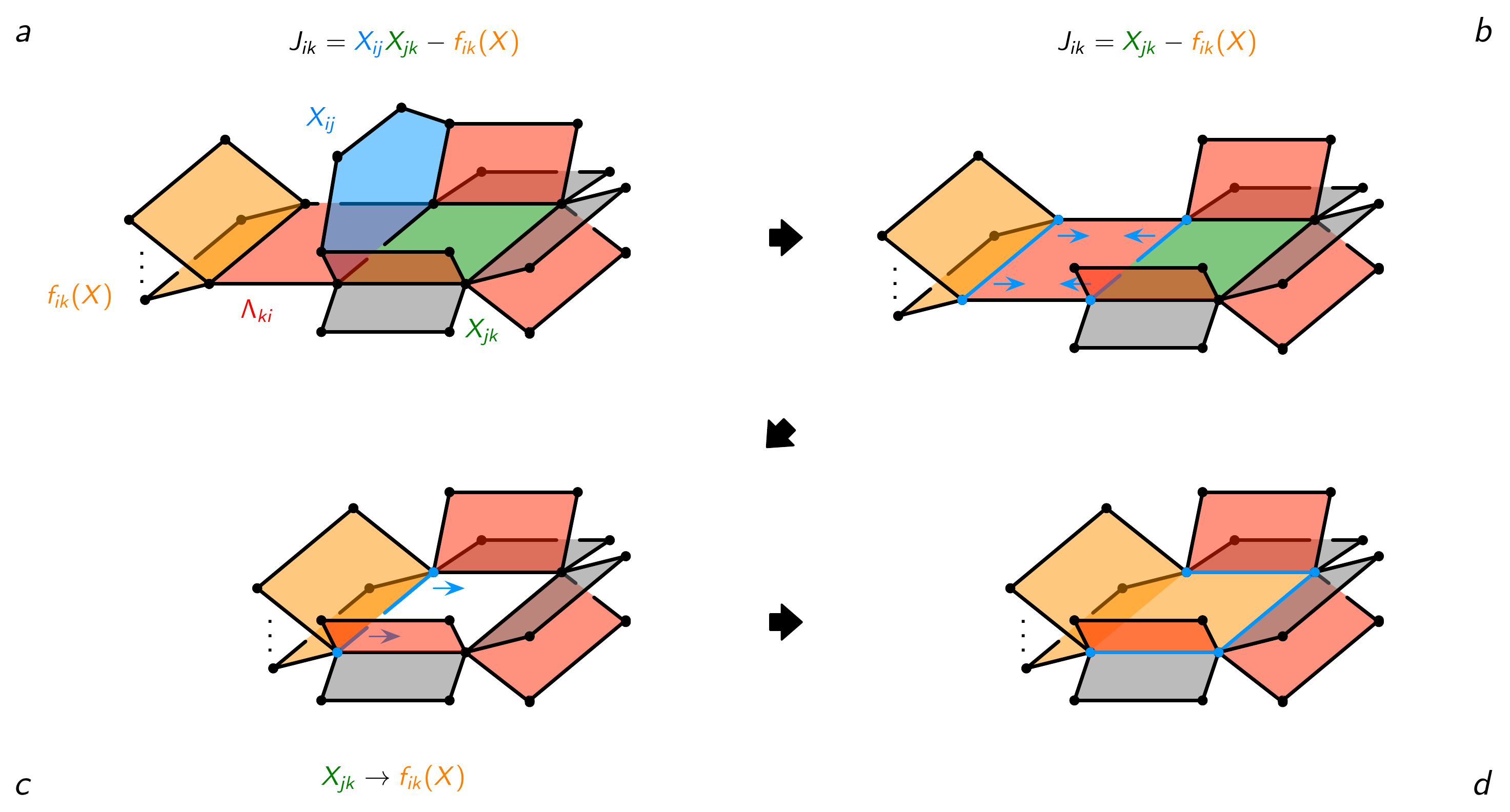}
}  
\caption{\textit{Generating a massive Fermi-chiral pair by higgsing and integrating it out.}
(a) Starting from $J_{ik} = X_{ij} X_{jk} - f_{ik}(X)$, a VEV for $X_{ij}$ generates a massive edge connecting $\Lambda_{ki}$ to $X_{jk}$, as shown in (b). (c) The massive edge and the opposite one in the face for $\Lambda_{ki}$ are merged, making the face disappear. (c) The replacement $X_{jk}\rightarrow f_{ik}(X)$ removes the face for $X_{jk}$ and (d) glues its edges to the one for $f_{ik}(X)$.
\label{fbrickhiggs2}}
 \end{center}
 \end{figure}

Massive fields can be generated in a variety of ways. Higgsing is one of them. In this case, a massive pair arises when an originally quadratic $J$- or $E$-term becomes linear after turning on a VEV. In the brane brick model, such terms correspond to edges that initially are attached to a Fermi and two chiral faces. When the face associated to the field acquiring the non-zero VEV is deleted, a massive edge is generated, as shown in \fref{fbrickhiggs2} (a).

As an example, let us consider a quadratic $J$-term for the Fermi field $\Lambda_{ki}$ that becomes linear when the chiral field $X_{ij}$ receives a VEV as follows\footnote{The case of a linear $E$-term is identical.}
\beal{es900a1}
J_{ik} = X_{ij} X_{jk} - f_{ik}(X)~~~ \longrightarrow~~~ X_{jk} - f_{ik}(X) ~.~
\eea
The gauge indices work properly, since nodes $i$ and $j$ are identified by the higgsing. Recall that all Fermi faces are squares. In \eref{es900a1}, $f_{ij}(X)$ indicates a product of chiral fields associated with one of the edges attached to $\Lambda_{ki}$. The linear term $X_{jk}$ in $J_{ik}$ corresponds to the opposite edge on the $\Lambda_{ki}$ face. This massive edge is indeed attached to the two massive fields: $\Lambda_{ki}$ and $X_{jk}$.

At low energies, $\Lambda_{ki}$ and $X_{jk}$ can be integrated out. In this process, the terms $J_{ik}$ and $E_{ki}$ associated to $\Lambda_{ki}$ are removed from the Lagrangian. This is nicely captured by the brane brick model as shown in \fref{fbrickhiggs2}, from steps (b) to (d). When integrating out the massive fields, $J_{ik}$ is set to zero and we replace $X_{jk}\rightarrow f_{ik}(X)$. For clarity, it is convenient to split this process into two stages, although no physical interpretation should be assigned to the intermediate step (c). The first step, shown in (c), corresponds to shrinking the face associated to $\Lambda_{ki}$ until the massive edge and the opposite one merge into a single one that we associate to $f_{ik}(X)$. When doing so, the two other edges of $\Lambda_{ki}$, which represent $E_{ki}$, also disappear. Finally, in step (d), the face for $X_{jk}$ is removed and the edges that formed its perimeter, with the exception of the massive edge, are glued to the one for $f_{ik}(X)$. This implements the replacement $X_{jk}\rightarrow f_{ik}(X)$ in all $J$- and $E$-terms.

Let us conclude this section with a few additional remarks regarding the brane brick models obtained by integrating out massive fields. The procedure summarized in \fref{fbrickhiggs2} faithfully incorporates all pertinent manipulations of the gauge theory. The two chiral faces shown in orange end up having three consecutive common edges. This might naively seem a little odd, since it may require curved brick faces. This, on its own, is not an issue at the level of discussion in the current paper, since we are only concerned with the combinatorial properties of brane brick models. More importantly, this feature can be simply avoided if the three edges are collinear or, as it occurs in several of the explicit examples we have studied, additional fields become simultaneously massive. In the latter case, integrating out all massive fields leads to configurations in which no pair of chiral faces is glued along three consecutive edges.

\section{Brane Brick Models from Geometry\label{sgeomtobrick}}

This section studies the geometry of the brane brick model in further detail. This analysis will result in a new method for constructing the brane brick model directly from the underlying Calabi-Yau 4-fold. This procedure is a natural generalization of the {\it fast inverse algorithm} for brane tilings, which constructs the tiling from zig-zag paths \cite{Hanany:2005ss,Feng:2005gw}. We refer to the algorithm for brane brick models in the same way. 

\subsection{Amoebas and Coamoebas}

As explained in section \sref{section_brane_brick_model_generalities}, the NS5-brane in a brane brick model wraps a holomorphic surface $\Sigma$ defined as the zero locus of the Newton polynomial of the toric CY$_4$ cone. Reproducing \eref{eq_Sigma} here for convenience, $\Sigma$ is hence defined by
\begin{align}
 \sum_{(a,b,c)\in V} c_{(a,b,c)} x^a y^b z^c = 0 \, .
\end{align}
$\Sigma$ is 2-complex dimensional, i.e. 4-real dimensional. 

\begin{figure}[ht!!]
\begin{center}
\resizebox{0.6\hsize}{!}{
\includegraphics[trim=0cm 0cm 0cm 0cm,totalheight=10 cm]{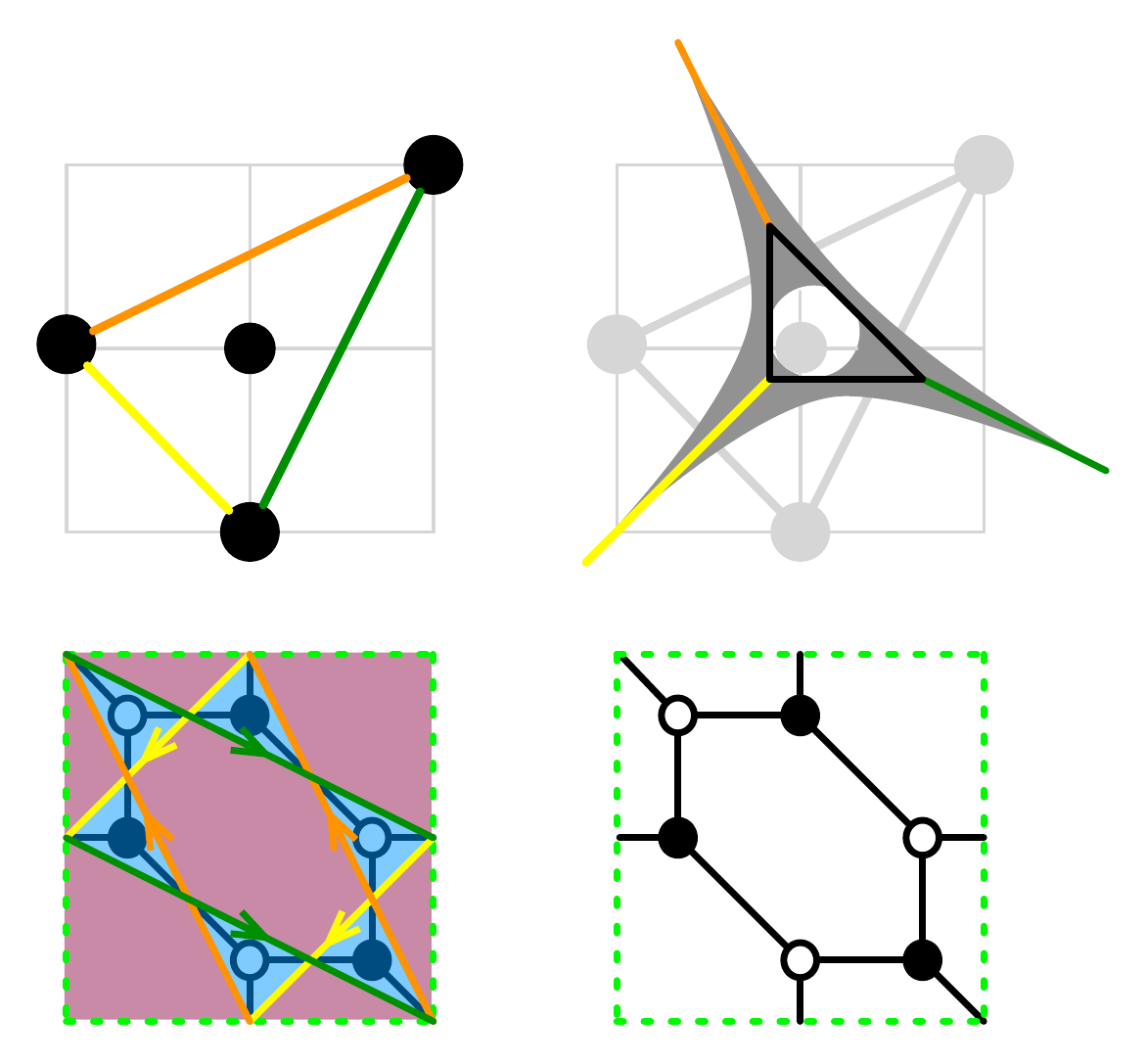}
}  
\caption{
The toric diagram, the amoeba, a singular ``approximation" to the coamoeba determined by the zig-zag paths (shown in color) and the brane tiling for $\text{dP}_0$.
\label{fdp0amoeba}}
 \end{center}
 \end{figure}

Two natural projections help us to visualize and to study $\Sigma$. The first one, called the {\it amoeba} \cite{2004math......3015M,2001math......8225M}, is a projection onto $(\log |x|, \log |y|, \log |z|) \in  \mathbb{R}^3$. The amoeba is a smooth geometric object dual to the toric diagram. The second projection, the {\it coamoeba}, maps $\Sigma$ onto $(\mathrm{arg}(x),\mathrm{arg}(y),\mathrm{arg}(z)) \in T^3$. The coamoeba captures the geometry of the NS5-brane in the internal $T^3$ and hence contains the non-trivial information necessary to define the corresponding quiver gauge theory. Both the amoeba and the coamoeba are 3-real dimensional objects. Visualizing them is challenging, but it is possible to get a flavor of their general structure by considering their analogues in the simple case of toric Calabi-Yau 3-folds with $2d$ toric diagrams. Such objects have been studied in the physics context in \cite{Feng:2005gw} and \fref{fdp0amoeba} presents a simple example. 

We refer to the points at the corners of toric diagrams, both for Calabi-Yau 3-folds and 4-folds, as {\it extremal points}. In order to simplify our presentation, in the rest of the paper we assume that toric diagrams do not contain additional intermediate points on the edges connecting pairs of extremal points. All our ideas, however, extend to the general situation in which such points are present.

The example in \fref{fdp0amoeba} illustrates the general fact that for Calabi-Yau 3-folds the amoeba goes to infinity along ``legs" that approach a linear behavior normal to the sides of the toric diagram. Similarly, in the Calabi-Yau 4-fold case the amoeba contains a leg for every edge of the $3d$ toric diagram, along which it asymptotes to a plane normal to the edge under consideration. More concretely, for two extremal points in the toric diagram with coordinates $(m_x,m_y,m_z)$ and $(n_x,n_y,n_z)$, the equation defining $\Sigma$ simplifies to an equation consisting of only two terms 
\beq
c_{(m_x,m_y,m_z)} x^{m_x} y^{m_y} z^{m_z}+ c_{(n_x,n_y,n_z)} x^{n_x} y^{n_y} z^{n_z}=0
\eeq
when (some of) $(|x|,|y|,|z|)$ go to infinity along the corresponding leg of the amoeba. In this limit, the amoeba projection of $\Sigma$ becomes the $2d$ plane given by
\beq
(n_x-m_x,n_y-m_y,n_z-m_z)\cdot (\log|x|,\log|y|,\log|y|)=0 \, ,
\eeq
up to a shift controlled by the values of the $c_{(m_x,m_y,m_z)}$ and $c_{(n_x,n_y,n_z)}$ coefficients.

Similarly, the coamoeba simplifies considerably when approaching infinity along each leg of the amoeba, also becoming a $2d$-plane orthogonal to the corresponding edge of the toric diagram, in this case living in $T^3$. We refer to each of these planes as a {\it phase boundary}.\footnote{Here we use a nomenclature that is closer to the standard one in the mathematical literature. In a previous work \cite{Franco:2015tna}, we referred to these objects as coamoeba boundaries or coamoeba planes.} Each phase boundary is thus controlled by a pair of extremal points in the toric diagram and given by
\beq
(n_x-m_x,n_y-m_y,n_z-m_z)\cdot (\mathrm{arg}(x),\mathrm{arg}(y),\mathrm{arg}(z))=0 \, .
\eeq
 Once again, each plane can be shifted by tuning the coefficients in the Newton polynomial. In other words, phase boundaries are planes in $T^3$ with winding numbers $(n_x-m_x,n_y-m_y,n_z-m_z)$. More generally, accounting for dual gauge theories sometimes requires deforming phase boundaries while preserving their homology. This possibility will be explored in a forthcoming publication \cite{topub2}. Phase boundaries are the brane brick model analogues of zig-zag paths in brane tilings.

\begin{figure}[ht!!]
\begin{center}
\includegraphics[trim=0cm 0cm 0cm 0cm,totalheight=4 cm]{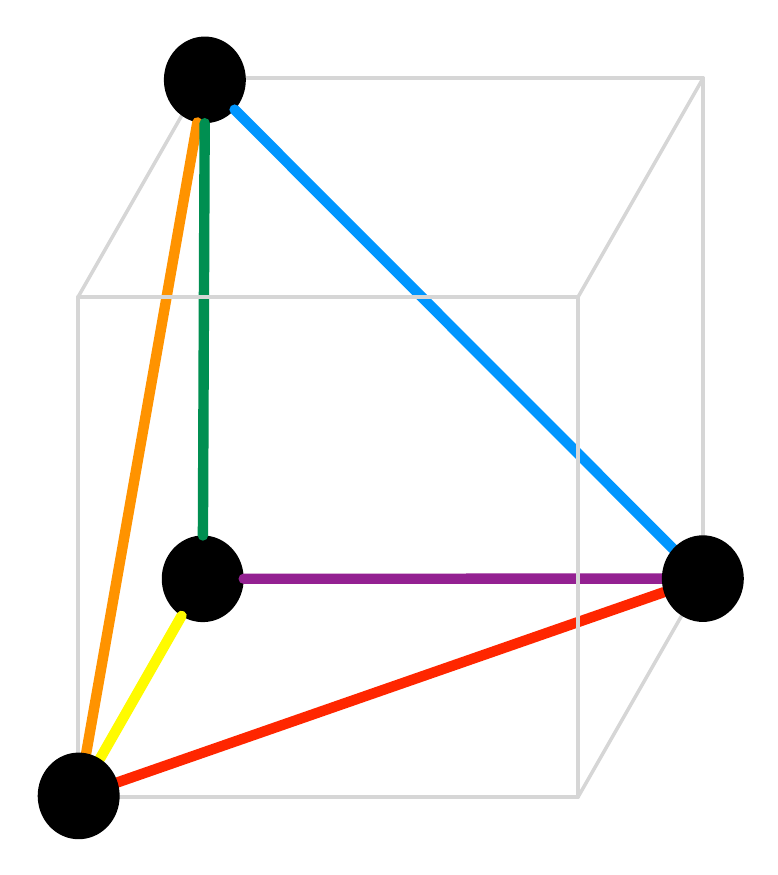}
\caption{Toric diagram for $\mathbb{C}^4$.
\label{c4toriccoamoeba2-copy}}
 \end{center}
 \end{figure}

The union of all phase boundaries contains a proxy for the boundary of the coamoeba, which gets replaced by a collection of planar facets. It is thus possible to discuss how phase boundaries divide $T^3$ into two types of regions, corresponding to the interior and the complement of the coamoeba. In order to illustrate our previous discussion, let us consider the $\mathbb{C}^4$ example, for which the toric diagram is shown in \fref{c4toriccoamoeba2-copy}. In this particular case, all coefficients in the Newton polynomial can be removed by rescalings and we have
\begin{align}
1+x+y+z=0 \,.
\end{align} 

\begin{figure}[ht!!]
\begin{center}
\resizebox{0.6\hsize}{!}{
\includegraphics[trim=0cm 0cm 0cm 0cm,totalheight=10 cm]{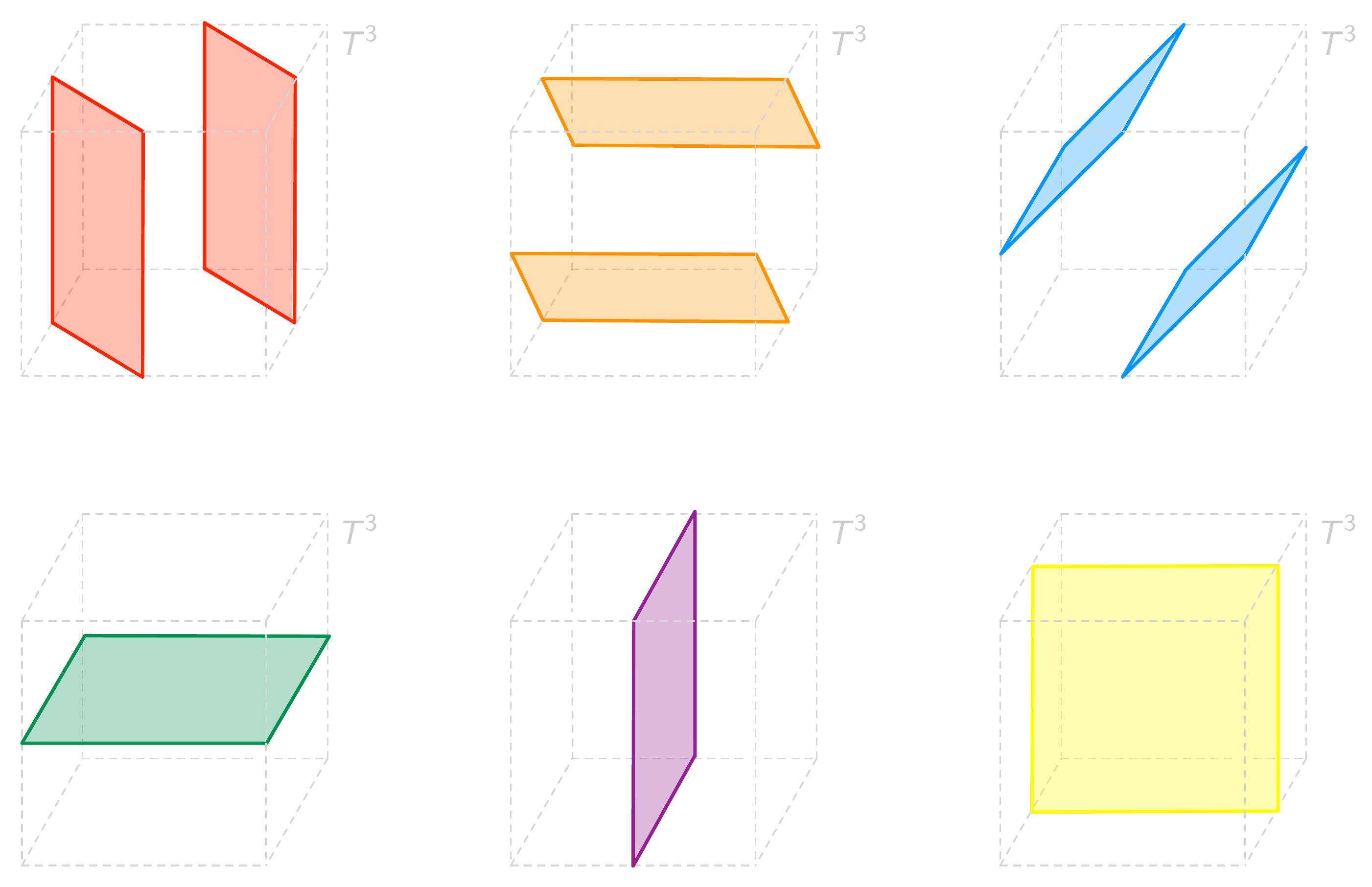}
}  
\caption{The six phase boundaries in $T^3$ corresponding to the six edges of the toric diagram of $\mathbb{C}^4$. We use the same colors for the planes and their normal edges in \fref{c4toriccoamoeba2-copy}. 
\label{fc4planesB}}
 \end{center}
 \end{figure}

Each of the six edges of the toric diagram in \fref{c4toriccoamoeba2-copy} gives rise to a phase boundary, as shown in \fref{fc4planesB}. Phase boundaries are presented in the same colors of the corresponding edges in the toric diagram.

The six phase boundaries divide $T^3$ into various regions, which correspond to either the interior or the complement of the coamoeba. In the example at hand, they carve out a single {\it rhombic dodecahedron} (RD)\footnote{Coamoeba and their phase boundaries on $T^3$ have been studied in the mathematical literature \cite{2011arXiv1110.1033N}.} in the complement, as shown in \fref{fc4coamoeba}.\footnote{Stating that the RD is in the complement of the coamoeba requires a criterion for, starting from knowledge of the phase boundaries, identifying the complement of the coamoeba from its interior. In this simple example, we can take a shortcut by exploiting our knowledge of the periodic quiver, since by definition its nodes live in the complement of the coamoeba. Below we will introduce an algorithm for making this identification in generic theories.} 

\subsection{Periodic Quivers from Coamoebas}

The periodic quiver, or equivalently the brane brick model, can be constructed from the phase boundaries of the corresponding Calabi-Yau 4-fold.  In short, there is a node in the quiver for every component into which the phase boundaries split the complement of the coamoeba and chiral and Fermi fields arise at every point intersection of three or more phase boundaries.\footnote{As for zig-zags in the case of brane tilings, phase boundaries need to be shifted in $T^3$ until reaching a configuration that can be interpreted as consistent gauge theory.} We refer to this procedure as the fast inverse algorithm for brane brick models. In the remainder of this section, we elaborate its detailed implementation.

\begin{figure}[ht!!]
\begin{center}
\resizebox{0.9\hsize}{!}{
\includegraphics[trim=0cm 0cm 0cm 0cm,totalheight=10 cm]{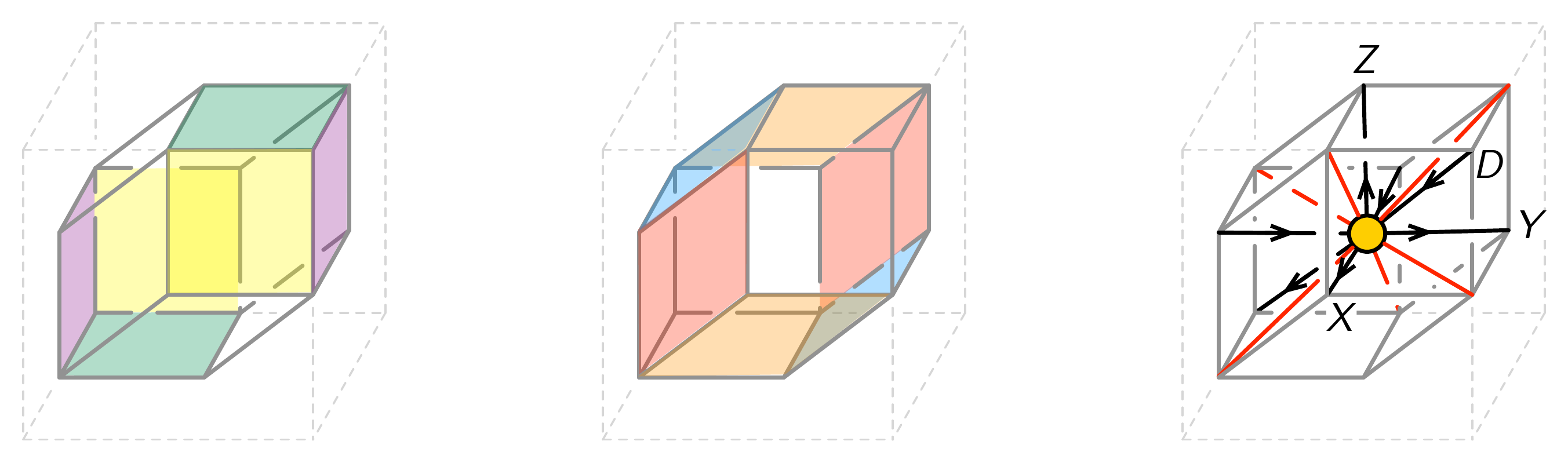}
}  
\caption{$\mathbb{C}^4$ has six phase boundaries that cut out a RD in $T^3$, which gives rise to the single gauge group of the corresponding gauge theory. The RD has eight 3-valent vertices and six 4-valent vertices that in this case are periodically identified in pairs and correspond to the four chiral and three Fermi fields of the $\mathbb{C}^4$ theory, respectively.
\label{fc4coamoeba}}
 \end{center}
 \end{figure}

\begin{figure}[ht!!]
\begin{center}
\resizebox{0.7\hsize}{!}{
\includegraphics[trim=0cm 0cm 0cm 0cm,totalheight=10 cm]{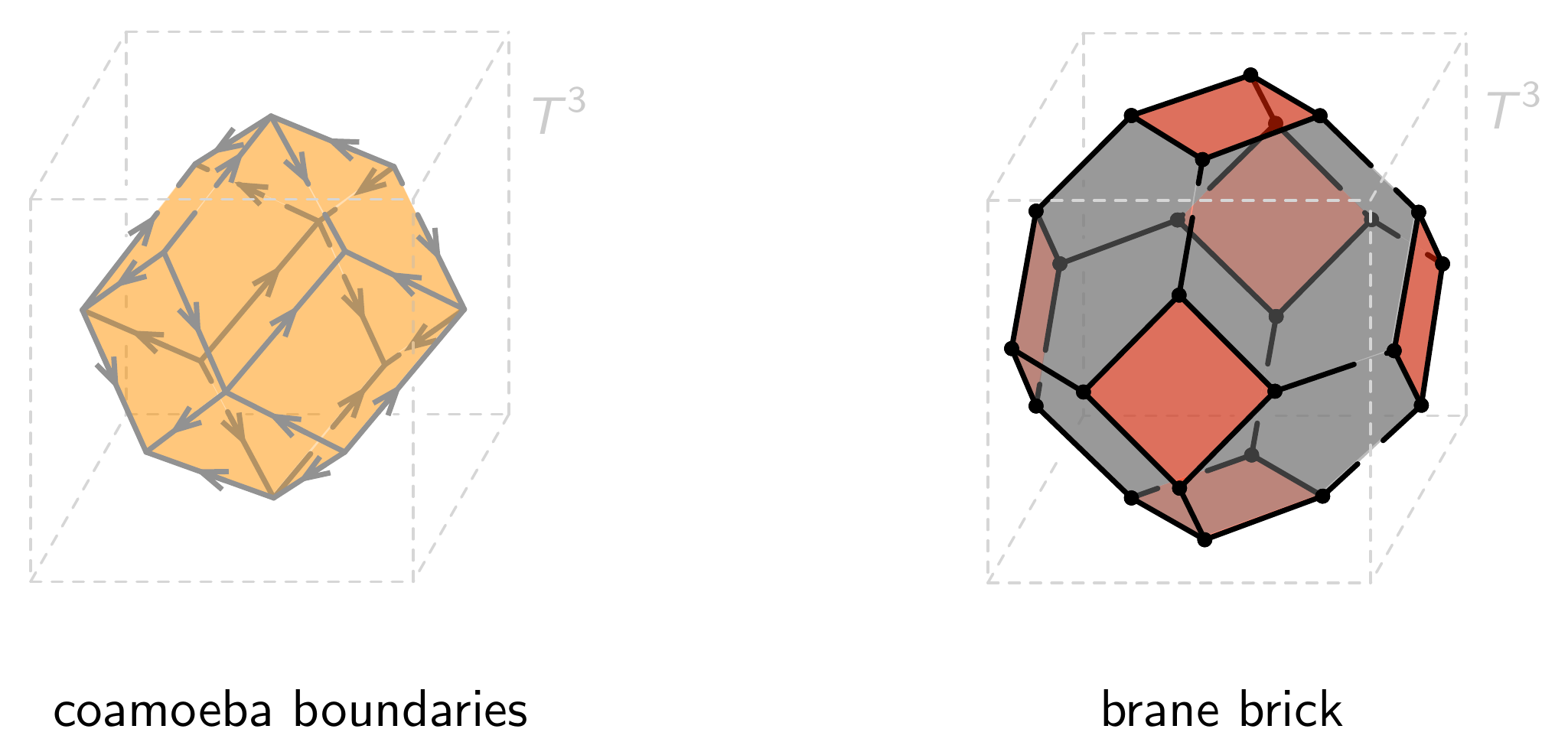}
}  
\caption{Complement of the coamoeba cut out in $T^3$ for $\mathbb{C}^4$ by the phase boundaries and the corresponding brane brick model. For convenience, the unit cell has been rotated with respect to the one used in \fref{fc4coamoeba}.}
\label{ft3diagrams2}
 \end{center}
 \end{figure}

Consider the $\mathbb{C}^4$ example as an illustration. The RD is identified with the single gauge group of the corresponding gauge theory,  as shown in \fref{fc4coamoeba}. We see that every point intersection of phase boundaries corresponds to a field in the periodic quiver. Equivalently, such intersections map to faces in the brane brick model, as illustrated in \fref{ft3diagrams2}.

\begin{figure}[ht!!]
\begin{center}
\resizebox{1\hsize}{!}{
\includegraphics[trim=0cm 0cm 0cm 0cm,totalheight=4.5 cm]{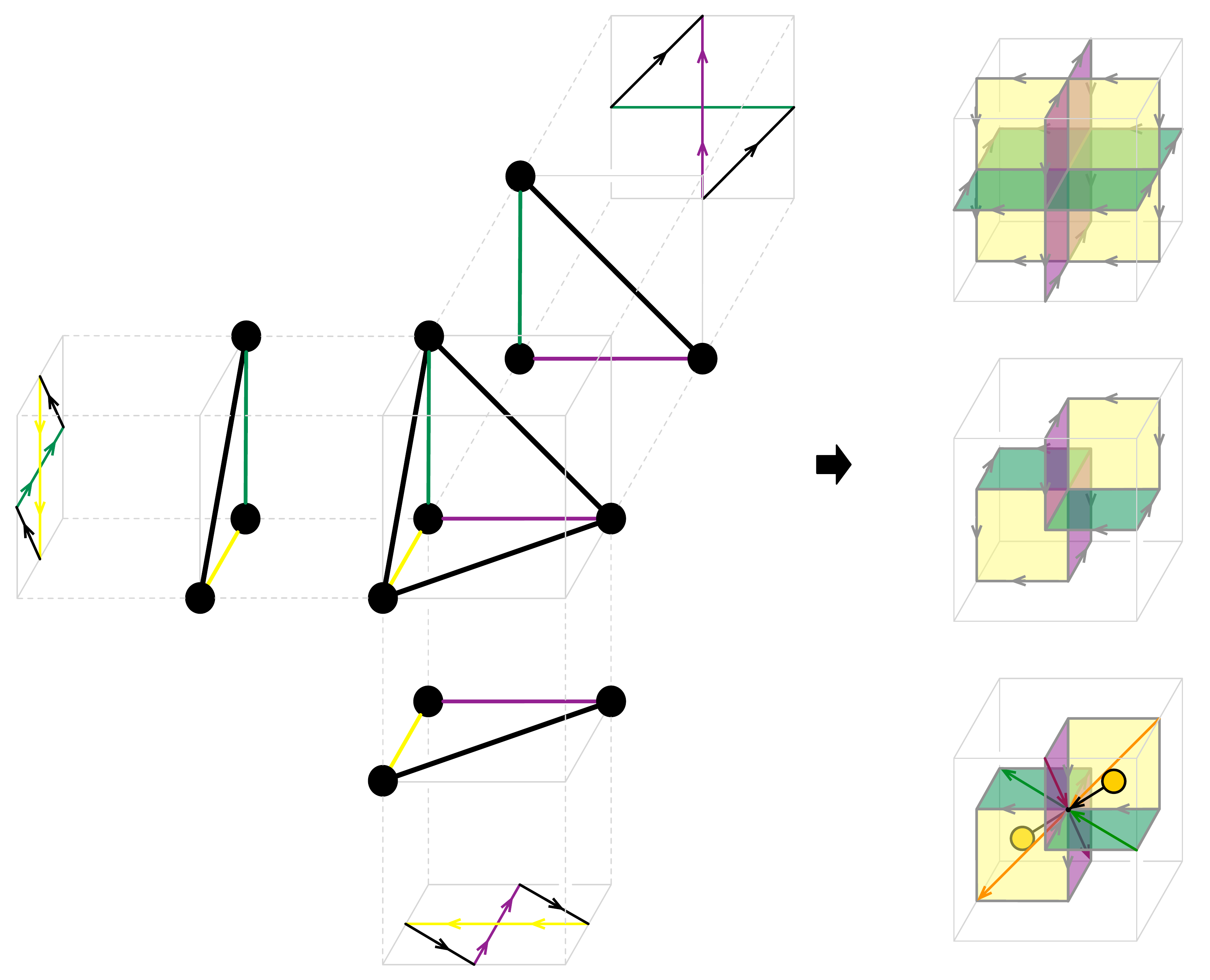}
}
\caption{An oriented intersection between the three phase boundaries associated to the green, yellow and purple edges in the toric diagram of $\mathbb{C}^4$. The pair of $2d$ faces in the toric diagram that share a given edge determine the orientation of the corresponding phase boundary. Each projection gives rise to a collection of oriented lines on $T^2$ that correspond to the intersection between the phase boundaries and a plane passing through the intersection. The chiral field extends along the oriented cones, connecting two gauge groups that live in the complement of the coamoeba.
\label{ft2t3project}}
 \end{center}
 \end{figure}

\begin{figure}[ht!!]
\begin{center}
\resizebox{1\hsize}{!}{
\includegraphics[trim=0cm 0cm 0cm 0cm,totalheight=4.5 cm]{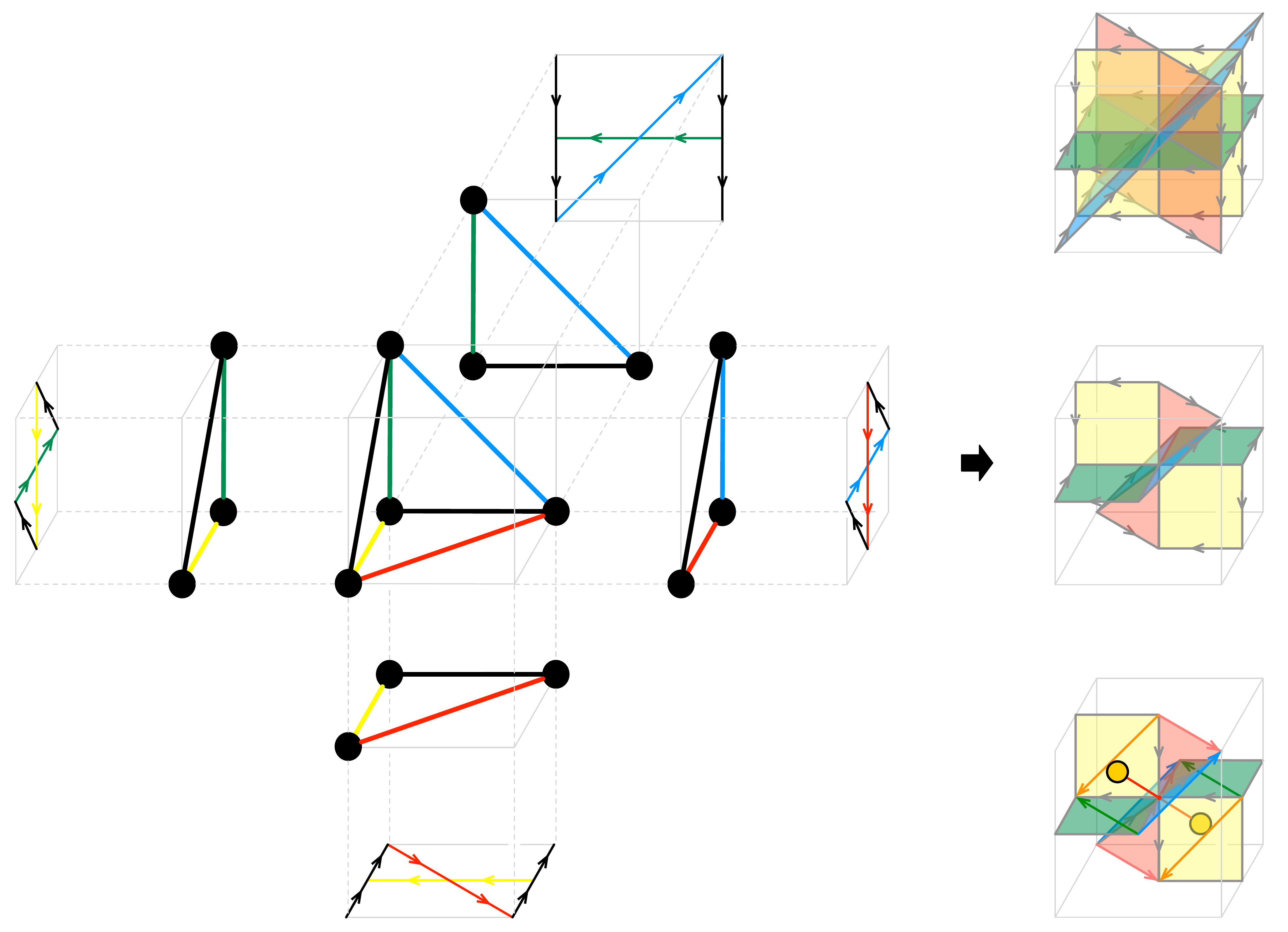}
}
\caption{
An alternating intersection between the four phase boundaries associated to the yellow, red, green and blue edges in the toric diagram of $\mathbb{C}^4$. The orientation of each phase boundary is established by considering the pair of $2d$ faces in the toric diagram associated to it. The Fermi field extends along the two cones with alternating orientations, connecting two gauge groups that live in the complement of the coamoeba.
\label{t2t3projectfermis}}
 \end{center}
 \end{figure}

Let us now study how the orientability of phase boundary intersections allows us to distinguish between chiral and Fermi fields. We will illustrate the ideas in the context of the $\mathbb{C}^4$ example, but they straightforwardly generalize to more complicated Calabi-Yau 4-folds. In order to study the orientation of phase boundaries, which are objects living in $3d$, it is useful to dissect the problem into a collection of $2d$ projections. To do so, we consider each of the faces of the toric diagram and assign an outward pointing normal vector to each of its edges, as shown in Figures \ref{ft2t3project} and \ref{t2t3projectfermis}.\footnote{This is precisely the procedure for determining zig-zag paths associated to the external edges of $2d$ toric diagrams in the case of Calabi-Yau 3-folds.}

Every edge in the toric diagram is contained in a pair of faces and hence the previous prescription assigns to it a pair of vectors. By construction both vectors live on the plane orthogonal to the edge, i.e. on the corresponding phase boundary. We would now like to combine these vectors into a single one determining the orientation of the phase boundary. For our purposes it is sufficient to take the sum of them.\footnote{Any linear combination of the two vectors with non-zero positive coefficients would also do the job.} The resulting vector points towards the exterior of the toric diagram. Notice that this is an orientation {\it on the phase boundary plane} and not one orthogonal to it. There is no natural orientation normal to phase boundaries, since this would correspond to an inexistent orientation of edges in the toric diagram.

Phase boundaries divide the neighborhood of any point intersection into a collection of polyhedral cones. Depending on the behavior of the orientations of phase boundaries in these cones, we can distinguish two types of intersections, which we now explain.

\medskip

\paragraph{Oriented Intersections: Chiral Fields.}

We define an {\it oriented intersection} as one containing two opposite {\it oriented cones}.  An oriented cone is one for which all phase boundaries are oriented towards the point intersection or away from it. Notice that the number of phase boundaries participating in the oriented cones might be lower than the total number of phase boundaries in the intersection. Every oriented intersection gives rise to a chiral field, whose orientation is determined by the oriented cones.

Since matter fields in the quiver stretch between gauge groups, which in turn map to components in the complement of the coamoeba, the previous prescription also serves for distinguishing the interior from the complement of the coamoeba. From all the cones meeting an oriented intersection, only the oriented ones correspond to corners of regions in the complement of the coamoeba. \fref{ft2t3project} illustrates the previous ideas for an oriented intersection in $\mathbb{C}^4$.\footnote{All examples considered in this section are such that for every pair of intersecting phase boundaries, the projections of their orientations onto their line intersection are parallel. This property is not generic. Our general discussion applies even when this is not the case.}

\medskip

\paragraph{Alternating Intersections: Fermi Fields.}

Every alternating intersection contains a pair of special {\it alternating cones}. Alternating cones are such that the orientations of the line intersections between consecutive pairs of phase boundaries alternate between going into and away from the intersection. This implies that a necessary condition for a cone to be alternating is to involve an even number of phase boundaries. It is natural to conjecture that alternating cones always comprise four phase boundaries, an observation supported by all the explicit examples we have considered. 

Every alternating intersection gives rise to a Fermi field, which extends along the two alternating cones. Similarly to the oriented intersection case, the corresponding pair of nodes in the quiver lay within the alternating cones, which can thus also be used to identify the complement of the coamoeba. \fref{t2t3projectfermis} shows an alternating intersection for $\mathbb{C}^4$ and the steps involved in determining the corresponding Fermi field.

To conclude this section, it is important to emphasize that the distinction between intersections leading to chiral and Fermi fields depends on their orientation or lack thereof and not on the number of intersecting phase boundaries. It is possible for a point intersection of phase boundaries to be neither oriented nor alternating. When this occurs, it does not correspond to any field in the gauge theory.

\section{Geometry from Brane Brick Models \label{sbricktogeom}}

\label{section_geometry_from_brane_bricks}

Brane brick models greatly simplify the determination of the probed toric Calabi-Yau 4-fold starting from the corresponding $2d$ $(0,2)$ quiver gauge theories, providing an alternative to the {\it forward algorithm} studied in \cite{Franco:2015tna}. A similar feat is achieved by brane tilings, which connect $4d$ $\mathcal{N}=1$ gauge theories to Calabi-Yau 3-folds, and it is at the heart of some of their most important applications. This combinatorial approach is often referred to as the {\it fast forward algorithm}, and we will employ the same name for its $2d$ analogue. A crucial point for achieving this, which will be elaborated in this section, is a correspondence between GLSM fields and certain objects in the brane brick model with simple combinatorial properties that we call {\it brick matchings}. They play a role analogous to perfect matchings for brane tilings.

In order to get some useful intuition for identifying what brick matchings are, it is convenient to review perfect matchings first. A perfect matching is a collection of edges in a brane tiling such that every node in the tiling is the endpoint of exactly one edge in the perfect matching. Every node in the tiling corresponds to a superpotential term in the $4d$ gauge theory or, equivalently, to a plaquette in the dual periodic quiver. We can thus alternatively define a perfect matching as a collection of chiral fields that contains exactly one field for every plaquette in the periodic quiver. An important property that follows from their definition is that all perfect matchings for a given brane tiling contain the same number of edges, which is equal to half the number of superpotential terms in the theory.\footnote{This is not the case, however, for {\it almost perfect matchings}, which generalize perfect matchings to bipartite graphs with boundaries \cite{Franco:2012mm}.} 

Let us now move to the $2d$ case and try to find a combinatorial interpretation for GLSM fields. A natural starting point is the $P$-matrix, which relates chiral fields to GLSM fields. Several explicit examples can be found in \cite{Franco:2015tna}. 

\begin{figure}[ht!]
\begin{center}
\resizebox{1\hsize}{!}{
\includegraphics[trim=0cm 0cm 0cm 0cm,totalheight=10 cm]{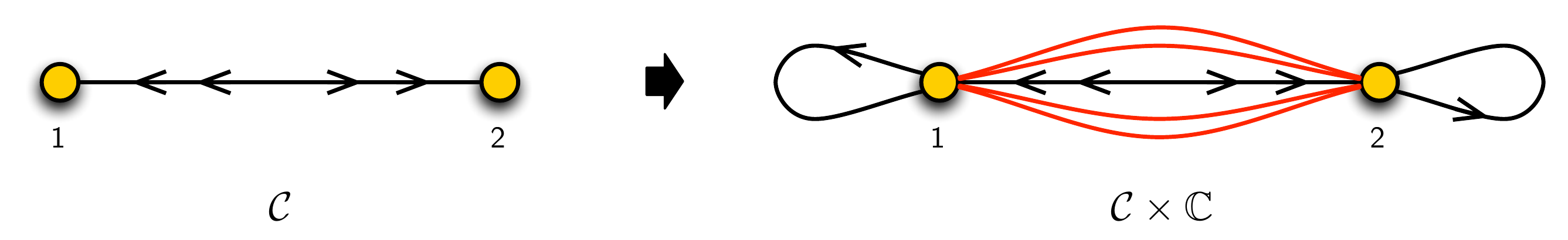}
}  
\caption{Quiver diagrams for the $4d$ $\mathcal{C}$ theory and for the $2d$ $ \mathcal{C}\times \mathbb{C}$ theory obtained from it by dimensional reduction.
\label{quiver_conifoldxC}}
 \end{center}
 \end{figure}

For concreteness, let us consider the $\mathcal{C}\times \mathbb{C}$ theory \cite{Franco:2015tna}. Its quiver diagram is shown in \fref{quiver_conifoldxC}. The $J$- and $E$-terms are 
\beq
\begin{array}{rclccclcc}
& & \ \ \ \ \ \ \ \ \ \ \ \ \ \ \ \ \ \ \ \ J & & & & \ \ \ \ \ \ \ \ \ \ \ \ \ \ E & & \\
 \Lambda_{12}^{1} : & \ \ \ & X_{21}\cdot X_{12}\cdot Y_{21} - Y_{21}\cdot X_{12}\cdot X_{21}& = & 0  & \ \ \ \ & \Phi_{11}\cdot Y_{12} - Y_{12}\cdot \Phi_{22} & = & 0  \\
 \Lambda_{12}^{2} : & \ \ \ & Y_{21}\cdot Y_{12}\cdot X_{21} -X_{21}\cdot Y_{12}\cdot Y_{21} & = & 0  & \ \ \ \ & \Phi_{11}\cdot X_{12} - X_{12}\cdot \Phi_{22} & = & 0 \\
 \Lambda_{21}^{1} : & \ \ \ & X_{12}\cdot Y_{21}\cdot Y_{12} - Y_{12}\cdot Y_{21}\cdot X_{12}& = & 0  & \ \ \ \ & \Phi_{22}\cdot X_{21} - X_{21}\cdot \Phi_{11} & = & 0  \\
 \Lambda_{21}^{2} : & \ \ \ &  Y_{12}\cdot X_{21}\cdot X_{12} -X_{12}\cdot X_{21}\cdot Y_{12}& = & 0  & \ \ \ \ & \Phi_{22}\cdot Y_{21} - Y_{21}\cdot \Phi_{11}& = & 0
\end{array}
\label{es200a1-new}
\eeq

The corresponding $P$-matrix is
\beal{P_conifoldxC}
P=
\left(
\begin{array}{c|ccccc}
\; & p_1 & p_2 & p_3 & p_4 & s \\
\hline
X_{21} & 1 & 0 & 0 & 0 & 0 \\
X_{12} & 0 & 1 & 0 & 0 & 0 \\
Y_{21} & 0 & 0 & 1 & 0 & 0 \\
Y_{12} & 0 & 0 & 0 & 1 & 0 \\
\Phi_{11} & 0 & 0 & 0 & 0 & 1 \\
\Phi_{22} & 0 & 0 & 0 & 0 & 1 \\
\end{array}
\right)
~.~ 
\eea
This simple example already exhibits a crucial difference between brick matchings and perfect matchings: brick matchings may involve different numbers of chiral fields. Here, every $p_i$ contains a single chiral field, while $s$ contains two chiral fields.

\subsection{Phase Boundaries and Brick Matchings}

\label{section_phase_boundaries_and_bms}

As already discussed in section \sref{sgeomtobrick}, phase boundaries are analogous to zig-zag paths in brane tilings. In this section, we will generalize the well-known connection between perfect matchings and zig-zag paths to a new relation between brick matchings and phase boundaries. This will finally allow us to define brick matchings.

\begin{figure}[H]
\begin{center}
\resizebox{0.6\hsize}{!}{
\includegraphics[trim=0cm 0cm 0cm 0cm,totalheight=10 cm]{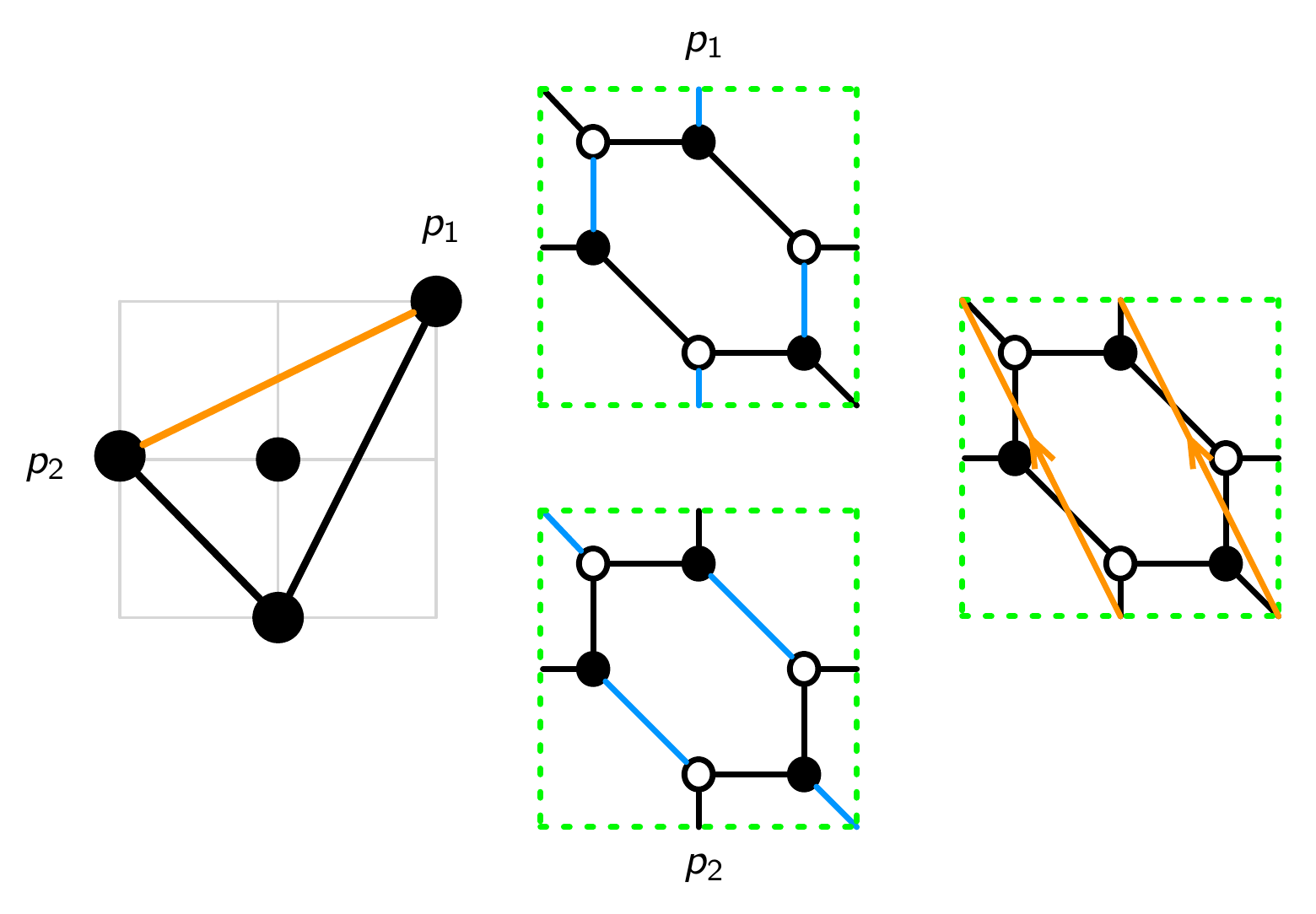}
}  
\caption{
The toric diagram and brane tiling for $\text{dP}_0$. The figure illustrates the construction of a zig-zag path (orange) as the difference of the two extremal perfect matchings $p_1$ and $p_2$.
\label{fdp0pmzz}}
 \end{center}
 \end{figure}

Perfect matchings are in one-to-one correspondence with GLSM fields for the $4d$ quiver theories associated to brane tilings and hence map to points in the $2d$ toric diagrams of the corresponding Calabi-Yau 3-folds \cite{Franco:2005rj,Franco:2006gc}. We refer to the perfect matchings located at extremal points of the toric diagram as {\it extremal perfect matchings}. Perfect matchings can be endowed with an orientation, e.g. by orienting all edges from white to black nodes. The zig-zag path associated to an external edge in the toric diagram corresponds to the difference between the two extremal perfect matchings connected by the edge \cite{GarciaEtxebarria:2006aq}. \fref{fdp0pmzz} illustrates the construction of zig-zag paths from extremal perfect matchings for the $\text{dP}_0$ example.

\begin{figure}[H]
\begin{center}
\resizebox{0.6\hsize}{!}{
\includegraphics[trim=0cm 0cm 0cm 0cm,totalheight=10 cm]{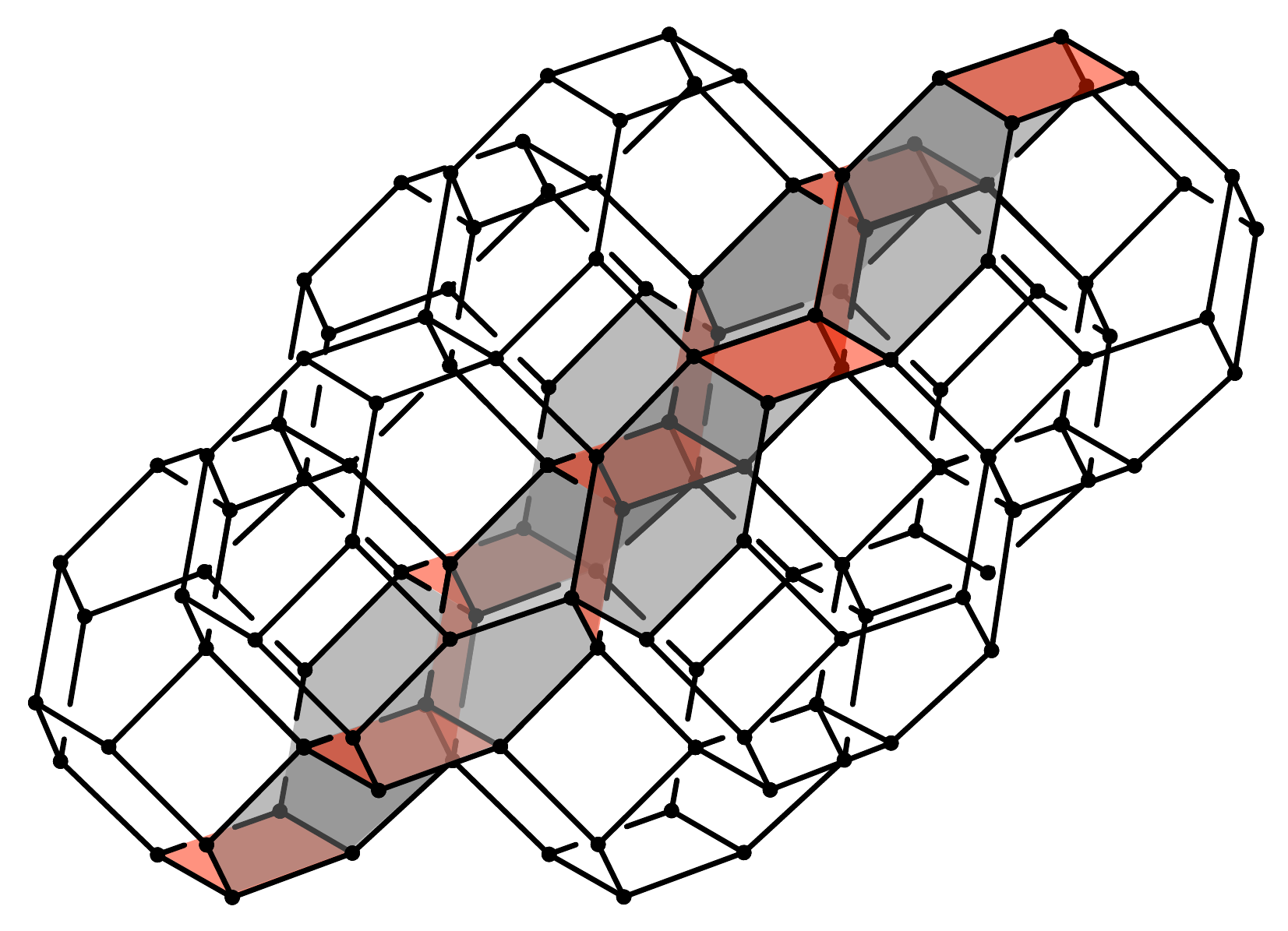}
}  
\caption{
The brane brick model for $\mathbb{C}^4$ with a collection of highlighted chiral and Fermi faces that form one of its phase boundaries.
\label{fc4brickplane}}
 \end{center}
 \end{figure}

We can think about phase boundaries as collections of faces on the brane brick model, as shown in \fref{fc4brickplane} for $\mathbb{C}^4$. It is useful to introduce the {\it phase boundary matrix} $H$ to encode this information. Columns in this matrix correspond to phase boundaries $\eta_\alpha$ and rows correspond to chiral and Fermi fields, i.e. to faces in the brane brick model. An entry in $H_{i\alpha}$ is equal to $\pm1$ if the face associated to the row $i$ is contained in the boundary represented by the column $\alpha$ (with the sign controlled by the relative orientation) and $0$ otherwise. For $\mathcal{C}\times \mathbb{C}$, we have
\beal{esa100}
H= 
\left(
\ba{c|cccc|cccc}
\; & \eta_{12} & \eta_{23} & \eta_{34} & \eta_{41} & \eta_{1s} & \eta_{2s} & \eta_{3s} & \eta_{4s}
\\
\hline
X_{12} & 1 & \, -1 \, & 0 & 0 & 0 & \, -1 \, & 0 & 0
\\
X_{21} & \, -1 \, & 0 & 0 & 1 & \, -1 \, & 0 & 0 & 0 
\\
Y_{12} & 0 & 0 & 1 & \, -1 \, & 0 & 0 & 0 & \, -1 \, 
\\
Y_{21} & 0 & 1 & \, -1 \, & 0 & 0 & 0 & \, -1 \, & 0
\\
\Phi_{11} & 0 & 0 & 0 & 0 & 1 & 1 & 1 & 1
\\
\Phi_{22} & 0 & 0 & 0 & 0 & 1 & 1 & 1 & 1
\\
\hline
\Lambda_{12}^{1} & 0 & 0 & 1 & -1 & 1 & 1 & 1 & 0
\\
\Lambda_{12}^{2} & 1 & -1 & 0 & 0 & 1 & 0 & 1 & 1
\\
\Lambda_{21}^{1}  & -1 & 0 & 0 & 1 & 0 & 1 & 1 & 1
\\
\Lambda_{21}^{2} & 0 & 1 & -1 & 0 & 1 & 1 & 0 & 1
\ea
\right) ~.~
\eea
We can also regard $H$ as summarizing the net intersection numbers, counted with orientation, between the phase boundaries and the fields in the periodic quiver within a unit cell.

Our goal is to establish a one-to-one correspondence between brick matchings and GLSM fields. This, in turn, will determine that brick matchings are mapped to points in the $3d$ toric diagram of the underlying Calabi-Yau 4-fold. We expect brick matchings to correspond to collections of fields in the quiver and hence to collections of faces in the brane brick model. In analogy with the brane tiling case, it is natural to envisage that the phase boundary associated to an edge in the toric diagram is given by the difference between the two extremal brick matchings connected by the edge, i.e. $\eta_{\mu\nu}=p_\mu-p_\nu$. It then becomes clear that if brick matchings consisted only of chiral fields, then the resulting surfaces would have holes corresponding to the Fermi fields. We conclude that brick matching must contain both chiral and Fermi fields. 

Based on this reasoning, let us generalize the $P$-matrix to include Fermi fields.  Allowing only for $1$ and $0$ entries depending on whether a brick matching contains a field or not, \eref{esa100} uniquely determines 
\beal{P_lambda_conifold}
P_{\Lambda}= 
\left(
\ba{c|ccccc}
\; & p_1 & p_2 & p_3 & p_4 & s 
\\
\hline
X_{12} & 0 & \, 1 \, & 0 & 0 & 0 
\\
X_{21} & \, 1 \, & 0 & 0 & 0 & \, 0  
\\
Y_{12} & 0 & 0 & 0 & \, 1 \, & 0  
\\
Y_{21} & 0 & 0 & \, 1 \, & 0 & 0 
\\
\Phi_{11} & 0 & 0 & 0 & 0 & 1 
\\
\Phi_{22} & 0 & 0 & 0 & 0 & 1 
\\
\hline
\Lambda_{12}^{1} & 0 & 0 & 0 & 1 & 1 
\\
\Lambda_{12}^{2} & 0 & 1 & 0 & 0 & 1 
\\
\Lambda_{21}^{1}  & 1 & 0 & 0 & 0 & 1 
\\
\Lambda_{21}^{2} & 0 & 0 & 1 & 0 & 1 
\ea
\right) ~.~
\eea
The columns in $P_\Lambda$ do not correspond to the brick matchings we are after, yet.

Physically, Fermi fields $\Lambda_a$ and their conjugate $\bar{\Lambda}_a$ are on an equal footing. It is hence reasonable to consider a definition of brick matchings that treats them symmetrically. We thus also include rows for $\bar{\Lambda}_a$ in a new $P$-matrix, to which we refer as $P_{\Lambda \bar{\Lambda}}$. This new matrix contains exactly the same information as $P_\Lambda$. The entries for the $\bar{\Lambda}_a$ rows are determined such that they obey
\beq
P_{\Lambda\bar{\Lambda},\Lambda_a \mu} + P_{\Lambda\bar{\Lambda},\bar{\Lambda}_a \mu}=1 \, .
\eeq 
It will soon become clear that this choice also leads to brick matchings with nice combinatorial properties. It is important to emphasize, though, that $\Lambda_a$ and $\bar{\Lambda}_a$ do not correspond to independent degrees of freedom.

Starting from \eref{P_lambda_conifold} we obtain the following matrix for the $\mathcal{C}\times \mathbb{C}$ theory
\beal{P_lambda_lambdabar_conifold}
P_{\Lambda\bar{\Lambda}}= 
\left(
\ba{c|ccccc}
\; & p_1 & p_2 & p_3 & p_4 & s 
\\
\hline
X_{12} & 0 & \, 1 \, & 0 & 0 & 0 
\\
X_{21} & \, 1 \, & 0 & 0 & 0 & \, 0  
\\
Y_{12} & 0 & 0 & 0 & \, 1 \, & 0  
\\
Y_{21} & 0 & 0 & \, 1 \, & 0 & 0 
\\
\Phi_{11} & 0 & 0 & 0 & 0 & 1 
\\
\Phi_{22} & 0 & 0 & 0 & 0 & 1 
\\
\hline
\Lambda_{12}^{1} & 0 & 0 & 0 & 1 & 1 
\\
\bar{\Lambda}_{12}^{1} & 1 & 1 & 1 & 0 & 0 
\\
\Lambda_{12}^{2} & 0 & 1 & 0 & 0 & 1 
\\
\bar{\Lambda}_{12}^{2} & 1 & 0 & \, 1 \, & 1 & 0 
\\
\Lambda_{21}^{1}  & 1 & 0 & 0 & 0 & \, 1 \, 
\\
\bar{\Lambda}_{21}^{1}  & 0 & 1 & 1 & 1 & 0 
\\
\Lambda_{21}^{2} & 0 & 0 & 1 & 0 & 1 
\\
\bar{\Lambda}_{21}^{2} & \, 1 \, & \, 1 \, & 0 & \, 1 \, & 0 
\ea
\right) ~.~
\eea

The subtraction of brick matchings that leads to the $H$-matrix is defined in terms of the corresponding columns of $P_{\Lambda\bar{\Lambda}}$ as follows
\beq
\begin{array}{ccl}
H_{X_i \eta_{\mu\nu}} & = & P_{\Lambda\bar{\Lambda},X_i \mu}-P_{\Lambda\bar{\Lambda},X_i \nu} \\ [0.25cm]
H_{\Lambda_a \eta_{\mu\nu}}& =& {1\over 2}\left[\left(P_{\Lambda\bar{\Lambda},\Lambda_a \mu}-P_{\Lambda\bar{\Lambda},\bar{\Lambda}_a \mu}\right)-\left(P_{\Lambda\bar{\Lambda},\Lambda_a \nu}-P_{\Lambda\bar{\Lambda},\bar{\Lambda}_a \nu}\right)\right] 
\end{array}
\eeq
for all chiral fields $X_i$ and Fermi fields $\Lambda_a$.

We are now ready to define a brick matching as the collection of fields in the quiver determined by the corresponding column in the $P_{\Lambda\bar{\Lambda}}$-matrix. Strictly speaking, our previous discussion only allows the determination of extremal brick matchings. The following section introduces a combinatorial definition of general brick matchings.

\subsection{A Combinatorial Definition of Brick Matchings}

\label{section_combinatorics_brick_matchings}

In is possible to provide a combinatorial definition of brick matchings that is highly reminiscent to the one of perfect matchings for $4d$ theories. To do so, it is convenient to complete $J_a$- and $E_a$-terms into pairs of plaquettes, by multiplying them by the corresponding $\Lambda_a$ or $\bar{\Lambda}_a$. A brick matching is then defined as a collection of chiral, Fermi and conjugate Fermi fields contributing to every plaquette in the theory exactly once as follows:

\begin{itemize}
\item[1.] For every Fermi field pair $(\Lambda_a,\bar{\Lambda}_a)$, the chiral fields in the brick matching cover {\it either} each of the two $J_a$-term plaquettes {\it or} each of the two $E_a$-term plaquettes exactly once. 

\item[2.] 
If the chiral fields in the brick matching cover the plaquettes associated to the $J_a$-term, then $\bar{\Lambda}_a$ is included in the brick matching. As a result, each of the two plaquettes associated to the $E_a$-term is covered exactly once. 
$\Lambda_a$ is not included in the brick matching, since it would produce an additional contribution to the plaquettes associated to $J_a$.

\item[3.] 
If the chiral fields in the brick matching cover the plaquettes associated to the $E_a$-term, then $\Lambda_a$ is included in the brick matching. As a result, each of the two plaquettes associated to the $J_a$-term is covered exactly once.  
$\bar{\Lambda}_a$ is not included in the brick matching, since it would produce an additional contribution to the plaquettes associated to $E_a$.
\end{itemize}

\comment{
The above can be expressed for a chiral field $X$ in perfect matching $p^{\Lambda\bar{\Lambda}}_i$ for E-term plaquettes $E_{(\Lambda,\bar{\Lambda})}$ and J-term plaquettes $J_{(\Lambda,\bar{\Lambda})}$ for a Fermi pair $(\Lambda,\bar{\Lambda})$ as follows
\beal{esb1}
X\in p^{\Lambda\bar{\Lambda}}_i ~ 
\left\{\ba{ccc}
X\in J_{(\Lambda,\bar{\Lambda})}~,~
X\notin E_{(\Lambda,\bar{\Lambda})}
&
\Rightarrow 
&
\Lambda\in p^{\Lambda\bar{\Lambda}}_i
\\
X\notin J_{(\Lambda,\bar{\Lambda})}~,~
X\in E_{(\Lambda,\bar{\Lambda})}
&
\Rightarrow 
&
\Lambda\in p^{\Lambda\bar{\Lambda}}_i
\ea
\right.
~~~
\forall (\Lambda,\bar{\Lambda})
~.~
\eea}
The set of all brick matchings correspond to all collections of fields satisfying these three properties. Determining them is a non-trivial combinatorial problem, but can be efficiently implemented in a computer. It would certainly be desirable to find an analytic method for finding brick matchings, analogous to the one based on the Kasteleyn matrix for perfect matchings \cite{Franco:2005rj}.
 
Following the definition, the total number of Fermi plus conjugate Fermi fields in all brick matchings is the same and it is equal to the number of Fermi fields in the theory. The number of chiral fields in brick matchings, as noted earlier, might vary.

For illustration, let us explicitly verify how some of the brick matchings encoded by \eref{P_lambda_lambdabar_conifold} satisfy the definition above. To do so, we first complete the $J$- and $E$-terms presented in \eref{es200a1-new} into plaquettes by multiplying them by the corresponding Fermi fields. Let us consider $p_1=\{X_{21}, \bar{\Lambda}_{12}^{1}, \bar{\Lambda}_{12}^{2}, \Lambda_{21}^{1},\bar{\Lambda}_{21}^{2}\}$. The contributions to the plaquettes are
\beq
\begin{array}{ccc}
J & & E \\
 \Lambda_{12}^{1} \cdot \textcolor{blue}{X_{21}}\cdot X_{12}\cdot Y_{21} -  \Lambda_{12}^{1} \cdot Y_{21}\cdot X_{12}\cdot \textcolor{blue}{X_{21}}  & \ \ \ \ &  \textcolor{forestgreen}{\bar{\Lambda}_{12}^{1}} \cdot \Phi_{11}\cdot Y_{12} -  \textcolor{forestgreen}{\bar{\Lambda}_{12}^{1}} \cdot Y_{12}\cdot \Phi_{22}   \\
\Lambda_{12}^{2} \cdot  Y_{21}\cdot Y_{12}\cdot \textcolor{blue}{X_{21}} - \Lambda_{12}^{2} \cdot  \textcolor{blue}{X_{21}}\cdot Y_{12}\cdot Y_{21}  & \ \ \ \ &  \textcolor{forestgreen}{\bar{\Lambda}_{12}^{2}} \cdot  \Phi_{11}\cdot X_{12} -  \textcolor{forestgreen}{\bar{\Lambda}_{12}^{2}} \cdot  X_{12}\cdot \Phi_{22}  \\
 \textcolor{forestgreen}{\Lambda_{21}^{1}} \cdot X_{12}\cdot Y_{21}\cdot Y_{12} -  \textcolor{forestgreen}{\Lambda_{21}^{1}} \cdot Y_{12}\cdot Y_{21}\cdot X_{12}  & \ \ \ \ &  \bar{\Lambda}_{21}^{1} \cdot \Phi_{22}\cdot \textcolor{blue}{X_{21}} -  \bar{\Lambda}_{21}^{1} \cdot \textcolor{blue}{X_{21}}\cdot \Phi_{11}   \\
\Lambda_{21}^{2} \cdot Y_{12}\cdot \textcolor{blue}{X_{21}}\cdot X_{12} - \Lambda_{21}^{2} \cdot X_{12}\cdot \textcolor{blue}{X_{21}}\cdot Y_{12}  & \ \ \ \ &  \textcolor{forestgreen}{\bar{\Lambda}_{21}^{2}} \cdot \Phi_{22}\cdot Y_{21} -  \textcolor{forestgreen}{\bar{\Lambda}_{21}^{2}} \cdot Y_{21}\cdot \Phi_{11}
\end{array}
\label{}
\eeq
where we indicate chiral and Fermi fields in blue and green, respectively. Similarly, for $s=\{ \Phi_{11}, \Phi_{22},\Lambda_{12}^{1},\Lambda_{12}^{2},\Lambda_{21}^{1},\Lambda_{21}^{2} \}$, we have
\beq
\begin{array}{ccc}
J & & E \\
 \textcolor{forestgreen}{\Lambda_{12}^{1}} \cdot X_{21}\cdot X_{12}\cdot Y_{21} -  \textcolor{forestgreen}{\Lambda_{12}^{1}} \cdot Y_{21}\cdot X_{12}\cdot X_{21}  & \ \ \ \ &  \bar{\Lambda}_{12}^{1} \cdot \textcolor{blue}{\Phi_{11}}\cdot Y_{12} -  \bar{\Lambda}_{12}^{1} \cdot Y_{12}\cdot \textcolor{blue}{\Phi_{22}}   \\
 \textcolor{forestgreen}{\Lambda_{12}^{2}} \cdot  Y_{21}\cdot Y_{12}\cdot X_{21} - \textcolor{forestgreen}{\Lambda_{12}^{2}} \cdot  X_{21}\cdot Y_{12}\cdot Y_{21}  & \ \ \ \ &  \bar{\Lambda}_{12}^{2} \cdot  \textcolor{blue}{\Phi_{11}}\cdot X_{12} -  \bar{\Lambda}_{12}^{2} \cdot  X_{12}\cdot \textcolor{blue}{\Phi_{22}}  \\
\textcolor{forestgreen}{\Lambda_{21}^{1}} \cdot X_{12}\cdot Y_{21}\cdot Y_{12} -  \textcolor{forestgreen}{\Lambda_{21}^{1}} \cdot Y_{12}\cdot Y_{21}\cdot X_{12}  & \ \ \ \ &  \bar{\Lambda}_{21}^{1} \cdot \textcolor{blue}{\Phi_{22}}\cdot X_{21} -  \bar{\Lambda}_{21}^{1} \cdot X_{21}\cdot \textcolor{blue}{\Phi_{11}}   \\
 \textcolor{forestgreen}{\Lambda_{21}^{2}} \cdot Y_{12}\cdot X_{21}\cdot X_{12} - \textcolor{forestgreen}{\Lambda_{21}^{2}} \cdot X_{12}\cdot X_{21}\cdot Y_{12}  & \ \ \ \ & \bar{\Lambda}_{21}^{2} \cdot \textcolor{blue}{\Phi_{22}}\cdot Y_{21} - \bar{\Lambda}_{21}^{2} \cdot Y_{21}\cdot \textcolor{blue}{\Phi_{11}}
\end{array}
\label{}
\eeq

The chiral field content of the $P_{\Lambda\bar{\Lambda}}$-matrix precisely agrees with the $P$-matrix of the forward algorithm \cite{Franco:2015tna}.\footnote{This agreement holds modulo extra GLSM fields, which are discussed below.} The forward algorithm is insensitive to the Fermi fields in brick matchings since they do not contain scalar components and hence do not participate in the classical mesonic moduli space. Our discussion above emphasizes, however, that incorporating Fermi fields into brick matchings is crucial for connecting them to phase boundaries and for their combinatorial interpretation.

\paragraph{Extra GLSM Fields.} In \cite{Franco:2015tna}, it was observed that the forward algorithm sometimes makes use of additional GLSM fields. Mesonic gauge invariant operators parameterize the mesonic moduli space and can be expressed in terms chiral fields or GLSM fields. When {\it extra GLSM fields} are present, they can be neglected when studying the geometry of the moduli space because they do not affect the spectrum of gauge invariant operators but rather correspond to an over-parameterization of it. In other words, the generators and relations amongst generators of the mesonic moduli space are unaffected by the presence of extra GLSM fields. In appendix \ref{shilbert}, we explicitly study the algebraic structure of the mesonic moduli spaces of certain brane brick models by computing their Hilbert series \cite{Benvenuti:2006qr}. By doing so, we illustrate the over-parameterization by extra GLSM fields. Additional examples can be found in \cite{Franco:2015tna}.

Various criteria for recognizing extra GLSM fields were provided in \cite{Franco:2015tna}. Remarkably, they can be identified combinatorially: our study of numerous explicit examples suggests that extra GLSM fields are combinations of fields in the quiver that do not satisfy the brick matching definition. It is sufficient to restrict our attention to ordinary GLSM fields corresponding to brick matchings, since they are sufficient for fully parameterizing the mesonic moduli space.

\subsection{A Correspondence Between GSLM Fields and Brick Matchings}

We claim that there is a one-to-one correspondence between the brick matchings we have combinatorially defined in the previous section and the GLSM fields describing the classical mesonic moduli space of the gauge theory. Here we will provide strong evidence supporting this claim, by showing the brick matchings automatically satisfy vanishing $J$- and $E$-terms. The proof is similar to the one that shows that perfect matchings satisfy $F$-terms of $4d$ toric theories.

To do so, we introduce the following map between chiral fields and brick matchings
\beq
X_i=\prod_\mu p_\mu^{P_{\Lambda\bar{\Lambda},i \mu} } \, .
\label{map_X_p}
\eeq
This is precisely the map between chiral fields and GLSM fields provided, as mentioned earlier, the chiral field part of $P_{\Lambda\bar{\Lambda}}$ is interpreted as the $P$-matrix of the forward algorithm. We are not interested in Fermi fields at this point, since they do not contain scalars contributing to the mesonic moduli space.

Let us consider the $J$- and $E$-terms associated to a Fermi field $\Lambda_a$ and express them in terms of brick matchings using \eref{map_X_p}. We obtain
\beq
\begin{array}{lcl}
J_a=0 & \ \ \ \Leftrightarrow \ \ \ & {\displaystyle \prod_{X_i\in J_{a,1}}\prod_\mu  p_\mu^{P_{\Lambda\bar{\Lambda},i \mu} }  = \prod_{X_i\in J_{a,2}}\prod_\mu  p_\mu^{P_{\Lambda\bar{\Lambda},i \mu}}} \\[.6cm]
E_a=0 & \ \ \ \Leftrightarrow \ \ \ & {\displaystyle \prod_{X_i\in E_{a,1}}\prod_\mu  p_\mu^{P_{\Lambda\bar{\Lambda},i \mu} }  = \prod_{X_i\in E_{a,2}}\prod_\mu  p_\mu^{P_{\Lambda\bar{\Lambda},i \mu}}} 
\end{array}
\eeq
where $J_{a,n}$ and $E_{a,n}$, $n=1,2$, indicate each of the two terms in $J_a$ and $E_a$, respectively. Each of these terms is a monomial in chiral fields and can be completed to form a plaquette by multiplying it by $\Lambda_a$ or $\bar{\Lambda}_a$. $J_a$ and $E_a$ vanish if every brick matching appearing in $J_{a,1}$ also appears in $J_{a,2}$, and every brick matching appearing in $E_{a,1}$ also appears in $E_{a,2}$.

Let us consider an arbitrary brick matching $p_\mu$. From the definition of a brick matching, we know it contains either $\Lambda_a$ or $\bar{\Lambda}_a$. Let us assume it contains $\Lambda_a$. Point 3 in section \sref{section_combinatorics_brick_matchings} combined with \eref{map_X_p} implies that $p_\mu$ appears in exactly one chiral field in $E_{a,1}$ and one chiral field in $E_{a,2}$. Point 3 also implies that $p_\mu$ does not contain any chiral field in either $J_{a,1}$ or $J_{a,2}$. If the brick matching contains $\bar{\Lambda}_a$ instead of $\Lambda_a$, the same proof holds upon exchanging $J_a \leftrightarrow E_a$. Our arguments apply to all brick matchings and all Fermi fields, so we conclude that the map in \eref{map_X_p} in conjunction with the combinatorial properties of brick matchings automatically satisfy all vanishing $J$- and $E$-terms.

\subsection{The Fast Forward Algorithm for Brane Brick Models}

\label{section_fast_forward_algorithm} 

The variables provided by brick matchings not only bypass the need to solve for vanishing $J$- and $E$-terms by automatically satisfying them, but are also ideally suited for the toric description of the Calabi-Yau 4-fold. Achieving the latter is usually the most computationally demanding part of the standard forward algorithm, since it involves the calculation of dual cones. The combinatorial interpretation of GLSM fields in terms of brick matchings hence leads to a substantial simplification and speed up of the computation of the mesonic moduli space. We refer to the resulting approach, which we finalize developing in this section, as the {\it fast forward algorithm}.\footnote{For the $4d$ analog of the fast forward algorithm connecting brane tilings to toric CY$_3$ singularities by means of perfect matchings, we refer the reader to \cite{Franco:2005rj,Franco:2006gc}.} In order to identify the geometry of the moduli space, the remaining task after finding the brick matchings is to assign coordinates in $\mathbb{Z}^3$ to them, such that they generate the toric diagram. Below we present two methods for doing so. Appendix \ref{sfast} explains in detail how the two procedures are implemented in an explicit illustrative example.

\paragraph{Toric Diagram from Face Intersections.}

Let us call $\gamma_x$, $\gamma_y$ and $\gamma_z$ the edges of the unit cell along its three fundamental directions.\footnote{In fact, in analogy to the discussion of {\it flux lines} in brane tilings \cite{2003math.....10326K,2003math.ph..11005K}, it is not necessary to consider unit cells whose boundaries are planes or whose edges are straight lines. Our discussion applies to these general situations without changes.} The position of the brick matching in the toric diagram is a vector $(n_x,n_y,n_z)\in \mathbb{Z}^3$ that is fully determined by the chiral field faces in the brick matching intersecting $\gamma_x$, $\gamma_y$ and $\gamma_z$ as follows
\beq
\label{es600a1}
n_a = \sum \langle X _{ij}, \gamma_a\rangle~, \ \ \ \ \ \ \ \ \ a=x,y,z~,
\eeq
where angle brackets indicate the usual intersection number between an oriented surface and an oriented line. The coordinate $n_a$ is thus a sum of $\pm 1$ contributions, where the sign depends on the relative orientation between the field in the periodic quiver and $\gamma_a$. As usual, multiple brick matchings might correspond to the same point in the toric diagram.

We have tested this proposal in numerous examples. In particular, we have checked that it works for general abelian orbifolds of $\mathbb{C}^4$. By this we not only mean that the correct toric diagram is reproduced, but also that the position of every brick matching agrees with the one of the corresponding GLSM field resulting from the standard forward algorithm. Any toric CY$_4$ singularity, and the corresponding brane brick model, can be obtained from an appropriate abelian orbifold of $\mathbb{C}^4$ by partial resolution \cite{Franco:2015tna}. Partial resolution only removes some of the brick matchings, without altering the positions of the remaining ones in the toric diagram. The surviving brick matchings are those that do not contain the chiral field getting a non-zero VEV. We thus conclude that the positions of the brick matchings are correctly established by our prescription or, equivalently, that it works for arbitrary brane brick models. 

\paragraph{Toric Diagram from the Height Function.}

An alternative, and admittedly more formal, way of determining the position of a brick matching in a toric diagram is as follows. Given a brick matching $p_\mu$, it is possible to define an integer-valued {\it height function} $h_\mu$ over the brane brick model. To do so, we pick a reference brick matching $p_0$ and a brick $b_0$. In analogy to the discussion in section \sref{section_phase_boundaries_and_bms}, the difference $p_\mu-p_0$ defines a set of closed oriented surfaces on the brane brick model. The height function jumps by $\pm 1$ when traversing these surfaces, with the sign determined by the orientation of the crossing. The height for the reference brick $b_0$ is set to zero. This definition is the natural generalization to brane brick models of the height function for brane tilings \cite{2003math.....10326K,2003math.ph..11005K}. Interestingly, the difference between the height functions for two brick matchings is well defined and independent of $p_0$ and $b_0$. The position of the point in the toric diagram associated to a brick matching $p_\mu$ is given by the {\it slope of its height function}. It is defined as a vector in $\mathbb{Z}^3$ made out of the variations of the height function $(\Delta_x h_\mu,\Delta_y h_\mu,\Delta_z h_\mu)$ between adjacent unit cells along the three fundamental directions of $T^3$. The net effect of changing $p_0$ is simply an overall shift of the slopes of all brick matchings and hence does not modify the resulting geometry. 

\section{Partial Resolution}
\label{section_partial_resolution}

Different toric CY$_4$ can be connected by {\it partial resolution}, which corresponds to eliminating points in the toric diagram. From the perspective of the gauge theory, partial resolution translates into higgsing. In \cite{Franco:2015tna}, a systematic procedure for identifying the chiral fields that acquire non-zero VEVs in order to achieve a desired partial resolution was introduced. It relies on the map between chiral and GLSM fields encoded by the $P$-matrix. We refer the reader to \cite{Franco:2015tna} for a thorough presentation of these ideas. In addition, in the previous section we saw that the $P$-matrix admits a combinatorial interpretation by identifying GLSM fields with brick matchings. The purpose of this section is to develop a complementary viewpoint on partial resolution: its implementation in terms of phase boundaries.

\subsection{CY$_3$ Partial Resolution and Zig-Zag Paths}

Before studying partial resolutions of Calabi-Yau 4-folds from the viewpoint of phase boundaries, it is instructive to review partial resolutions of toric Calabi-Yau 3-folds in terms of zig-zag paths of the corresponding brane tilings. A generic partial resolution can be understood as a sequence of removals of single extremal points in the toric diagram. It is then sufficient to focus on the case in which a single extremal point in the toric diagram is eliminated. When this occurs, the two edges of the toric diagram that terminate on the deleted point disappear and are replaced by a new edge. As explained in section \sref{sgeomtobrick}, external edges in the toric diagram are in one-to-one correspondence with zig-zag paths in the brane tiling. Partial resolution thus corresponds to a recombination of two zig-zag paths into a new one.

\begin{figure}[ht]
\begin{center}
\resizebox{0.7\hsize}{!}{
\includegraphics[trim=0cm 0cm 0cm 0cm,totalheight=10 cm]{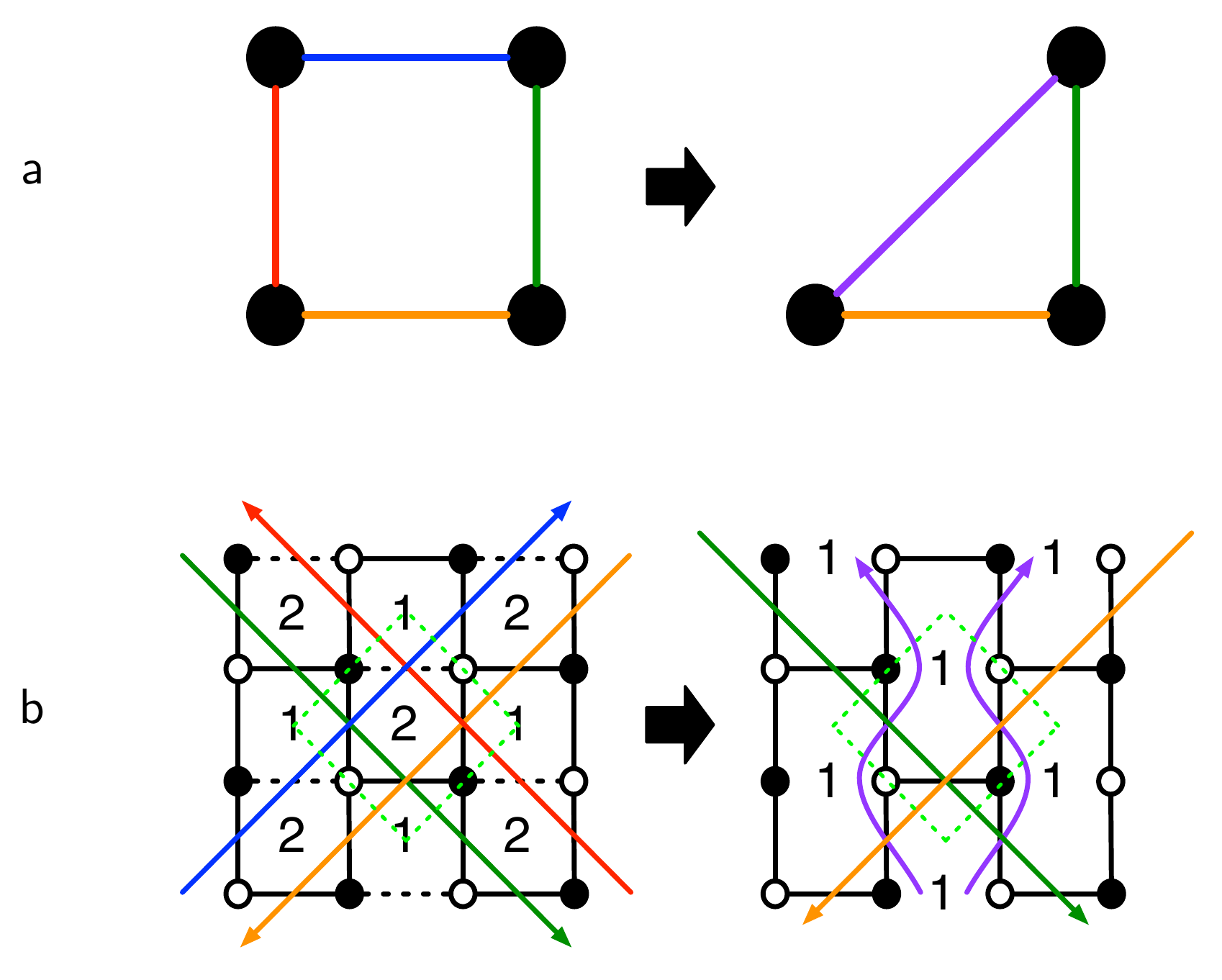}
}  
\caption{Partial resolution from the conifold to $\mathbb{C}^3$. a) Toric diagrams. b) Brane tilings showing the recombination of zig-zag paths. The edge associated to the bifundamental field getting a non-zero VEV, shown as a dotted line, sits at the intersection of the zig-zags that are recombined. The slope of every edge in the toric diagram is determined by the winding numbers along the two fundamental directions of the unit cell of the corresponding zig-zag path.
\label{toric_tiling_conifold_C3}}
 \end{center}
 \end{figure}

To illustrate these ideas, let us consider the example of the partial resolution from the conifold to $\mathbb{C}^3$. As shown in \fref{toric_tiling_conifold_C3}, when the top left corner of the conifold toric diagram is deleted, the two edges terminating on it are substituted by a new one. In the gauge theory, this partial resolution corresponds to turning on a non-zero VEV for a chiral field, which is located at an intersection between the two removed zig-zag paths. Giving a VEV to this field translates into deleting the corresponding edge in the brane tiling, which in turn results in a modification of the zig-zag paths. As shown in \fref{toric_tiling_conifold_C3}, the two zig-zag paths under consideration are recombined into one with winding numbers being given by the sum of the original ones. This is in agreement with the slope of the new edge in the toric diagram.

The number of gauge groups in the $4d$ gauge theory is equal to $2 \times A$, with $A$ the area of the toric diagram (see e.g. \cite{Franco:2005sm}). In our example this area decreases by 1/2 and hence the quiver loses a single node. This is in agreement with the higgsing by a single bifundamental VEV. More generally, removing an extremal point in the toric diagram can lead to a larger decrease in the area. When this occurs, more VEVs need to be simultaneously turned on in order to appropriately account for the reduction in the gauge symmetry.  

\begin{figure}[h]
\begin{center}
\resizebox{0.8\hsize}{!}{
\includegraphics[trim=0cm 0cm 0cm 0cm,totalheight=10 cm]{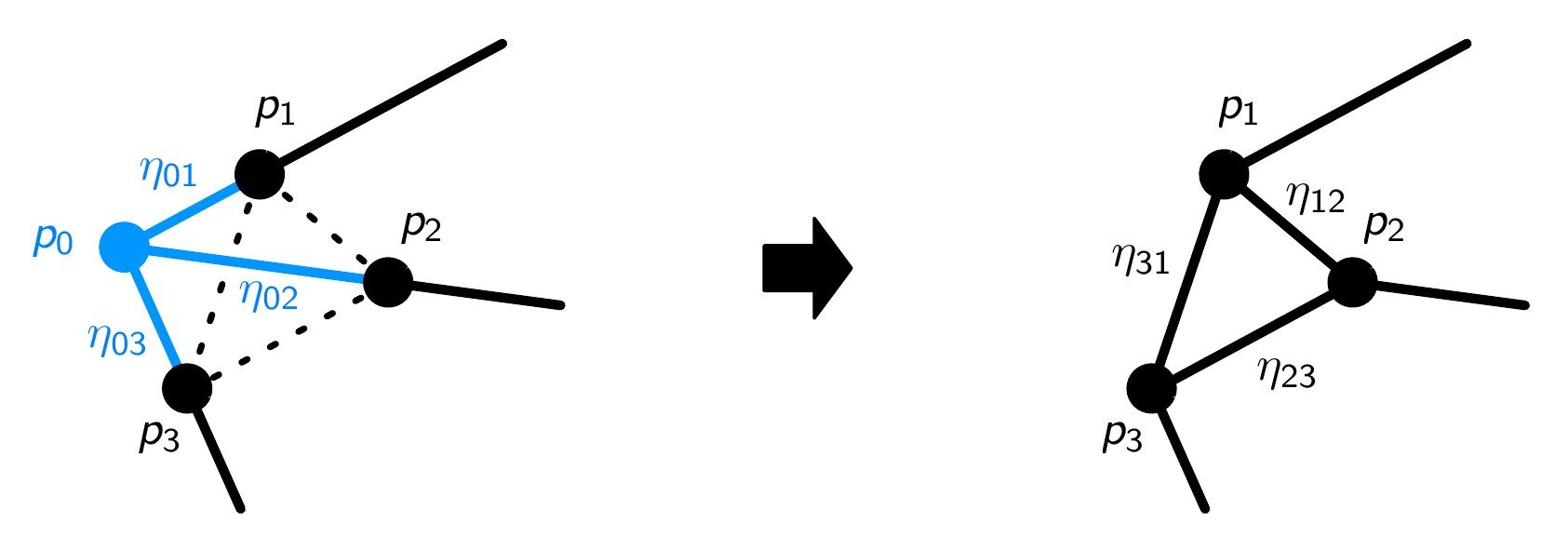}
}  
\caption{Partial resolution by removing a single extremal point in the toric diagram of a Calabi-Yau 4-fold. The edges connected to the removed point disappear and give rise to new edges, which correspond to their pairwise recombination. \label{general_PR_toric_diagram}}
 \end{center}
 \end{figure}

\subsection{CY$_4$ Partial Resolution and Phase Boundaries}

\begin{figure}[h]
\begin{center}
\resizebox{0.75\hsize}{!}{
\includegraphics[trim=0cm 0cm 0cm 0cm,totalheight=10 cm]{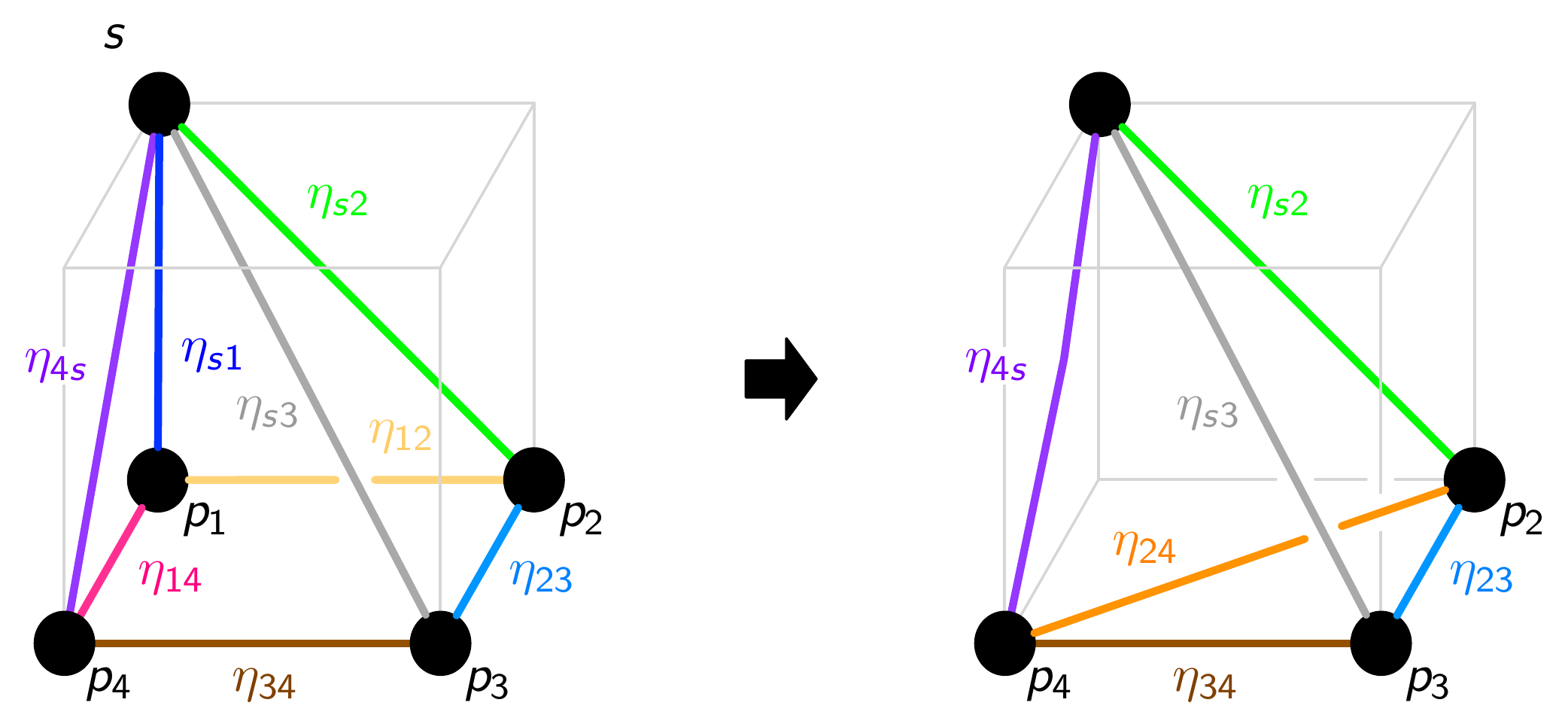}
}  
\caption{Toric diagrams for the partial resolution from $\mathcal{C}\times\mathbb{C}$ to $\mathbb{C}^4$.
\label{toric_PR_conifocxC_C4}}
 \end{center}
 \end{figure}

Let us now study the partial resolution of toric Calabi-Yau 4-folds in terms of phase boundaries. Once again, it is sufficient to consider the case in which a single extremal point in the toric diagram is eliminated. General partial resolutions can be achieved by iterating this process.

 \begin{figure}[ht!!]
\begin{center}
\resizebox{1\hsize}{!}{
\includegraphics[trim=0cm 0cm 0cm 0cm,totalheight=10 cm]{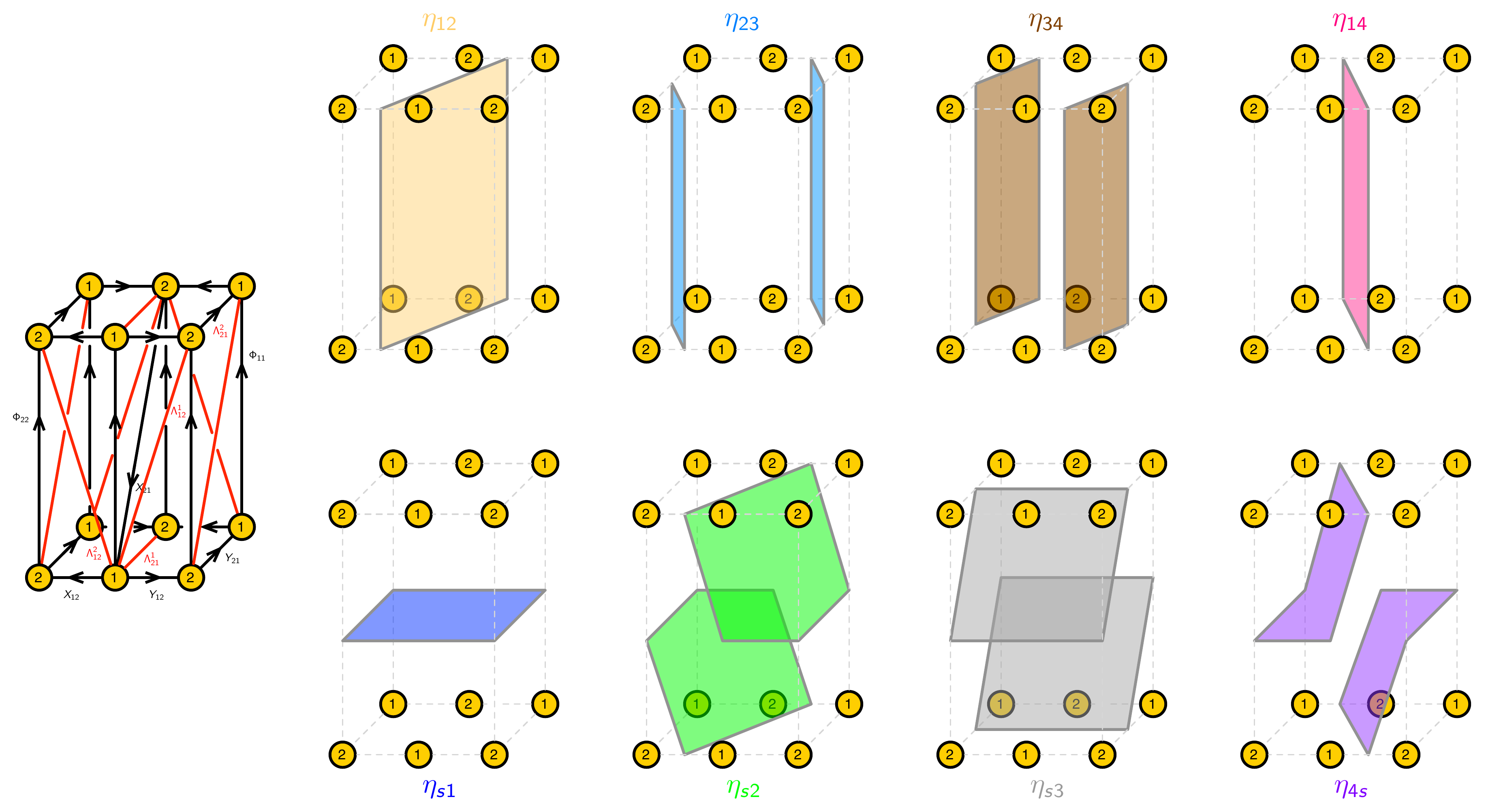}
}  
\caption{
Phase boundaries for $\mathcal{C}\times\mathbb{C}$.
\label{fconifoldcoamoeba}}
 \end{center}
 \end{figure}

Removing an extremal point in the toric diagram affects all external edges connected to it and generates new ones. Partial resolution leads to a pairwise recombination of phase boundaries that are in one-to-one correspondence with the affected edges. This can be understood in full generality as follows. Consider a general partial resolution that removes a point $p_0$, as shown in \fref{general_PR_toric_diagram}. Let us further assume that $p_0$ is the endpoint of three external edges in the toric diagram, which connect it to the points $p_1$, $p_2$ and $p_3$. The general case in which the removed point has a higher valence will be discussed at the end of this section. When $p_0$ is eliminated, the edges $\eta_{01}$, $\eta_{02}$ and $\eta_{03}$ disappear, while $\eta_{12}$, $\eta_{23}$ and $\eta_{31}$ emerge. Here we label edges according to their corresponding phase boundaries $\eta_{ij}=p_i-p_j$. Given this expression, it is straightforward to see that every new edge is given by the recombination of a pair of adjacent removed ones, i.e. $\eta_{ab}=\eta_{a0}-\eta_{b0}=\eta_{a0}+\eta_{0b}$, for $a,b=1,2,3$. 

For concreteness, let us consider the partial resolution of $\mathcal{C}\times\mathbb{C}$ to $\mathbb{C}^4$. The corresponding toric diagrams are shown in \fref{toric_PR_conifocxC_C4}. In general, the chiral field acquiring a VEV during a partial resolution sits precisely at the intersection of the phase boundaries associated to the edges connected to the removed point. The two toric diagrams in \fref{toric_PR_conifocxC_C4} are related by the removal of $p_1$. The affected edges are hence those related to the phase boundaries $\eta_{12}$, $\eta_{14}$ and $\eta_{1s}$. Following the general discussion in \cite{Franco:2015tna} and using the $P$-matrix \eref{P_conifoldxC}, we conclude that eliminating $p_1$ corresponds to giving a VEV to the chiral field $X_{21}$. As anticipated, this chiral field indeed lives at the intersection of $\eta_{12}$, $\eta_{14}$ and $\eta_{1s}$. 

\begin{figure}[ht!]
\begin{center}
\resizebox{0.8\hsize}{!}{
\includegraphics[trim=0cm 0cm 0cm 0cm,totalheight=10 cm]{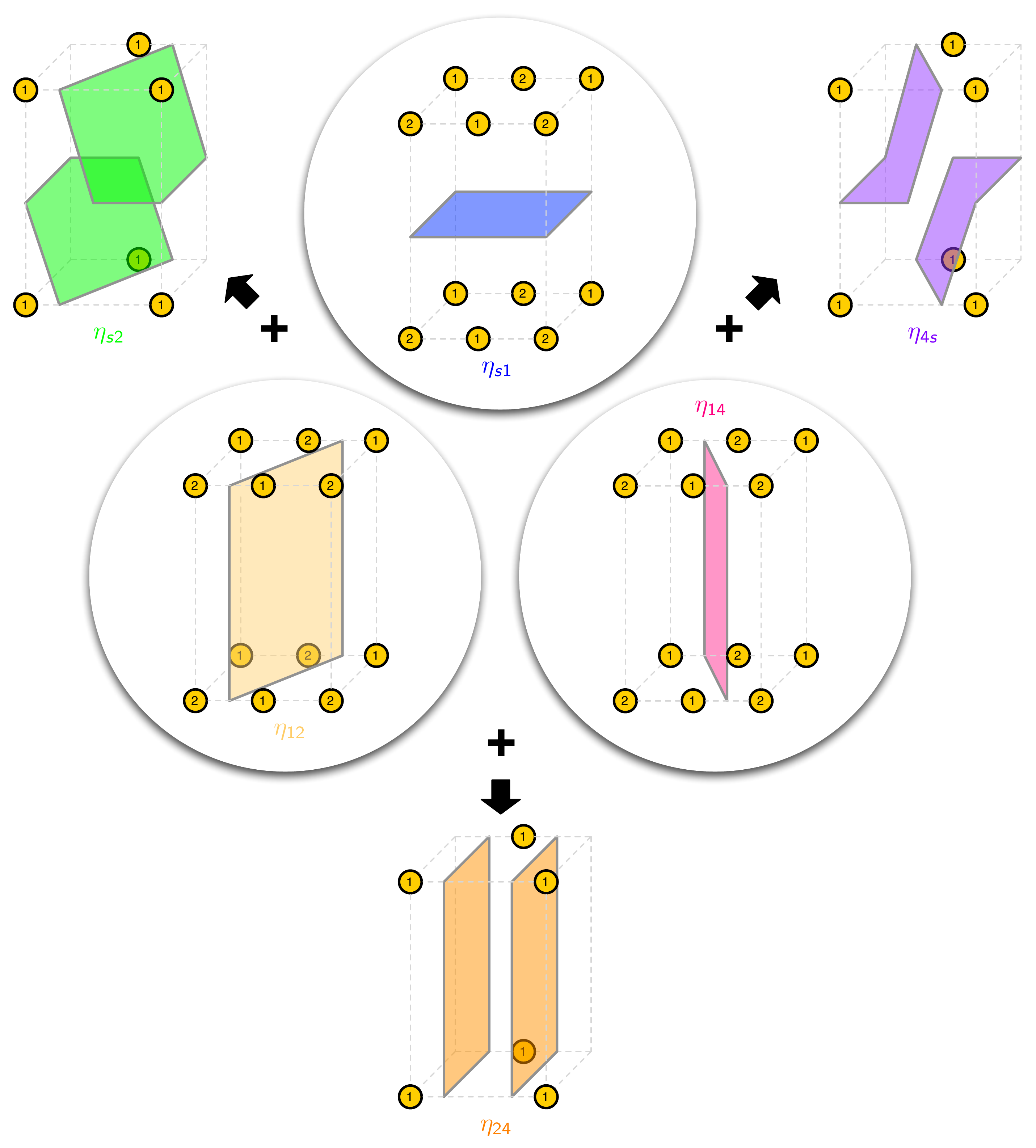}
}  
\caption{
The three phase boundaries intersecting at the chiral field $X_{21}$ of $\mathcal{C}\times\mathbb{C}$ combine pairwise into a new phase boundary and two existing phase boundaries when $X_{21}$ gets a non-zero VEV.
\label{fboundarycomb}}
 \end{center}
 \end{figure}

\begin{figure}[ht!!]
\begin{center}
\resizebox{0.8\hsize}{!}{
\includegraphics[trim=0cm 0cm 0cm 0cm,totalheight=10 cm]{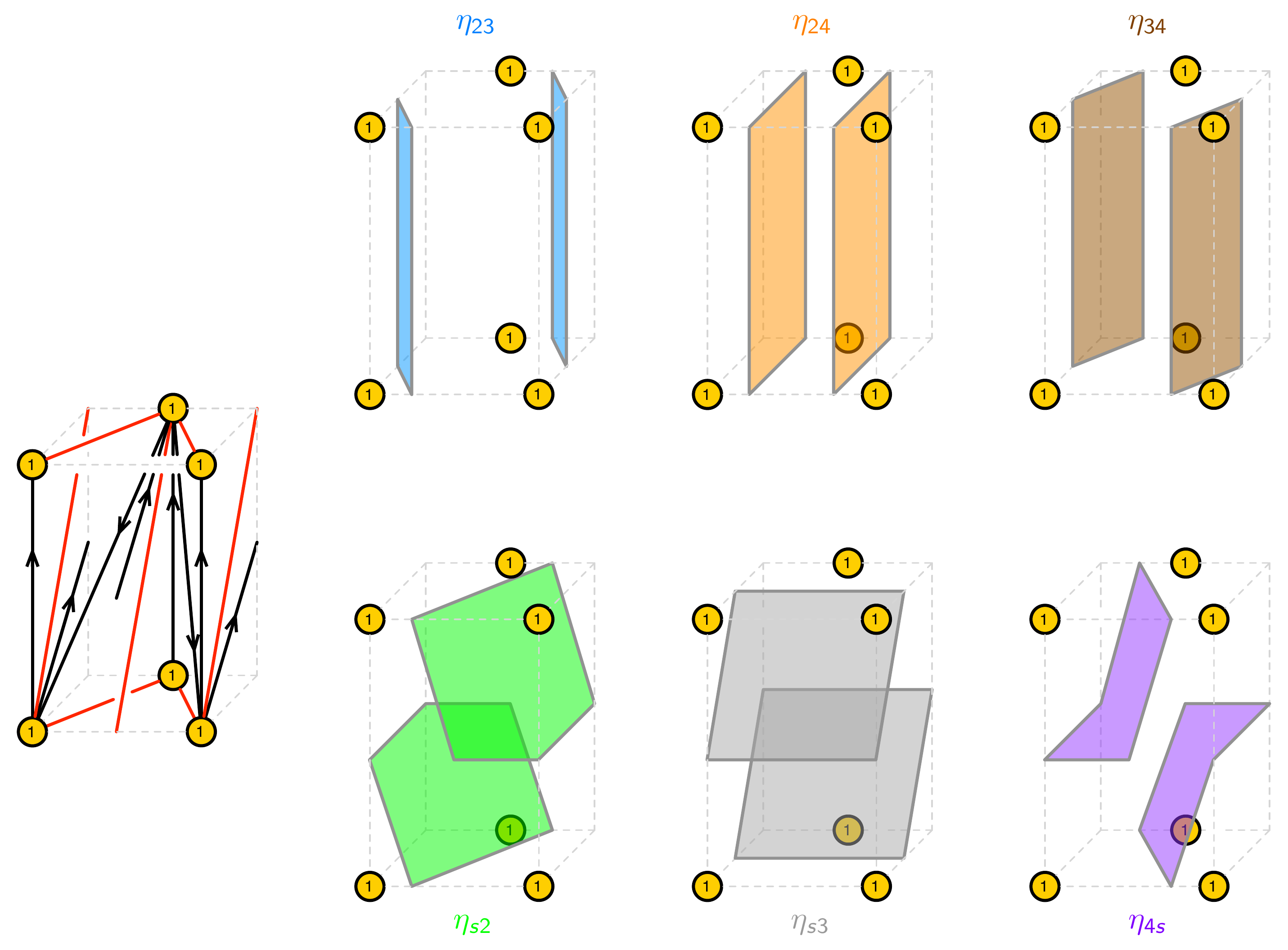}
}  
\caption{
The phase boundaries of the $\mathbb{C}^4$ brane brick model obtained by higgsing $\mathcal{C}\times \mathbb{C}$ in \fref{fconifoldcoamoeba}.
\label{fc4planes}}
 \end{center}
 \end{figure}

Let us now investigate in detail the fate of the phase boundaries, which are shown in \fref{fconifoldcoamoeba} for $\mathcal{C}\times\mathbb{C}$. When $X_{21}$ is removed, $\eta_{12}$, $\eta_{14}$ and $\eta_{1s}$ are pairwise recombined into $\eta_{24}$, $\eta_{4s}$ and $\eta_{s2}$ as shown in \fref{fboundarycomb}. The process parallels the transformation of edges in the toric diagram. This example shows that it is possible for some of the recombined phase boundaries to coincide with preexisting ones. In this case, $\eta_{4s}$ and $\eta_{s2}$ were present in the original theory and only $\eta_{24}$ gives rise to a new edge in the toric diagram. Note that, at the level of the brane brick model, the recombination of the phase boundaries is forced by the elimination of the face associated to $X_{21}$. \fref{fc4planes} shows the full set of phase boundaries for the final $\mathbb{C}^4$ theory.
 
The number of gauge groups in a general $2d$ $(2,0)$ toric gauge theory is equal to $6 \times V$, with $V$ the volume of the toric diagram \cite{Franco:2015tna}. When the removed point is connected to three edges, as in the previous example, a single tetrahedron is eliminated from the toric diagram. This corresponds to a reduction in the number of gauge groups by one and agrees with the higgsing by a single bifundamental VEV. More generally, when the removed extremal point in the toric diagram has a higher valence, more tetrahedra disappear. This implies that more than one node in the quiver is eliminated and hence the corresponding partial resolution involves turning on VEVs for more than one bifundamental field.

\section{Brane Brick Models for CY$_3 \times \mathbb{C}$ Theories}
\label{section_CY3_x_C}

In this section, we consider theories on D1-branes probing toric Calabi-Yau 4-folds of the form CY$_3 \times \mathbb{C}$. The resulting $2d$ theories generically have $(2,2)$ SUSY  (although further SUSY enhancement is possible in special cases) and can be derived by dimensional reduction of the $4d$ $\mathcal{N}=1$ theories on D3-branes over the corresponding toric CY$_3$. 

\subsection{Dimensional Reduction}

It is useful to very briefly review the basics of dimensional reduction of general $4d$ $\mathcal{N}=1$ to $2d$ $(2,2)$. Similar discussions can be found in \cite{GarciaCompean:1998kh,Franco:2015tna}. Let us denote by $\mathcal{V}_i$ and $\mathcal{X}_{ij}$ the vector and chiral multiplets of the $4d$ theory, respectively. Under dimensional reduction, they turn into $2d$ $(2,2)$ vector and chiral multiplets. In terms of $2d$ $(0,2)$ multiplets, we thus have:

\smallskip

\begin{itemize}
\item \underline{$4d$ $\mathcal{N}=1$ vector $\mathcal{V}_i$} $\rightarrow$ $2d$ $(0,2)$ vector $V_i$ + $2d$ $(0,2)$ adjoint chiral $\Phi_{ii}$
\item \underline{$4d$ $\mathcal{N}=1$ chiral $\mathcal{X}_{ij}$} $\rightarrow$ $2d$ $(0,2)$ chiral $X_{ij}$ + $2d$ $(0,2)$ Fermi $\Lambda_{ij}$
\end{itemize}  

\smallskip

All $J$-terms in $2d$ descend from $4d$ $F$-terms
 \bea
J_{ji} = \frac{\partial W}{\partial X_{ij}} ~,
\label{J_dim_red}
\eea
where $W$ is the $4d$ superpotential. Here we understand the $J$-terms and $W$ as functions of the $2d$ $(0,2)$ chiral multiplets coming from the $4d$ chiral multiplets.
 
Finally, the $E$-terms arise from the gauge interactions of the $4d$ theory and take the form
\beq
E_{ij} = \Phi_{ii} X_{ij} - X_{ij} \Phi_{jj}~.
\label{E_dim_red}
\eeq
Even though there is no invariant distinction between $J$- and $E$-terms, the dimensional reduction prescription outlined above naturally distinguishes between them according to their $4d$ origin.

\subsection{Brane Brick Models from Brane Tilings}

The $4d$ theories on D3-branes over toric Calabi-Yau 3-folds are fully captured by brane tilings on $T^2$ \cite{Franco:2005rj}. Here we introduce a {\it lifting algorithm} that, starting from brane tilings, constructs the brane brick models on $T^3$ for the dimensional reduced theories associated to CY$_4=\mathrm{CY}_3 \times \mathbb{C}$. A closely related prescription, phrased in terms of the dual periodic quivers, was introduced in \cite{Franco:2015tna}. The algorithm is introduced below, by explaining how the basic elements of brane tilings are transformed. It automatically implements all aspects of the dimensional reduction. 

\smallskip

\paragraph{Brane tiling faces.} 
A $4d$ vector multiplet $\mathcal{V}_i$ maps to a $2d$ $(0,2)$ vector multiplet $V_i$ and a $2d$ $(0,2)$ adjoint chiral multiplet $\Phi_{ii}$ under dimensional reduction. Faces in the brane tiling correspond to unitary gauge groups in the $4d$ theory. When lifting to the brane brick model, each of them therefore gives rise to a brick and a face, which correspond to a gauge group and a chiral adjoint $\Phi_{ii}$ in the $2d$ theory, respectively.

\begin{figure}[ht!]
\begin{center}
\resizebox{0.8\hsize}{!}{
\includegraphics[trim=0cm 0cm 0cm 0cm,totalheight=10 cm]{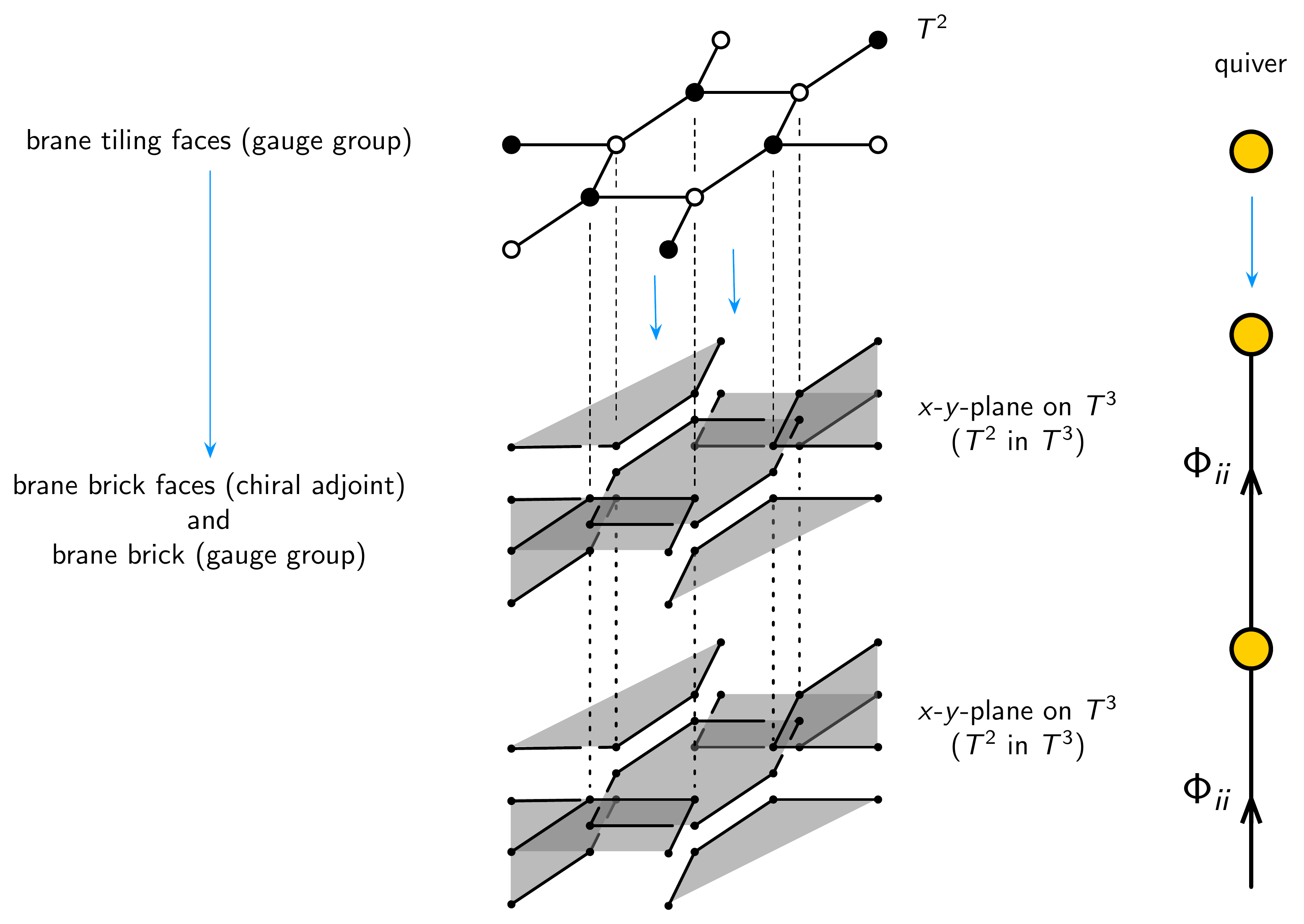}
}  
\caption{A brane tiling face is lifted to a brane brick and a face in the brane brick model. This process implements the dimensional reduction of a $4d$ vector multiplet into a $2d$ $(0,2)$ vector multiplet and a $2d$ $(0,2)$ adjoint chiral multiplet. 
\label{ftilingfaces}}
 \end{center}
 \end{figure}

Every brane tiling face directly becomes the brane brick face for $\Phi_{ii}$. This face lies on a $T^2$ within the $T^3$ of the brane brick model, which for convenience we call the $x$-$y$-plane, as illustrated in \fref{ftilingfaces}. The face for $\Phi_{ii}$ separates two copies of the same brane brick along the third cycle of $T^3$, the $z$-direction, as shown in \fref{ftilingfaces}. Since the brane brick corresponds to a gauge group in the $2d$ theory, $\Phi_{ii}$ transforms in the adjoint representation, as wanted.

\paragraph{Brane tiling edges.} 
Every edge in the brane tiling, which represents a $4d$ $\mathcal{N}=1$ chiral multiplet $\mathcal{X}_{ij}$, gets mapped to a pair of faces in the brane brick model, associated to the corresponding $2d$ $(0,2)$ chiral $X_{ij}$ and Fermi $\Lambda_{ij}$ multiplets. Both faces sit between the bricks associated to nodes $i$ and $j$.

\begin{figure}[ht!]
\begin{center}
\resizebox{0.78\hsize}{!}{
\includegraphics[trim=0cm 0cm 0cm 0cm,totalheight=10 cm]{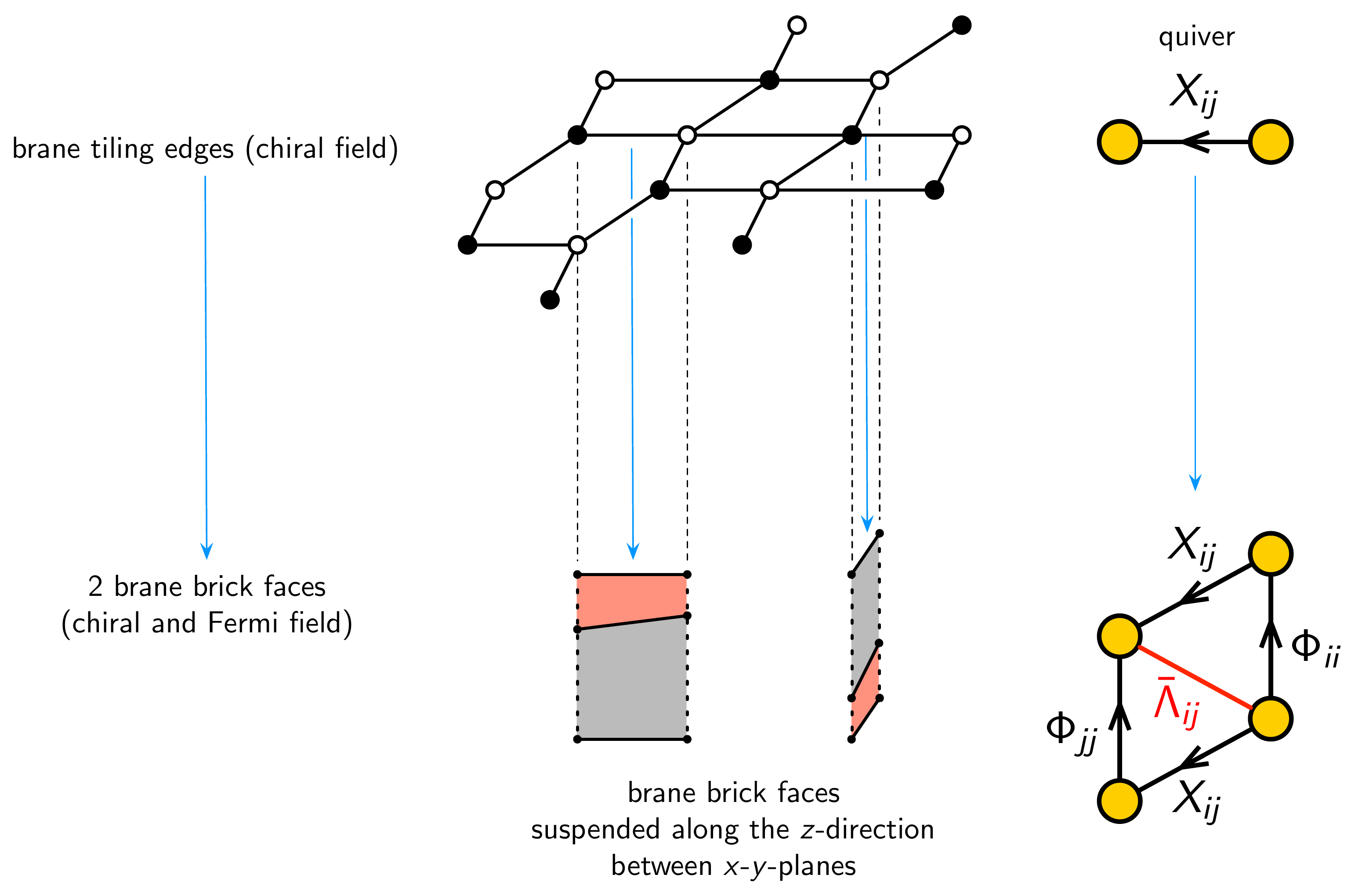}
}  
\caption{A brane tiling edge is lifted to two faces in the brane brick model, which correspond to a chiral multiplet $X_{ij}$ and a Fermi multiplet $\Lambda_{ij}$ in $2d$ $(0,2)$ language. $X_{ij}$ and $\Lambda_{ij}$ extend vertically between the same pair of edges, which intersect the horizontal faces for the adjoint fields $\Phi_{ii}$ and $\Phi_{jj}$, giving rise to the two terms in $E_{ij}=\Phi_{ii}\cdot X_{ij}-X_{ij} \cdot \Phi_{jj}$.
\label{ftilingedges}}
 \end{center}
 \end{figure}

The precise map is further constrained in order to generate the desired structure for $E$-terms. Due to the periodicity of $T^3$, the faces for $X_{ij}$ and $\Lambda_{ij}$ stretch along the $z$ direction between the same pair of edges. At each of these edges, they intersect with the horizontal faces associated to the adjoint chiral fields $\Phi_{ii}$ and $\Phi_{jj}$, which come from dimensional reduction of $4d$ $\mathcal{N}=1$ vector multiplets. The two edges thus give rise to $E_{ij}^{+} =\Phi_{ii}\cdot X_{ij}$ and $E_{ij^{-}}= X_{ij} \cdot \Phi_{jj}$, which together produce $E_{ij}= E_{ij}^{+}- E_{ij}^{-}$. The lift of a brane tiling edge is illustrated in \fref{ftilingedges}.

\paragraph{Brane tiling nodes.} 
As explained earlier, the $J$-terms of the $2d$ theory are in one-to-one correspondence with $F$-terms in $4d$. The $4d$ superpotential $W$ is understood as a function of $2d$ $(0,2)$ chiral multiples $X_{ji}$ under dimensional reduction. The  $J$-terms take the form $J_{ji} = \frac{\partial W}{\partial X_{ji}}= J_{ji}^{+} - J_{ji}^{-}$, since for toric $4d$ theories every $X_{ij}$ appears in the superpotential in two terms with opposite signs. These two superpotential terms are encoded by the white and black nodes at the endpoints of the associated edge of the brane tiling. For every $X_{ij}$, each of these two nodes gives rise to an edge in the brane brick model, generating the two contributions $J_{ji}^{+}$ and $J_{ji}^{-}$. As a result, every node in the brane tilings gives rise to a collection of edges in the brane brick model, one per each chiral field participating in the corresponding superpotential term. All these brane brick edges extend in the $z$ direction between two copies of the same node. A $4d$ chiral field $\mathcal{X}_{ji}$ contains a Fermi field $\Lambda_{ij}$ in its dimensional reduction. The brane brick face associated to $\Lambda_{ij}$ must terminate on the edge for $J_{ji}^\pm$. 

\begin{figure}[ht!]
\begin{center}
\resizebox{1\hsize}{!}{
\includegraphics[trim=0cm 0cm 0cm 0cm,totalheight=10 cm]{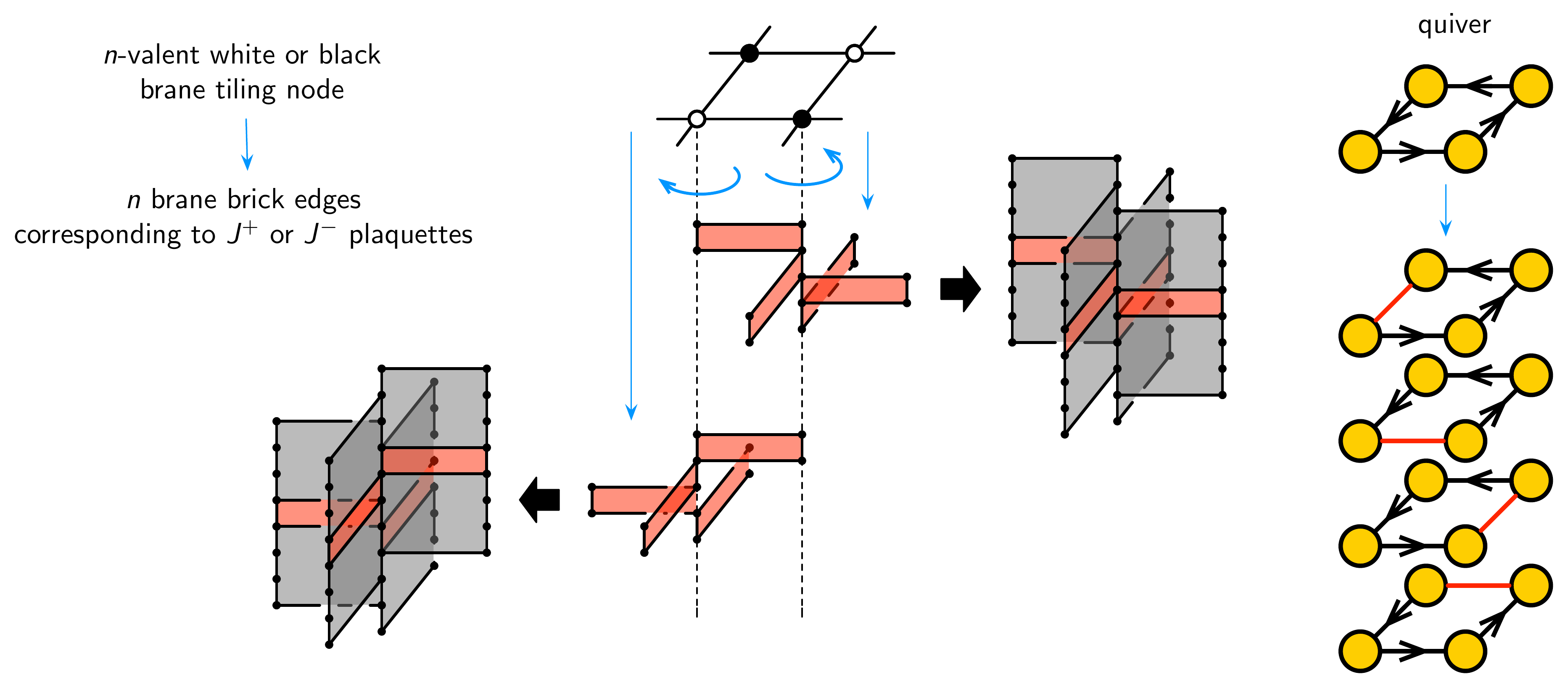}
}  
\caption{A white or black brane tiling node is lifted to a vertical collection of edges corresponding to $J^{+}$ or $J^{-}$ plaquettes, respectively. Every edge of the brane tiling terminating on the node corresponds to a $4d$ chiral field $\mathcal{X}_{ij}$ and gives rise to a pair of faces associated to a chiral multiplet $X_{ij}$ and a Fermi multiplet $\Lambda_{ij}$ in $2d$ $(0,2)$ language. The faces associated to the Fermi fields are attached to the vertical edges and form a helix whose orientation depends on the color of the parent node in the brane tiling. 
\label{ftilingnodes}}
 \end{center}
 \end{figure}

This structure of $J$-terms can be automatically incorporated in the brane brick model by the following construction. The faces associated to the Fermi fields are organized into helices emanating from each brane tiling node, with an orientation on the $x$-$y$-plane given by the one of the brane tiling edges for $\mathcal{X}_{ij}$. By convention, we consider helices that go clockwise/counterclockwise as we move to lower $z$ for white/black brane tiling nodes. The Fermi fields in the helices are glued to the vertical edges generated by the corresponding node. Along the $z$-direction, the gaps between consecutive faces for $\Lambda_{ij}$ are filled by the faces for the $2d$ chiral field $X_{ij}$, which also follow from dimensional reduction of the brane tiling edge for $\mathcal{X}_{ij}$. This guarantees that the vertical edges give rise to all the $J$-term plaquettes. \fref{ftilingnodes} illustrates the lift of nodes in the brane tiling.

\begin{figure}[ht!]
\begin{center}
\resizebox{0.7\hsize}{!}{
\includegraphics[trim=0cm 0cm 0cm 0cm,totalheight=10 cm]{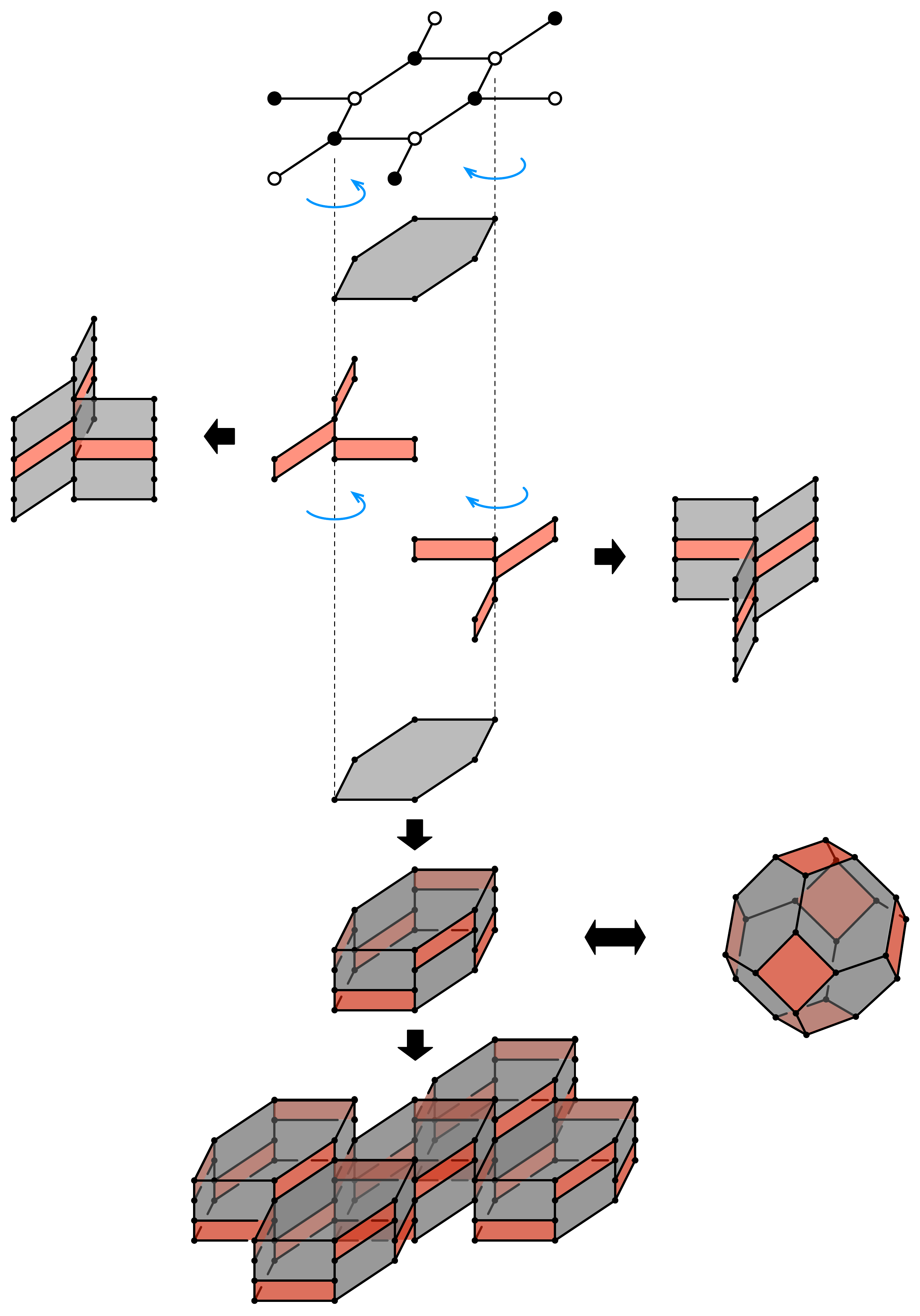}
}  
\caption{
Dimensional reduction of the $\mathbb{C}^3$ brane tiling to the $\mathbb{C}^4$ brane brick model, which consists of a single truncated octahedron brick.
\label{fred_c3}}
 \end{center}
 \end{figure}

The lifting algorithm that we introduced uniquely determines the dimensional reduction. Every brane tiling can be dimensionally reduced to a brane brick model using this procedure. \fref{fred_c3} illustrates the dimensional reduction of the $\mathbb{C}^3$ brane tiling to the brane brick model for $\mathbb{C}^4$. It is clear that the resulting brane brick model can be deformed into the one presented earlier in \fref{ft3diagrams}, which consists of a single truncated octahedron brick.

\subsection{Examples}

Here we present further explicit examples of dimensionally reduced theories. We use the lifting algorithm to produce their brane brick models and collect additional useful information for the theories, including their brick matchings and field content of the phase boundaries.

\subsubsection{$\mathcal{C} \times \mathbb{C}$ \label{sconifold}}

\begin{figure}[ht!]
\begin{center}
\resizebox{1\hsize}{!}{
\includegraphics[trim=0cm 0cm 0cm 0cm,totalheight=10 cm]{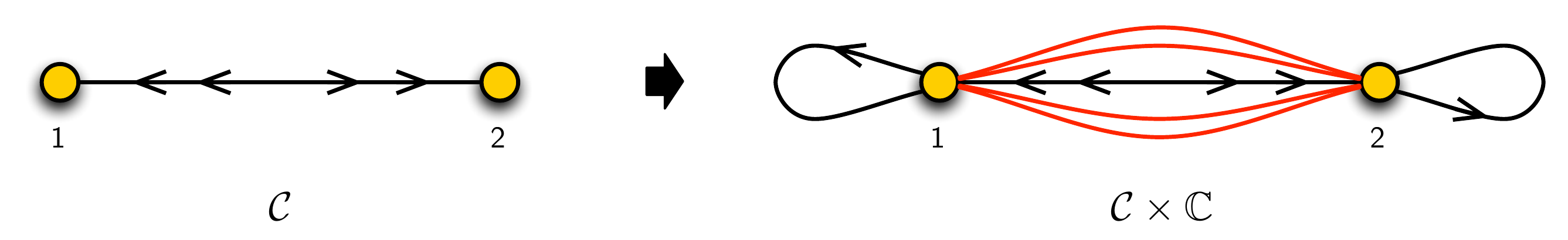}
}  
\caption{Dimensional reduction of the quiver for the conifold $\mathcal{C}$ to the one for $\mathcal{C}\times \mathbb{C}$.
\label{fzconfioldaaa}}
 \end{center}
 \end{figure}

The brane brick model for $\mathcal{C}\times \mathbb{C}$ can be obtained by dimensionally reducing the $4d$ conifold $\mathcal{C}$ theory \cite{Klebanov:1998hh}. The quiver for this theory is given on the left of \fref{fzconfioldaaa} and its superpotential is 
\beal{es200a1c}
W = X_{12} \cdot Y_{21} \cdot Y_{12} \cdot X_{21} - X_{12} \cdot X_{21}  \cdot Y_{12} \cdot Y_{21}  ~.~
\eea
The quiver for the dimensionally reduced $\mathcal{C} \times \mathbb{C}$ theory is shown on the right of \fref{fzconfioldaaa}. The $J$- and $E$-terms are determined by the general expressions \eref{J_dim_red} and \eref{E_dim_red}, and become
\beq
\begin{array}{rclccclcc}
& & \ \ \ \ \ \ \ \ \ \ \ \ \ \ \ \ \ \ \ \ J & & & & \ \ \ \ \ \ \ \ \ \ \ \ \ \ E & & \\
 \Lambda_{12}^{1} : & \ \ \ & X_{21}\cdot X_{12}\cdot Y_{21} - Y_{21}\cdot X_{12}\cdot X_{21}& = & 0  & \ \ \ \ & \Phi_{11}\cdot Y_{12} - Y_{12}\cdot \Phi_{22} & = & 0  \\
 \Lambda_{21}^{1} : & \ \ \ & X_{12}\cdot Y_{21}\cdot Y_{12} - Y_{12}\cdot Y_{21}\cdot X_{12}& = & 0  & \ \ \ \ & \Phi_{22}\cdot X_{21} - X_{21}\cdot \Phi_{11} & = & 0  \\
 \Lambda_{12}^{2} : & \ \ \ & Y_{21}\cdot Y_{12}\cdot X_{21} -X_{21}\cdot Y_{12}\cdot Y_{21} & = & 0  & \ \ \ \ & \Phi_{11}\cdot X_{12} - X_{12}\cdot \Phi_{22} & = & 0 \\
 \Lambda_{21}^{2} : & \ \ \ &  Y_{12}\cdot X_{21}\cdot X_{12} -X_{12}\cdot X_{21}\cdot Y_{12}& = & 0  & \ \ \ \ & \Phi_{22}\cdot Y_{21} - Y_{21}\cdot \Phi_{11}& = & 0
\end{array}
\label{es200a1-new2}
\eeq

The brick matchings can be determined from these $J$- and $E$-terms and are summarized by the matrix
\beal{es200a3}
\small
P_\Lambda=
\left(
\begin{array}{c|ccccc}
\; & p_1 & p_2 & p_3 & p_4 & s \\
\hline
X_{21} & 1 & 0 & 0 & 0 & 0 \\
X_{12} & 0 & 1 & 0 & 0 & 0 \\
Y_{21} & 0 & 0 & 1 & 0 & 0 \\
Y_{12} & 0 & 0 & 0 & 1 & 0 \\
\Phi_{11} & 0 & 0 & 0 & 0 & 1 \\
\Phi_{22} & 0 & 0 & 0 & 0 & 1 \\
\hline
\Lambda_{12}^{1} & 0 & 0 & 0 & 1 & 1
\\
\Lambda_{21}^{1} & 1 & 0 & 0 & 0 & 1
\\
\Lambda_{12}^{2} & 0 & 1 & 0 & 0 & 1
\\
\Lambda_{21}^{2} & 0 & 0 & 1 & 0 & 1
\\
\end{array}
\right)
~.~
\eea

\begin{figure}[ht!!]
\begin{center}
\resizebox{0.3\hsize}{!}{
\includegraphics[trim=0cm 0cm 0cm 0cm,totalheight=10 cm]{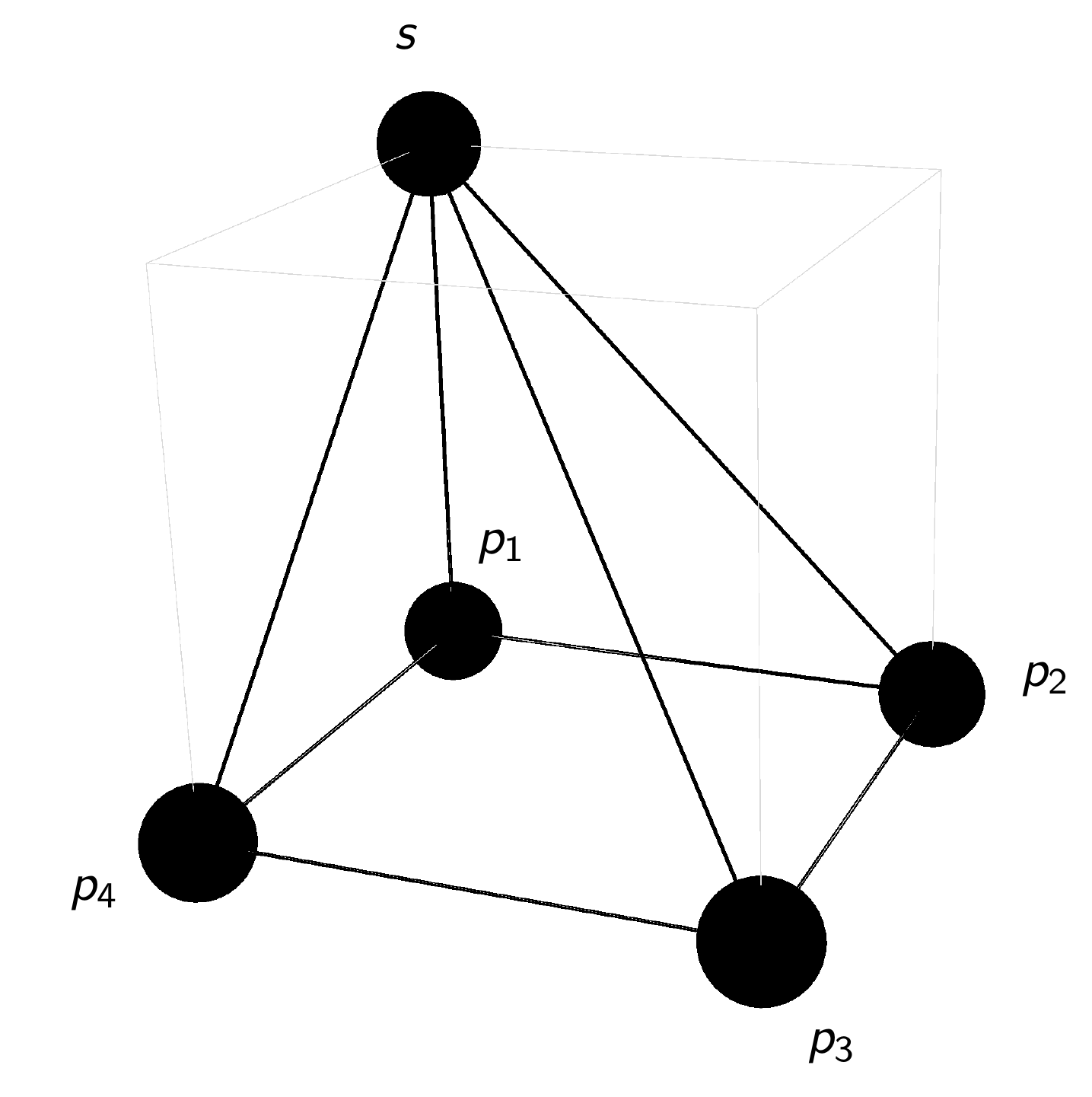}
}  
\caption{Toric diagram for $\mathcal{C} \times \mathbb{C}$, obtained as the classical mesonic moduli space of the dimensionally reduced conifold gauge theory.
\label{fzconfiold}}
 \end{center}
 \end{figure}

Using either the standard forward algorithm of \cite{Franco:2015tna} or the fast forward algorithm of section \sref{section_fast_forward_algorithm} based on the brick matchings, it is straightforward to verify that the classical mesonic moduli space of this theory is indeed $\mathcal{C}\times \mathbb{C}$, as shown in \fref{fzconfiold}.

\begin{figure}[ht!!]
\begin{center}
\resizebox{0.9\hsize}{!}{
\includegraphics[trim=0cm 0cm 0cm 0cm,totalheight=10 cm]{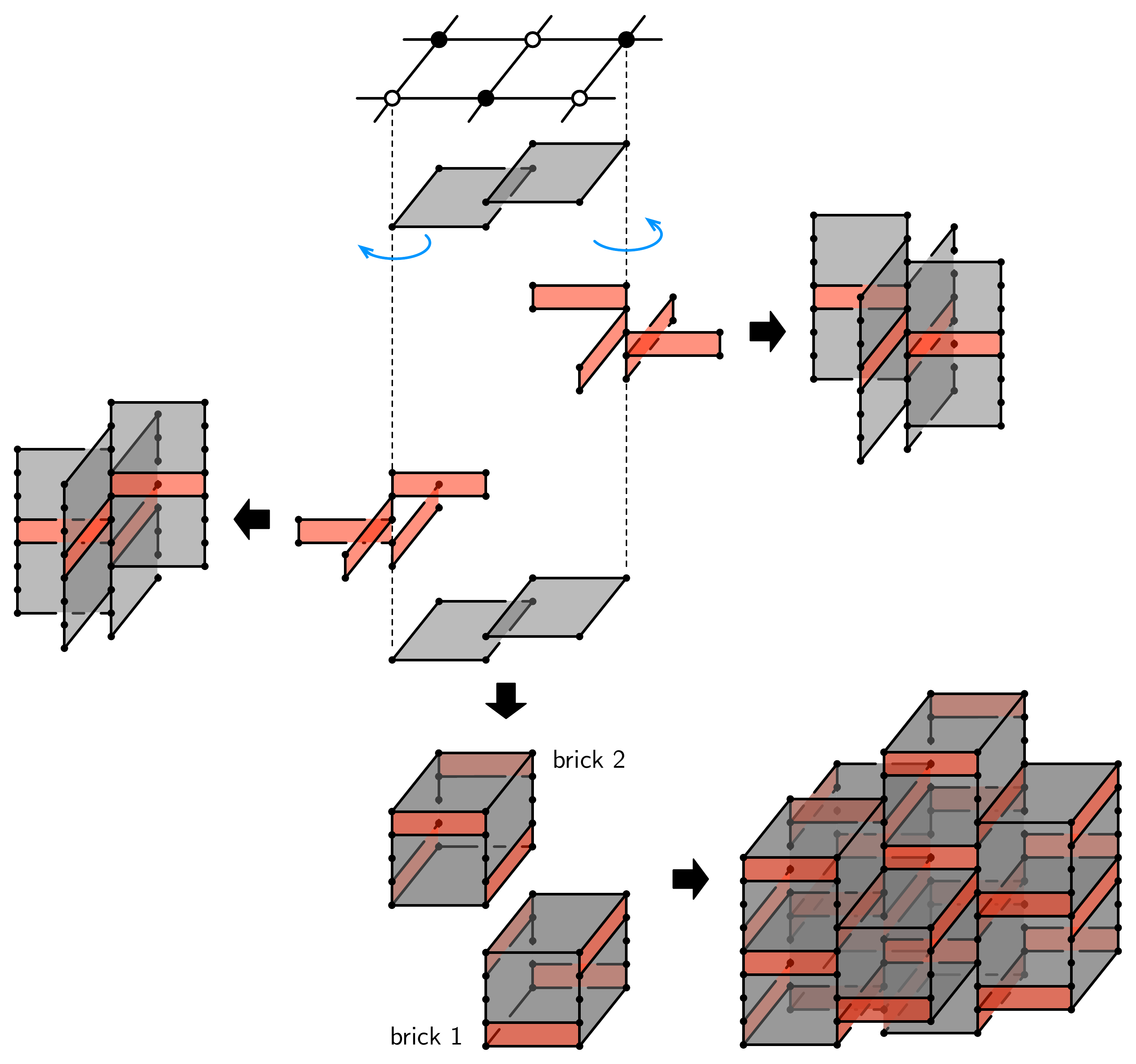}
}  
\caption{
Dimensional reduction of the conifold brane tiling to the $\mathcal{C}\times\mathbb{C}$ brane brick model.
\label{fred_con}}
 \end{center}
 \end{figure}

The phase boundaries can be determined from the brick matchings and are encoded by the following matrix
\beal{es200a4}
\small
H =
\left(
\begin{array}{c|cccccccc}
\; &\eta_{12} & \eta_{34} & \eta_{14} & \eta_{23} & \eta_{s1} & \eta_{s2} & \eta_{s3} & \eta_{s4}
\\
\hline
X_{21} & 1 & 0 & 1 & 0 & -1 & 0 & 0 & 0 \\
X_{12} &  -1 & 0 & 0 & 1 & 0 & -1 & 0 & 0 \\
Y_{21} & 0 & 1 & 0 & -1 & 0 & 0 & -1 & 0 \\
Y_{12} & 0 & -1 & -1 & 0 & 0 & 0 & 0 & -1 \\
\Phi_{11} & 0 & 0 & 0 & 0 & 1 & 1 & 1 & 1 \\
\Phi_{22} & 0 & 0 & 0 & 0 & 1 & 1 & 1 & 1 \\
\hline
\Lambda_{12}^{1} & 0 & -1 & -1 & 0 & 1 & 1 & 1 & 0 \\
\Lambda_{21}^{1} & 1 & 0 & 1 & 0 & 0 & 1 & 1 & 1 \\
\Lambda_{12}^{2} & -1 & 0 & 0 & 1 & 1 & 0 & 1 & 1 \\
\Lambda_{21}^{2} & 0 & 1 & 0 & -1 & 1 & 1 & 0 & 1 \\
\end{array}
\right)
~.~
\eea
Applying the lifting algorithm to the brane tiling of the conifold theory, we obtain the brane brick model for $\mathcal{C}\times \mathbb{C}$ shown in \fref{fred_con}. It consists of two copies of the same type of brick, which are rotated by $90^\circ$ with respect to each other, representing the two gauge groups of the theory. Each of the bricks contains four 8-sided faces representing bifundamental chiral fields, two 4-sided faces representing adjoint chiral fields and four 4-sided faces representing Fermi fields. Every edge in the brane brick model is adjacent to a single Fermi field and all the $J$- and $E$-terms in \eref{es200a1-new2} are nicely generated. The brane brick model is in precise agreement with the dual periodic quiver constructed in \cite{Franco:2015tna}. 

\subsubsection{$\text{SPP} \times \mathbb{C}$ \label{sspp}}

\begin{figure}[ht!!]
\begin{center}
\resizebox{0.9\hsize}{!}{
\includegraphics[trim=0cm 0cm 0cm 0cm,totalheight=10 cm]{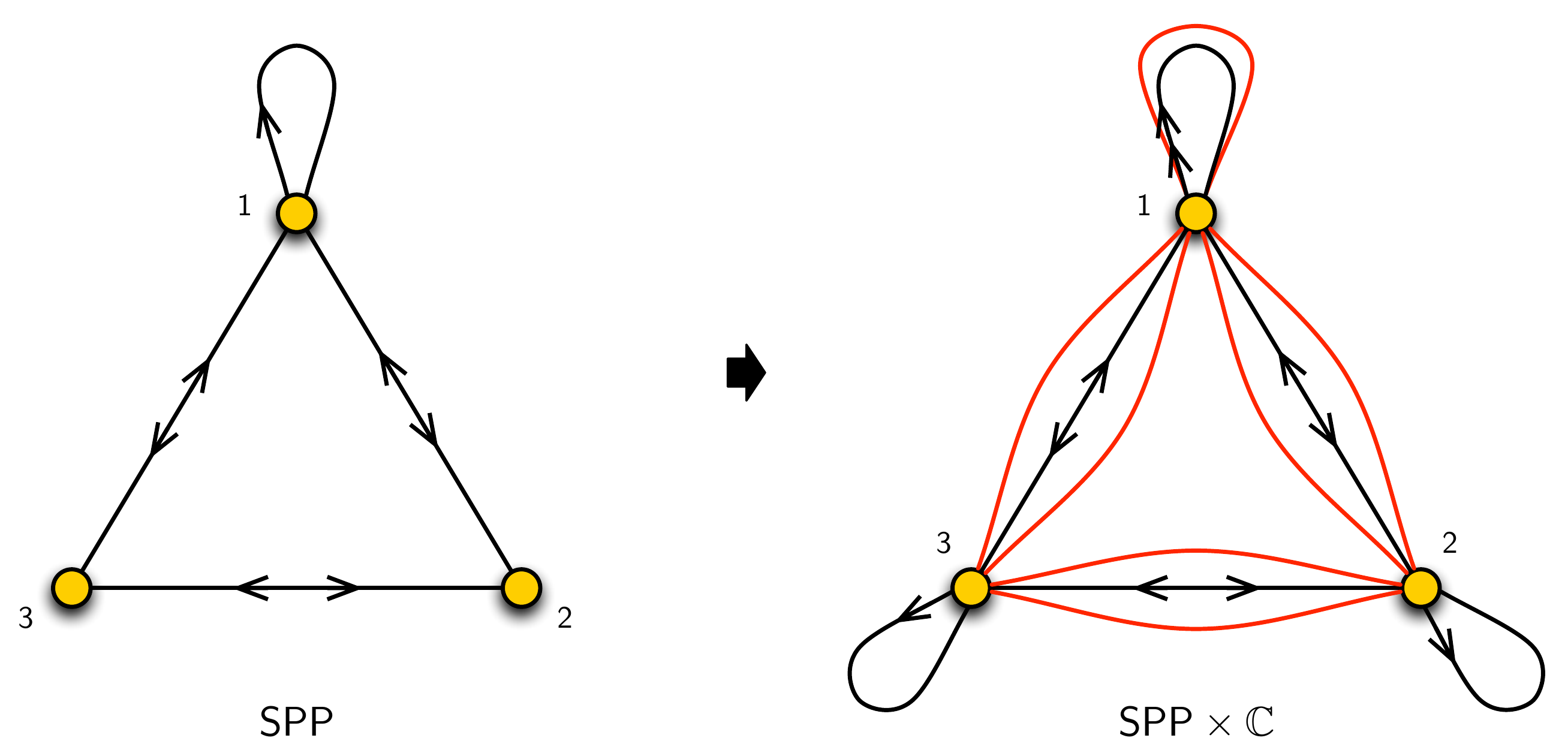}
}  
\caption{Dimensional reduction of the quiver for $\text{SPP}$ to the one for $\text{SPP}\times\mathbb{C}$. 
\label{fzspp}}
 \end{center}
 \end{figure}

By dimensionally reducing the suspended pinch point (SPP) theory \cite{Morrison:1998cs}, whose brane tiling was originally introduced in \cite{Franco:2005rj}, it is possible to obtain the brane brick model for $\text{SPP}\times\mathbb{C}$. The quivers for the parent $4d$ theory and its dimensional reduction are shown in \fref{fzspp}. The $4d$ superpotential is
\beal{es500a1}
W =  X_{13} \cdot X_{31}  \cdot X_{11} + X_{12}\cdot X_{23}\cdot X_{32}\cdot X_{21} - X_{12}\cdot X_{21}\cdot X_{11} - X_{13}\cdot X_{32}\cdot X_{23}\cdot X_{31}  ~.~ \nonumber \\
\eea

The $J$- and $E$-terms of the $2d$ theory take the form
\beq
\begin{array}{rcrccclcc}
& & J  \ \ \ \ \ \ \ \ \ \ \ \ \ & & & & \ \ \ \ \ \ \ \ \ \ \ \ \ \ \ \ \ \ \ E & & \\
\Lambda_{11 } : & \ \ \ & X_{13} \cdot X_{31} - X_{12} \cdot X_{21}& = & 0 & \ \ \ \ &\Phi_{11} \cdot X_{11} - X_{11} \cdot \Phi_{11}&= & 0 \\
\Lambda_{21} : & \ \ \ & X_{12} \cdot X_{23} \cdot X_{32} - X_{11} \cdot X_{12} & = & 0 & \ \ \ \ &\Phi_{22} \cdot X_{21} - X_{21} \cdot \Phi_{11}& = & 0 \\
\Lambda_{12} : & \ \ \ & X_{21} \cdot X_{11} - X_{23} \cdot X_{32} \cdot X_{21}& = & 0 & \ \ \ \ &\Phi_{11} \cdot X_{12} - X_{12} \cdot \Phi_{22}& = & 0 \\
\Lambda_{31} : & \ \ \ & X_{13} \cdot X_{32} \cdot X_{23}- X_{11} \cdot X_{13}& = & 0 & \ \ \ \ & \Phi_{33} \cdot X_{31} - X_{31} \cdot \Phi_{11}& = & 0 \\
\Lambda_{13} : & \ \ \ & X_{31} \cdot X_{11} - X_{32} \cdot X_{23} \cdot X_{31}& = & 0 & \ \ \ \ &\Phi_{11} \cdot X_{13} - X_{13} \cdot \Phi_{33}& = & 0 \\
\Lambda_{32} : & \ \ \ & X_{21} \cdot X_{12} \cdot X_{23} -X_{23} \cdot X_{31} \cdot X_{13}  & = & 0 & \ \ \ \ &\Phi_{33} \cdot X_{32} - X_{32} \cdot \Phi_{22} & = & 0 \\
\Lambda_{23} : & \ \ \ & X_{32} \cdot X_{21} \cdot X_{12} - X_{31} \cdot X_{13} \cdot X_{32}& = & 0 & \ \ \ \ & \Phi_{22} \cdot X_{23} - X_{23} \cdot \Phi_{33}& = & 0 
\end{array}
\label{es500a2}
\eeq
The brick matchings are then given by

\beal{es500a3}
\small
P_\Lambda=
\left(
\begin{array}{c|ccccccc}
\; & p_1 & p_2 & p_3 & p_4 & q_1 & q_2 & s \\
\hline
X_{23} & 1 & 0 & 0 & 0 & 0 & 0 & 0 \\
X_{11} & 1 & 1 & 0 & 0 & 0 & 0 & 0 \\
X_{32} & 0 & 1 & 0 & 0 & 0 & 0 & 0 \\
X_{13} & 0 & 0 & 1 & 0 & 1 & 0 & 0 \\
X_{21} & 0 & 0 & 1 & 0 & 0 & 1 & 0 \\
X_{12} & 0 & 0 & 0 & 1 & 1 & 0 & 0 \\
X_{31} & 0 & 0 & 0 & 1 & 0 & 1 & 0 \\
\Phi_{11} & 0 & 0 & 0 & 0 & 0 & 0 & 1 \\
\Phi_{22} & 0 & 0 & 0 & 0 & 0 & 0 & 1 \\
\Phi_{33} & 0 & 0 & 0 & 0 & 0 & 0 & 1 \\
\hline
\Lambda_{11} & 
1 & 1 & 0 & 0 & 0 & 0 & 1 \\
\Lambda_{21} &
 0 & 0 & 1 & 0 & 0 & 1 & 1 \\
\Lambda_{12} &
 0 & 0 & 0 & 1 & 1 & 0 & 1 \\
\Lambda_{31} &
 0 & 0 & 0 & 1 & 0 & 1 & 1 \\
\Lambda_{13} &
 0 & 0 & 1 & 0 & 1 & 0 & 1 \\
\Lambda_{32} &
 0 & 1 & 0 & 0 & 0 & 0 & 1 \\
\Lambda_{23} &
 1 & 0 & 0 & 0 & 0 & 0 & 1 \\
\end{array}
\right)~,~
\eea
from which we determine the phase boundaries
\beal{es500a4}
\small
H=
\left(
\begin{array}{c|ccccccccccc}
\; & \eta_{12} & \eta_{14} & \eta_{23} & \eta_{3q_1} & \eta_{3q_2} & \eta_{4q_1} & \eta_{3q_2} & \eta_{s1} & \eta_{s2} & \eta_{s3} & \eta_{s4}\\ 
\hline
X_{23} &
 1 & 1 & 0 & 0 & 0 & 0 & 0 & -1 & 0 & 0 & 0 \\
X_{11} &
 0 & 1 & 1 & 0 & 0 & 0 & 0 & -1 & -1 & 0 & 0 \\
X_{32} &
 -1 & 0 & 1 & 0 & 0 & 0 & 0 & 0 & -1 & 0 & 0 \\
X_{13} &
 0 & 0 & -1 & 0 & 1 & -1 & 0 & 0 & 0 & -1 & 0 \\
X_{21} &
 0 & 0 & -1 & 1 & 0 & 0 & -1 & 0 & 0 & -1 & 0 \\
X_{12} &
 0 & -1 & 0 & -1 & 0 & 0 & 1 & 0 & 0 & 0 & -1 \\
X_{31} &
 0 & -1 & 0 & 0 & -1 & 1 & 0 & 0 & 0 & 0 & -1 \\
\Phi_{11} & 
 0 & 0 & 0 & 0 & 0 & 0 & 0 & 1 & 1 & 1 & 1 \\
\Phi_{22} &
 0 & 0 & 0 & 0 & 0 & 0 & 0 & 1 & 1 & 1 & 1 \\
\Phi_{33} & 
 0 & 0 & 0 & 0 & 0 & 0 & 0 & 1 & 1 & 1 & 1 \\
\hline
\Lambda_{11} & 
 0 & 1 & 1 & 0 & 0 & 0 & 0 & 0 & 0 & 1 & 1 \\
\Lambda_{21} &
 0 & 0 & -1 & 1 & 0 & 0 & -1 & 1 & 1 & 0 & 1 \\
\Lambda_{12} &
 0 & -1 & 0 & -1 & 0 & 0 & 1 & 1 & 1 & 1 & 0 \\
\Lambda_{31} &
 0 & -1 & 0 & 0 & -1 & 1 & 0 & 1 & 1 & 1 & 0 \\
\Lambda_{13} &
 0 & 0 & -1 & 0 & 1 & -1 & 0 & 1 & 1 & 0 & 1 \\
\Lambda_{32} &
 -1 & 0 & 1 & 0 & 0 & 0 & 0 & 1 & 0 & 1 & 1 \\
\Lambda_{23} &
 1 & 1 & 0 & 0 & 0 & 0 & 0 & 0 & 1 & 1 & 1 \\
\end{array}
\right)~.~
\eea

\medskip

\begin{figure}[ht!!]
\begin{center}
\resizebox{0.4\hsize}{!}{
\includegraphics[trim=0cm 0cm 0cm 0cm,totalheight=10 cm]{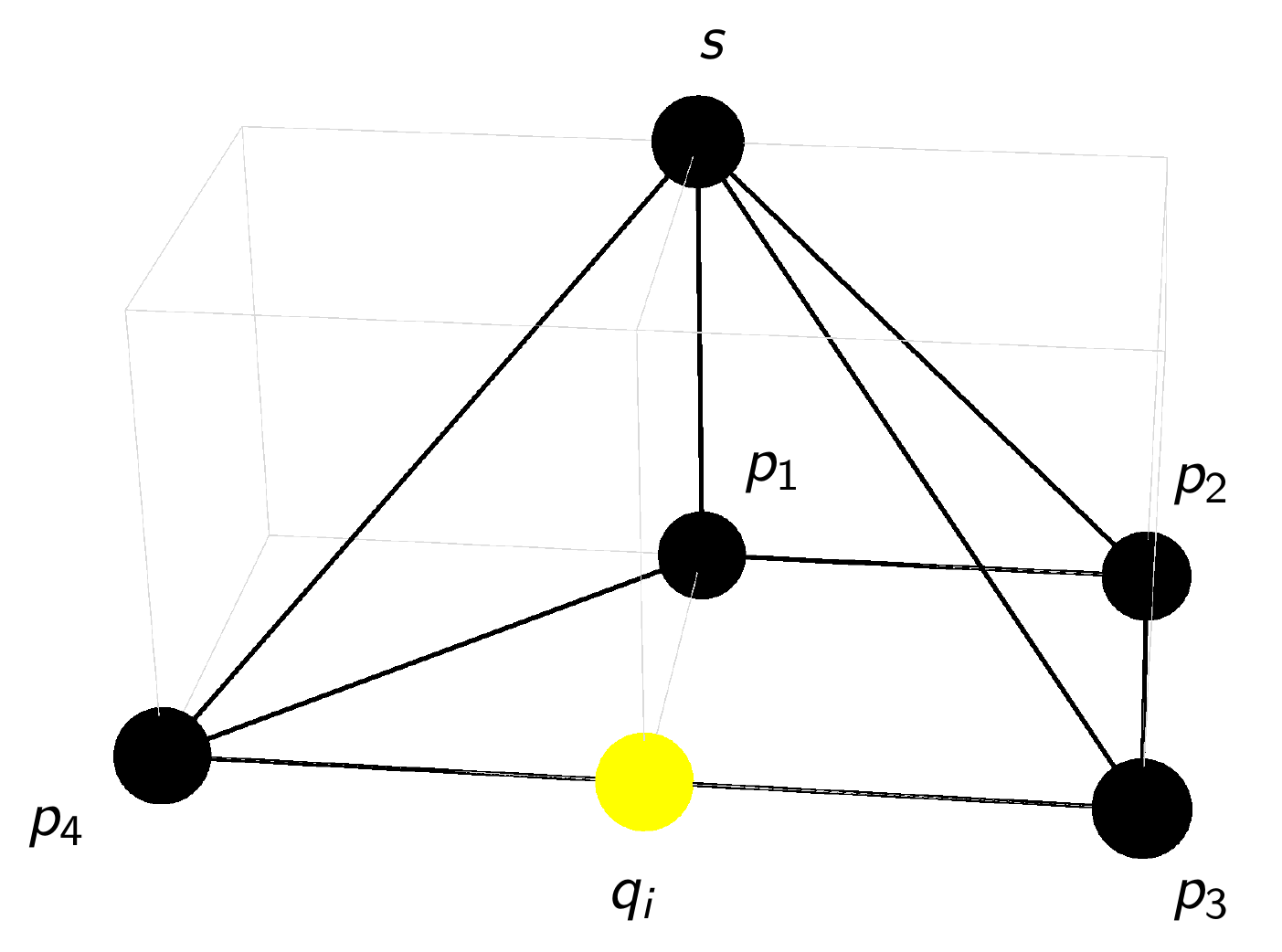}
}  
\caption{Toric diagram for $\text{SPP} \times \mathbb{C}$, obtained as the classical mesonic moduli space of the dimensionally reduced $\text{SPP}$ gauge theory.
\label{fzsppc}}
 \end{center}
 \end{figure}

Using the brick matchings in the fast forward algorithm, we confirm the mesonic moduli space is $\text{SPP} \times \mathbb{C}$, whose toric diagram is shown in \fref{fzsppc}.

\begin{figure}[ht!!]
\begin{center}
\resizebox{0.76\hsize}{!}{
\includegraphics[trim=0cm 0cm 0cm 0cm,totalheight=10 cm]{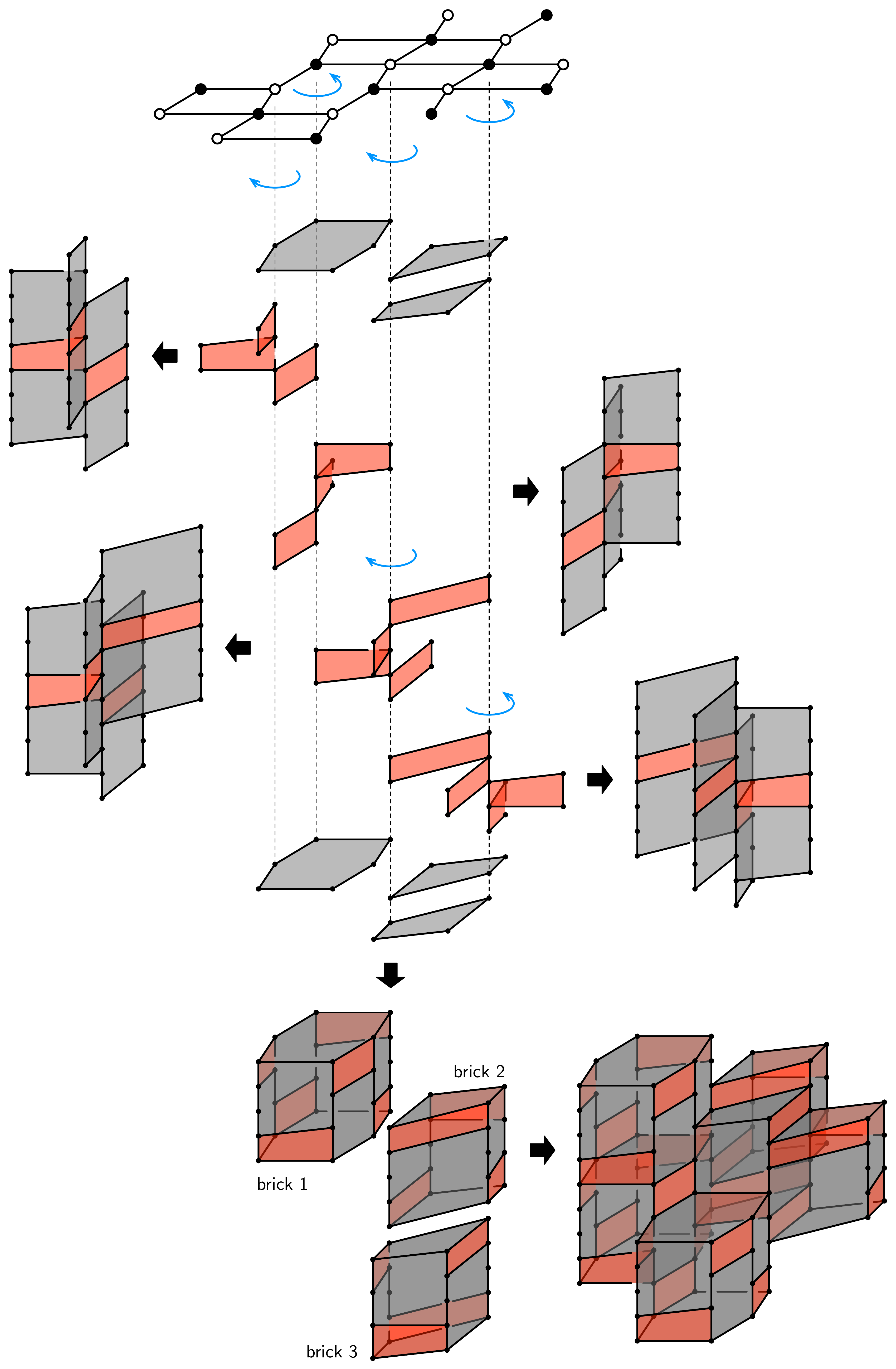}
}  
\caption{
Dimensional reduction of the $\text{SPP}$ brane tiling to the $\text{SPP}\times\mathbb{C}$ brane brick model.
\label{fred_spp}}
 \end{center}
 \end{figure}

\fref{fred_spp} shows the brane brick model obtained by acting with the lifting algorithm on the SPP brane tiling. In this case, there are three different bricks, corresponding to the three gauge groups of the theory. The first brick is made of two 8-sided faces corresponding to chiral adjoint fields, four 7-sided faces corresponding to chiral bifundamental fields and two 6-sided faces corresponding to two chiral adjoint fields. This brick also has six 4-sided Fermi faces. The two other bricks are identical up to a reflection with respect to their center. Both bricks are made of two 8-sided and two 7-sided faces corresponding to chiral bifundamental fields, and two 4-sided faces corresponding to chiral adjoint fields. They also contain four 4-sided faces corresponding to Fermi fields. Every edge is attached to a single Fermi face and all the terms in \eref{es500a2} are properly accounted for. Once again, the brane brick model agrees with the periodic quiver for this geometry constructed in \cite{Franco:2015tna}.

\section{Beyond Orbifolds and Dimensionally Reduced Theories}

\label{section_general_CY4}

We conclude this article with some examples of brane brick models for general toric singularities, which are neither abelian orbifolds of $\mathbb{C}^4$ nor of the form $\mathrm{CY}_3 \times \mathbb{C}$. As mentioned in earlier sections, there are alternative ways of constructing such models.

A first systematic algorithm for constructing the $2d$ gauge theory for an arbitrary toric CY$_4$ singularity was introduced in \cite{Franco:2015tna}. The procedure consists of obtaining the desired singularity by partial resolution of a parent one for which the gauge theory is known, which translates into higgsing in the gauge theory. Abelian orbifolds of $\mathbb{C}^4$ provide canonical starting points for partial resolution, since any toric diagram can be embedded into the one for a $\mathbb{C}^4/\mathbb{Z}_{n_1}\times \mathbb{Z}_{n_2} \times \mathbb{Z}_{n_3}$ orbifold with action $(1,0,0,-1)(0,1,0,-1)(0,0,1,-1)$ for sufficiently large $n_1$, $n_2$ and $n_3$. In this case, the orbifold toric diagram is a tetrahedron of length $n_1$, $n_2$ and $n_3$ along the three axes. The identification of the chiral fields whose scalar components acquire non-zero expectation values is straightforward thanks to the $P$-matrix reviewed in section \sref{sbricktogeom}, which provides a map between the points to be removed in the toric diagram and chiral fields. The full gauge theory associated to the CY$_4$ singularity under consideration, namely the quiver diagram and the $J$- and $E$-terms, is obtained by implementing the higgsing. If we are interested in the periodic quiver or brane brick model for such a generic theory, we simply follow how they transform under higgsing from the parent theory. The periodic quivers and brane brick models for general abelian orbifolds of $\mathbb{C}^4$ are very simple and were presented in section \sref{section_dictionary}. Additional aspects on the interplay between partial resolution and phase boundaries were presented in section \sref{section_partial_resolution}.

A second systematic procedure for constructing the gauge theory for a general toric CY$_4$ singularity, which in fact directly constructs the periodic quiver and brane brick model, is the fast inverse algorithm introduced in section \sref{sgeomtobrick}.

\subsection{Brane Brick Models from Global Symmetries}

In this section, we study a slightly different but closely related question: {\it if the gauge theory for a toric CY$_4$ singularity is known} in the form of an ordinary quiver diagram and its $J$- and $E$-terms, how do we construct the periodic quiver or, equivalently, the brane brick model. Of course this can always be done by ``brute force". However, as we explain below, the global symmetries of the gauge theory can be exploited for methodically constructing the periodic quiver.

A useful tool for this endeavor is the Hilbert series which, as reviewed in appendix \ref{shilbert}, is a generating function that counts chiral gauge invariant operators \cite{Benvenuti:2006qr,Hanany:2006uc,Feng:2007ur,Butti:2007jv,Hanany:2007zz}. It fully characterizes the algebraic structure of the classical mesonic moduli space $\mathcal{M}$. Using plethystics \cite{Benvenuti:2006qr,Hanany:2007zz,Feng:2007ur}, the algebraic structure of $\mathcal{M}$ given by the generators and relations of the algebraic variety can be identified from the Hilbert series.

The generators of $\mathcal{M}$ (and in fact any mesonic operator) are gauge invariant, hence they correspond to a collection of chiral fields in the periodic quiver forming a closed oriented path on $T^3$.\footnote{Since Fermi fields do not have scalar components, they do not contribute to moduli space generators.} Notice that, however, they generically possess non-trivial winding numbers. Every gauge invariant operator in $\mathcal{M}$ can be expressed in terms of its generators. Hence, identifying the generators, and in particular their homology on $T^3$, enables us to construct the entire periodic quiver.

The global symmetry of every $2d$ gauge theory associated to a toric CY$_4$ singularity automatically contains a $U(1)^3 \times U(1)_R$ Cartan subgroup that corresponds to the isometries of the underlying CY$_4$. Part of this symmetry is nicely geometrized by the periodic quiver. In particular, it is possible to identify a basis for the $U(1)^3$ piece such that the corresponding charges of mesonic operators are given by the winding numbers on $T^3$ of the corresponding paths in the periodic quiver. The winding number of any closed path on $T^3$ is a vector in $\mathbb{Z}^3$ that counts how many times it winds along the three fundamental directions. By identifying appropriate $U(1)^3$ charges of the generators and individual chiral fields, it is possible to use the winding numbers of the paths corresponding to the generators to construct the periodic quiver.

In the following section we will present the brane brick models for $D_3$ and $Q^{1,1,1}$. The full gauge theories for both geometries were originally derived in \cite{Franco:2015tna}, but the periodic quivers were not provided there. In order to explicitly construct these periodic quivers, we have used the ideas presented in this section in combination with the detailed analysis of the global symmetries and the mesonic moduli spaces of these theories presented in appendix \ref{shilbert}.

\subsection{Examples}

\subsubsection{$D_3$ Brane Brick Model}

\begin{figure}[ht!]
\begin{center}
\resizebox{0.8\hsize}{!}{
\includegraphics[trim=0cm 0cm 0cm 0cm,totalheight=10 cm]{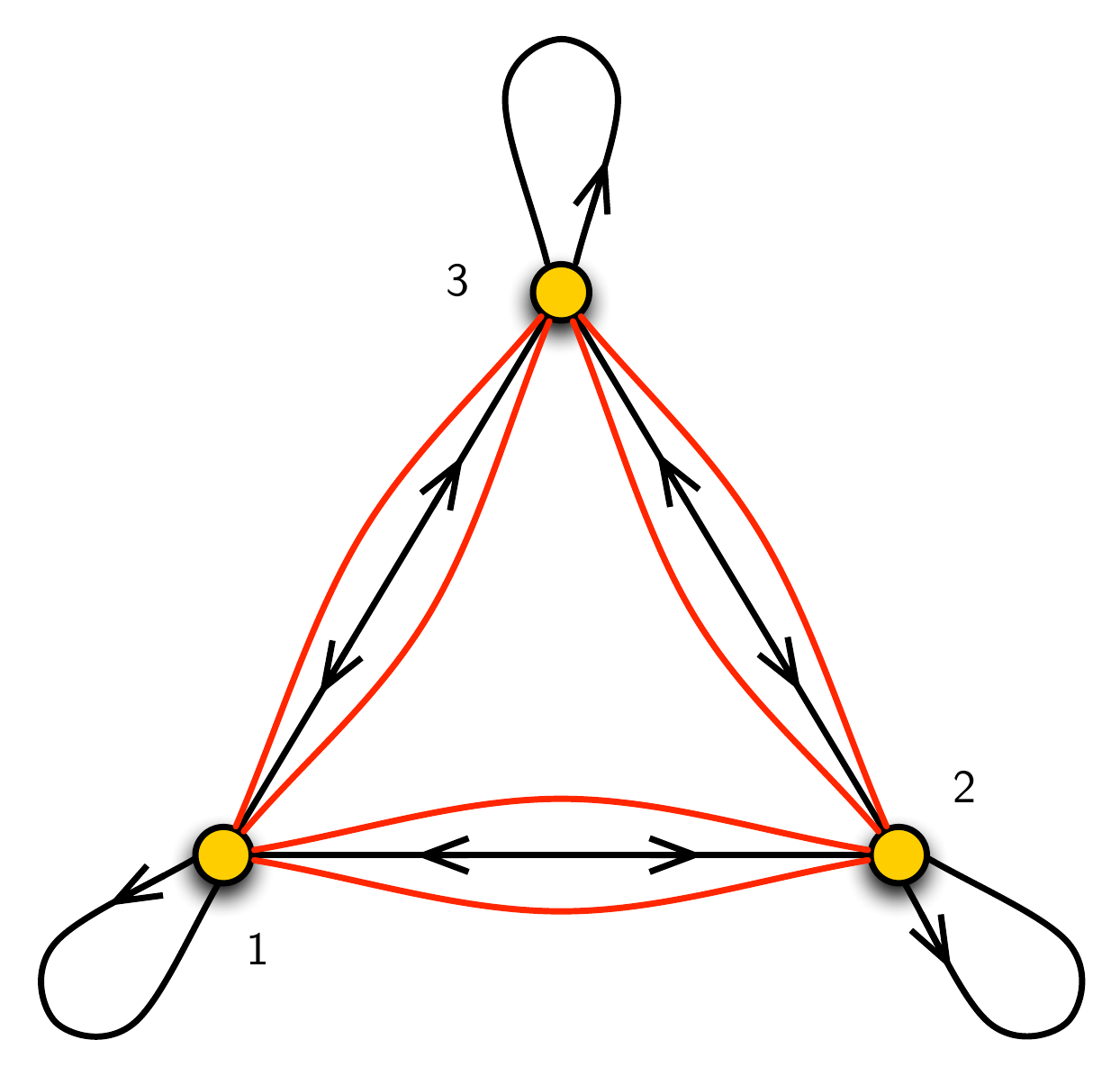}
\includegraphics[trim=0cm 0cm 0cm 0cm,totalheight=10 cm]{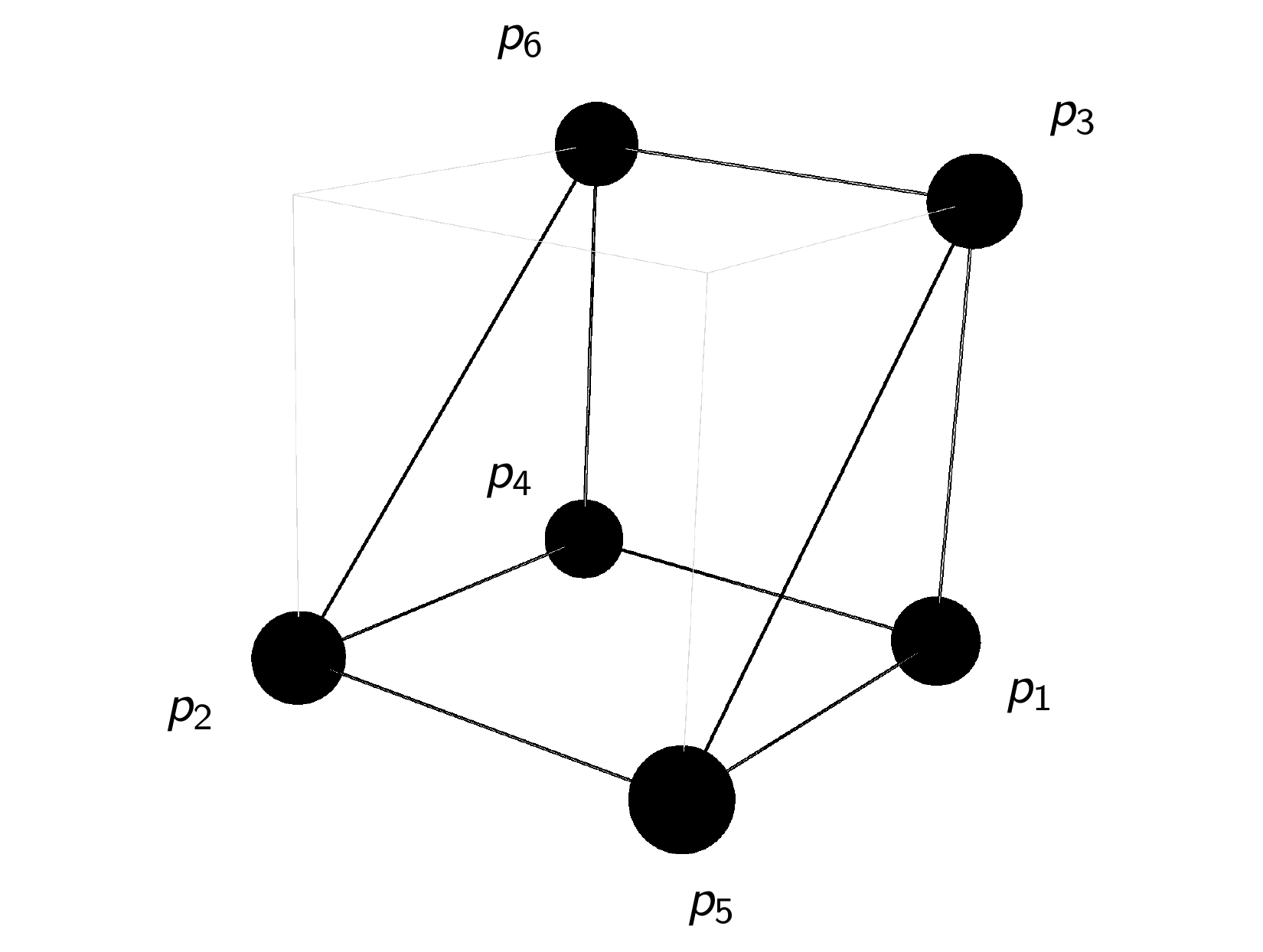}
}  
\caption{
Quiver and toric diagrams for $D_3$.
\label{quiver_toric_D3}}
 \end{center}
 \end{figure}

The quiver diagram for the $D_3$ theory is shown in \fref{quiver_toric_D3}. Its $J$- and $E$-terms are\footnote{The $X$, $Y$, $Z$ and $D$ labels for chiral fields used in this section for the $D_3$ and $Q^{1,1,1}$ theories arise when obtaining them by partial resolution of abelian orbifolds as in \cite{Franco:2015tna}.}
\beq
\begin{array}{rcrcccrcc}
& &  J \ \ \ \ \ \ \ \ \ \ \ \ \ & & & &  E \ \ \ \ \ \ \ \ \ \ \ \ \ & & \\
\Lambda_{21} : & \ \ \ & X_{13}\cdot X_{31}\cdot Y_{12}-Y_{12}\cdot X_{22 }& = & 0 & \ \ \ \ & D_{23}\cdot Z_{32}\cdot Z_{21}-Z_{21}\cdot D_{11}&= & 0 \\
 \Lambda_{12} : & \ \ \ & Z_{21}\cdot X_{13}\cdot X_{31} -X_{22}\cdot Z_{21}& = & 0 & \ \ \ \ &D_{11}\cdot Y_{12} - Y_{12}\cdot D_{23}\cdot Z_{32}&= & 0 \\
 \Lambda_{31}: & \ \ \ & X_{13}\cdot Y_{33} - Y_{12}\cdot Z_{21}\cdot X_{13}& = & 0 & \ \ \ \ & X_{31}\cdot D_{11} - Z_{32}\cdot D_{23}\cdot X_{31}&= & 0 \\
  \Lambda_{13} : & \ \ \ & X_{31}\cdot Y_{12}\cdot Z_{21} -Y_{33}\cdot X_{31}& = & 0 & \ \ \ \ &D_{11}\cdot X_{13} - X_{13}\cdot Z_{32}\cdot D_{23}&= & 0 \\
 \Lambda_{23}^{1 } : & \ \ \ & Y_{33}\cdot Z_{32} - Z_{32}\cdot Z_{21}\cdot Y_{12}& = & 0 & \ \ \ \ & D_{23}\cdot X_{31}\cdot X_{13} -X_{22}\cdot D_{23} &= & 0 \\
 \Lambda_{23}^{2 } : & \ \ \ & Z_{32}\cdot X_{22} - X_{31}\cdot X_{13}\cdot Z_{32}& = & 0 & \ \ \ \ &D_{23}\cdot Y_{33} - Z_{21}\cdot Y_{12}\cdot D_{23} &= & 0
\end{array}
\label{es130a1}
\eeq
In addition to the global $U(1)^3 \times U(1)_R$ symmetry, the theory has a discrete global $D_3$ symmetry.

From the $J$- and $E$-terms, we obtain the following brick matchings
\beal{es130aa2}
\small
P_\Lambda=
\left(
\begin{array}{c|cccccc}
\; & p_1 & p_2 & p_3 & p_4 & p_5 & p_6 
\\
\hline
D_{11} &1 & 0 & 0 & 1 & 0 & 0 \\
X_{22} & 0 & 1 & 0 & 0 & 1 & 0 \\
Y_{33} & 0 & 0 & 1 & 0 & 0 & 1 \\
D_{23} & 1 & 0 & 0 & 0 & 0 & 0 \\
Z_{32} & 0 & 0 & 0 & 1 & 0 & 0 \\
X_{13} & 0 & 1 & 0 & 0 & 0 & 0 \\
X_{31} & 0 & 0 & 0 & 0 & 1 & 0 \\
Y_{12} & 0 & 0 & 1 & 0 & 0 & 0 \\
Z_{21} & 0 & 0 & 0 & 0 & 0 & 1 \\
\hline
\Lambda_{21} & 1 & 0 & 0 & 1 & 0 &1\\
\Lambda_{12} & 1 & 0 & 1 & 1 & 0&0\\
\Lambda_{31} & 1 & 0 & 0 & 1 & 1&0\\
\Lambda_{13} & 1 & 1 & 0 & 1 & 0&0\\
\Lambda_{23}^{1} & 1 & 1 & 0 & 0 & 1&0\\
\Lambda_{23}^{2} & 1 & 0 & 1 & 0 & 0&1\\
\end{array}
\right)
~.~
\eea
The fast forward algorithm confirms the moduli space is $D_3$, whose toric diagram is shown in \fref{quiver_toric_D3}.

The phase boundary matrix is
\beal{es130cc1}
\small
H=
\left(
\begin{array}{c|ccccccccc}
\; & \eta_{14} & \eta_{36} & \eta_{52} & \eta_{13} & \eta_{35} & \eta_{51} & \eta_{46} & \eta_{62} & \eta_{24} \\
\hline
D_{11} &
 0 & 0 & 0 & 1 & 0 & -1 & 1 & 0 & -1 \\
X_{22} &
 0 & 0 & 0 & 0 & -1 & 1 & 0 & -1 & 1 \\
Y_{33} &
 0 & 0 & 0 & -1 & 1 & 0 & -1 & 1 & 0 \\
D_{23} &
 1 & 0 & 0 & 1 & 0 & -1 & 0 & 0 & 0 \\
Z_{32} &
 -1 & 0 & 0 & 0 & 0 & 0 & 1 & 0 & -1 \\
X_{13} &
 0 & 0 & -1 & 0 & 0 & 0 & 0 & -1 & 1 \\
X_{31} &
 0 & 0 & 1 & 0 & -1 & 1 & 0 & 0 & 0 \\
Y_{12} &
 0 & 1 & 0 & -1 & 1 & 0 & 0 & 0 & 0 \\
Z_{21} &
 0 & -1 & 0 & 0 & 0 & 0 & -1 & 1 & 0 \\
\hline
\Lambda_{21} & 
 0 & -1 & 0 & 1 & 0 & -1 & 0 & 1 & -1 \\
\Lambda_{12} & 
 0 & 1 & 0 & 0 & 1 & -1 & 1 & 0 & -1 \\
\Lambda_{31} & 
 0 & 0 & 1 & 1 & -1 & 0 & 1 & 0 & -1 \\
\Lambda_{13} & 
 0 & 0 & -1 & 1 & 0 & -1 & 1 & -1 & 0 \\
\Lambda_{23}^{1} &
 1 & 0 & 0 & 1 & -1 & 0 & 0 & -1 & 1 \\
\Lambda_{23}^{2} &
 1 & 0 & 0 & 0 & 1 & -1 & -1 & 1 & 0 
\end{array}
\right)
~.~
\eea
The periodic quiver can be constructed using the ideas introduced in the previous section, with the aid of the detailed symmetry analysis of appendix \ref{section_Hilbert_D3}. It is shown in \fref{fd3}, which also presents a few representative phase boundaries ($\eta_{14}$, $\eta_{36}$, $\eta_{52}$ and $\eta_{51}$)

\begin{figure}[ht!!!]
\begin{center}
\resizebox{0.9\hsize}{!}{
\includegraphics[trim=0cm 0cm 0cm 0cm,totalheight=10 cm]{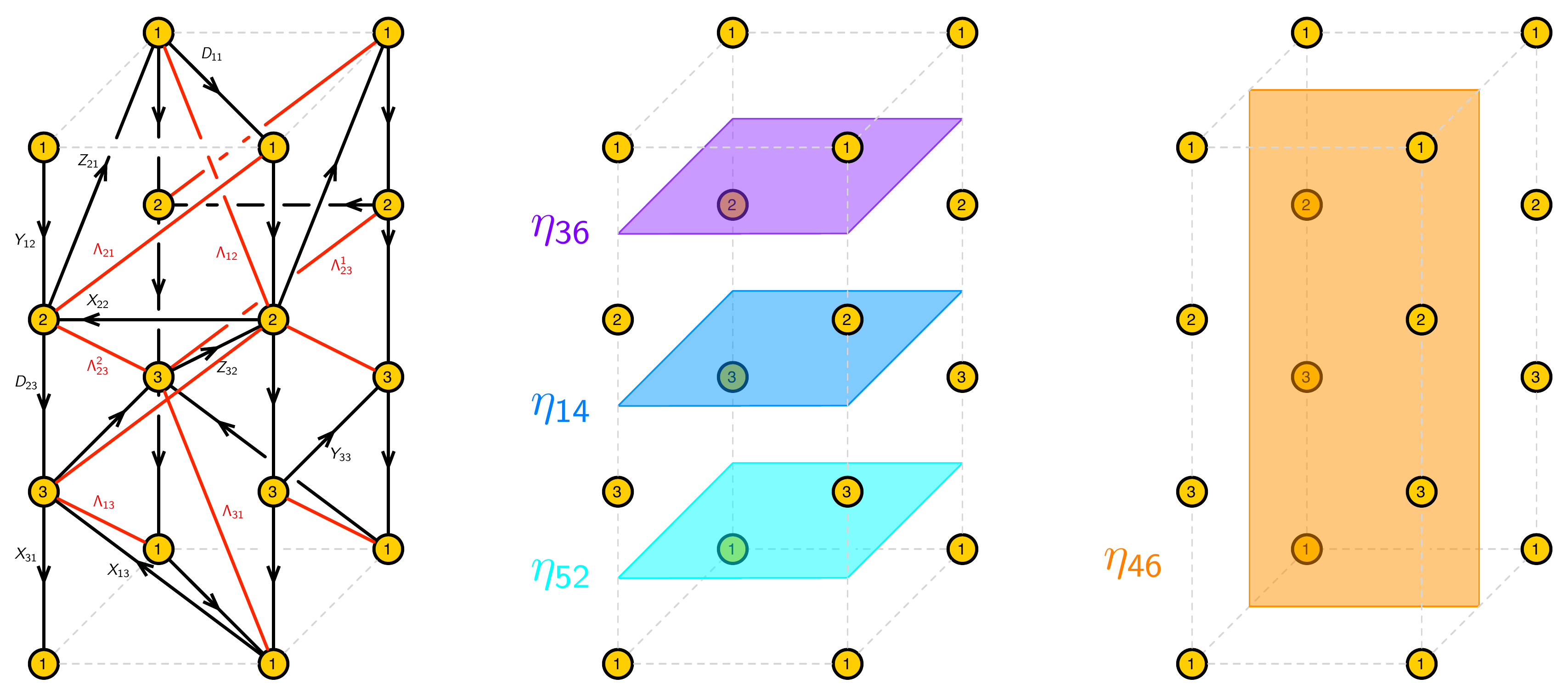}
}  
\caption{
The periodic quiver for $D_3$ and a few representative phase boundaries: $\eta_{14}$, $\eta_{36}$, $\eta_{52}$ and $\eta_{46}$.
\label{fd3}}
 \end{center}
 \end{figure}

\begin{figure}[H]
\begin{center}
\resizebox{0.6\hsize}{!}{
\includegraphics[trim=0cm 0cm 0cm 0cm,totalheight=10 cm]{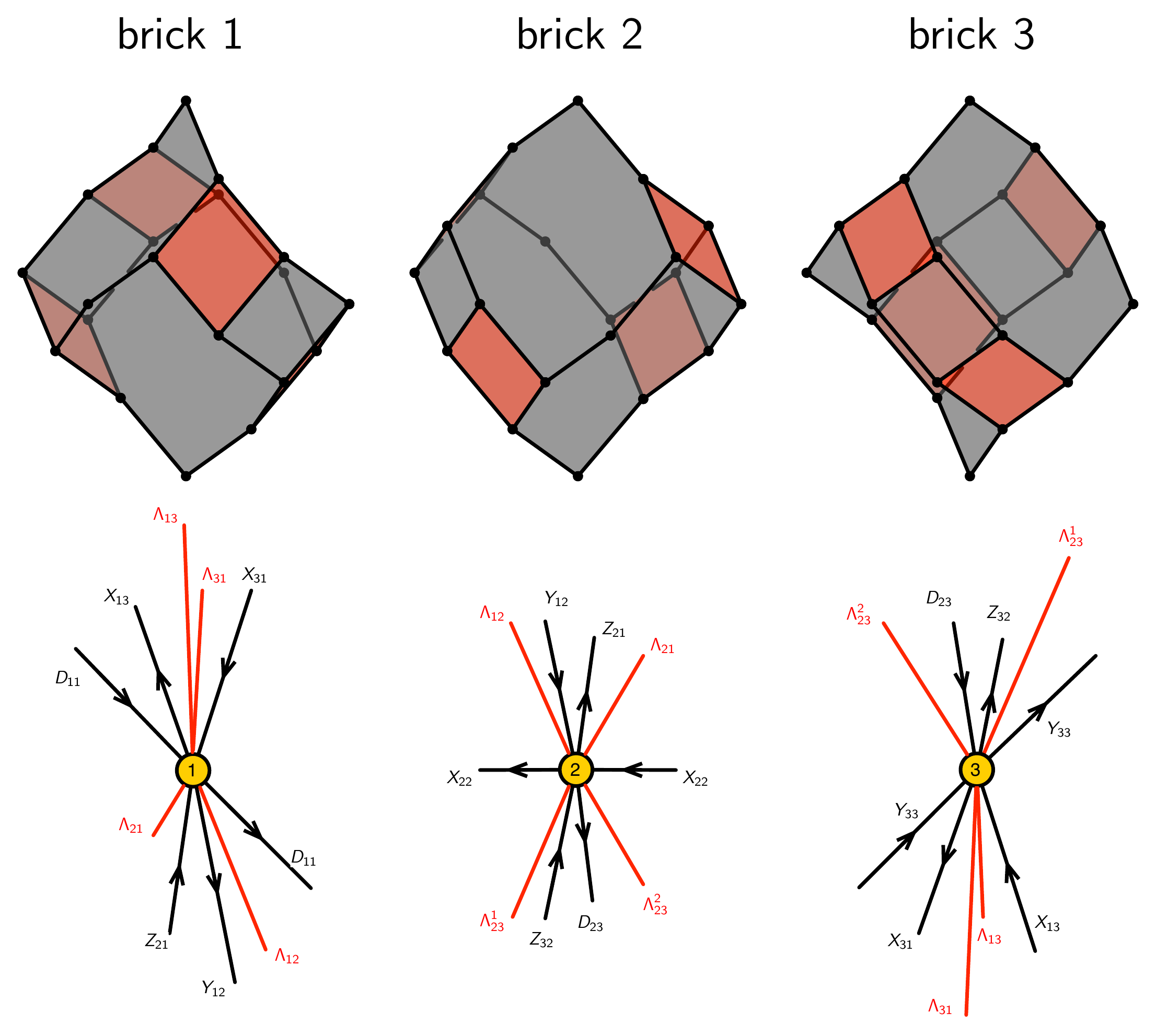}
}  
\caption{
The three bricks of the $D_{3}$ brane brick model and the corresponding nodes in the periodic quiver. They are identical up to $120^\circ$ rotations around the $z$ axis.
\label{fd3bricks}}
 \end{center}
 \end{figure}

We are now ready to construct the brane brick model, by simply dualizing the periodic quiver. The three gauge groups of the theory translate into the three bricks, shown in \fref{fd3bricks}, which are identical up to $120^\circ$ rotations around the $z$ axis. Each brane brick contains four 8-sided and two 4-sided faces that correspond to bifundamental and adjoint chiral fields, respectively. They also have four 4-sided Fermi faces. It is interesting to note that these bricks are identical to the two ones that are present in the brane brick model for $\mathcal{C}\times \mathbb{C}$ we presented in section \sref{sconifold}. This fact nicely matches the higgsing pattern that relates the two theories, which was studied in \cite{Franco:2015tna}.

\begin{figure}[ht!!]
\begin{center}
\resizebox{0.7\hsize}{!}{
\includegraphics[trim=0cm 0cm 0cm 0cm,totalheight=10 cm]{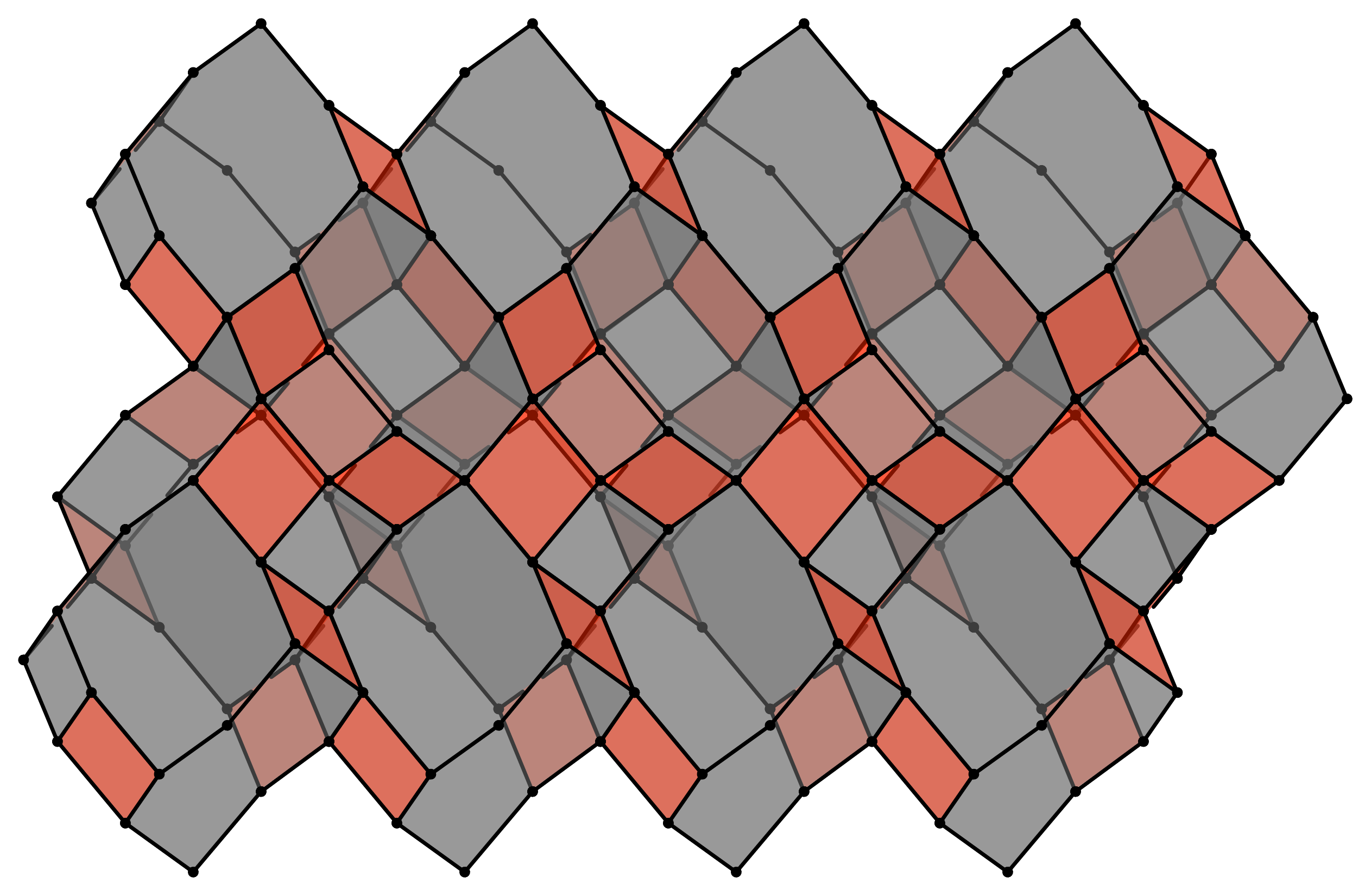}
}  
\caption{
The brane brick model for $D_{3}$.
\label{fd3branebricks}}
 \end{center}
 \end{figure}

 The full brane brick model is shown in \fref{fd3branebricks}.\footnote{The brick faces in \fref{fd3branebricks} are not planar. This is not an issue, since we are only concerned about combinatorial properties of the brane brick model. Having said that, it would be interesting to investigate whether it is possible to deform the brane brick model we presented into one in which all brick faces are planar.} While this figure is not particularly illuminating, it serves to illustrate how non-trivial and restrictive the $3d$ gluing of bricks can be. The resulting brane brick model must account not only for the matter content but also for all the $J$- and $E$-term plaquettes in terms of its edges. 

\begin{figure}[ht!!]
\begin{center}
\resizebox{0.7\hsize}{!}{
\includegraphics[trim=0cm 0cm 0cm 0cm,totalheight=10 cm]{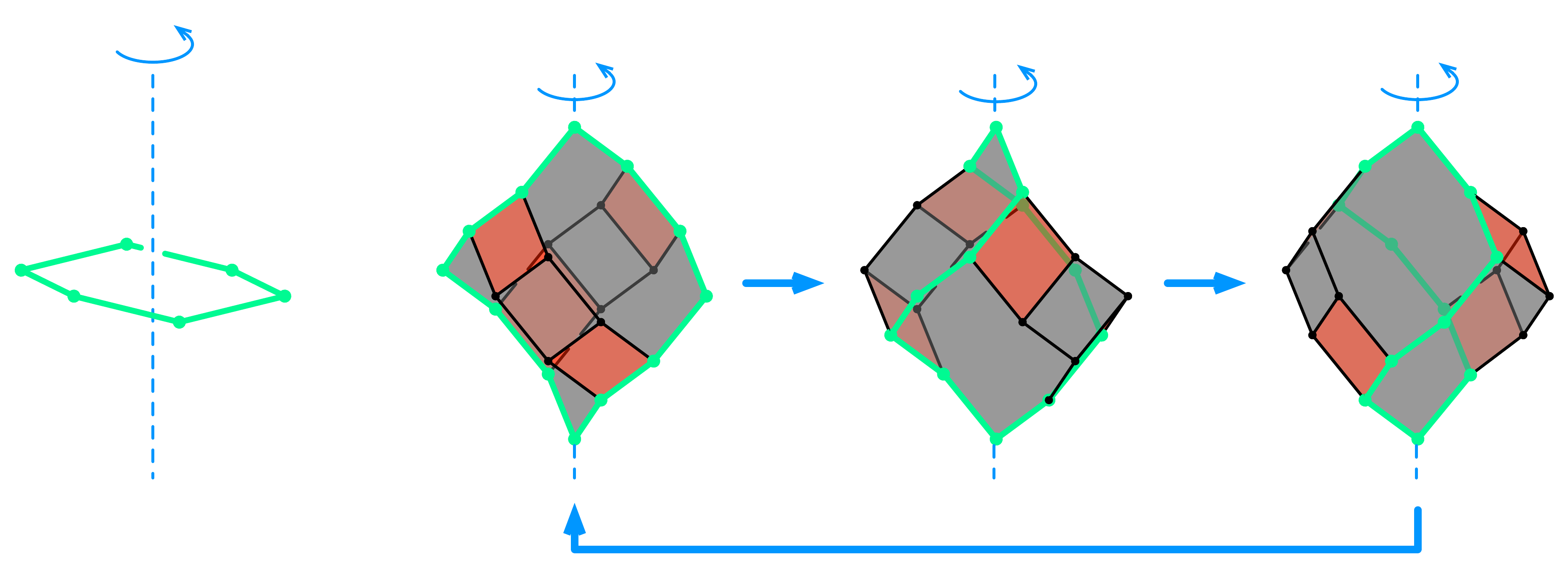}
}  
\caption{
The three bricks of the $D_3$ theory map into each other under $120^\circ$ rotations. These rotations combine with a vertical shift that moves between horizontal planes to realize the  
$\mathbb{Z}_{3}$ global symmetry at the level of the brane brick model.
\label{fd3rotate}}
 \end{center}
 \end{figure}

As it is very clear from the periodic quiver in \fref{fd3}, the brane brick model consists of horizontal planes, each of which is made out of copies of a single gauge group. The $D_3$ global symmetry of the theory is nicely captured by the brane brick model. Its $\mathbb{Z}_3$ subgroup maps to $120^\circ$ rotations, which as shown in \fref{fd3rotate} transforms the different types of bricks into one another, and an overall vertical shift by one level that moves between horizontal planes.

\subsubsection{$Q^{1,1,1}$ Brane Brick Model \label{sq111}}

\begin{figure}[H]
\begin{center}
\resizebox{0.7\hsize}{!}{
\includegraphics[trim=0cm 0cm 0cm 0cm,totalheight=10 cm]{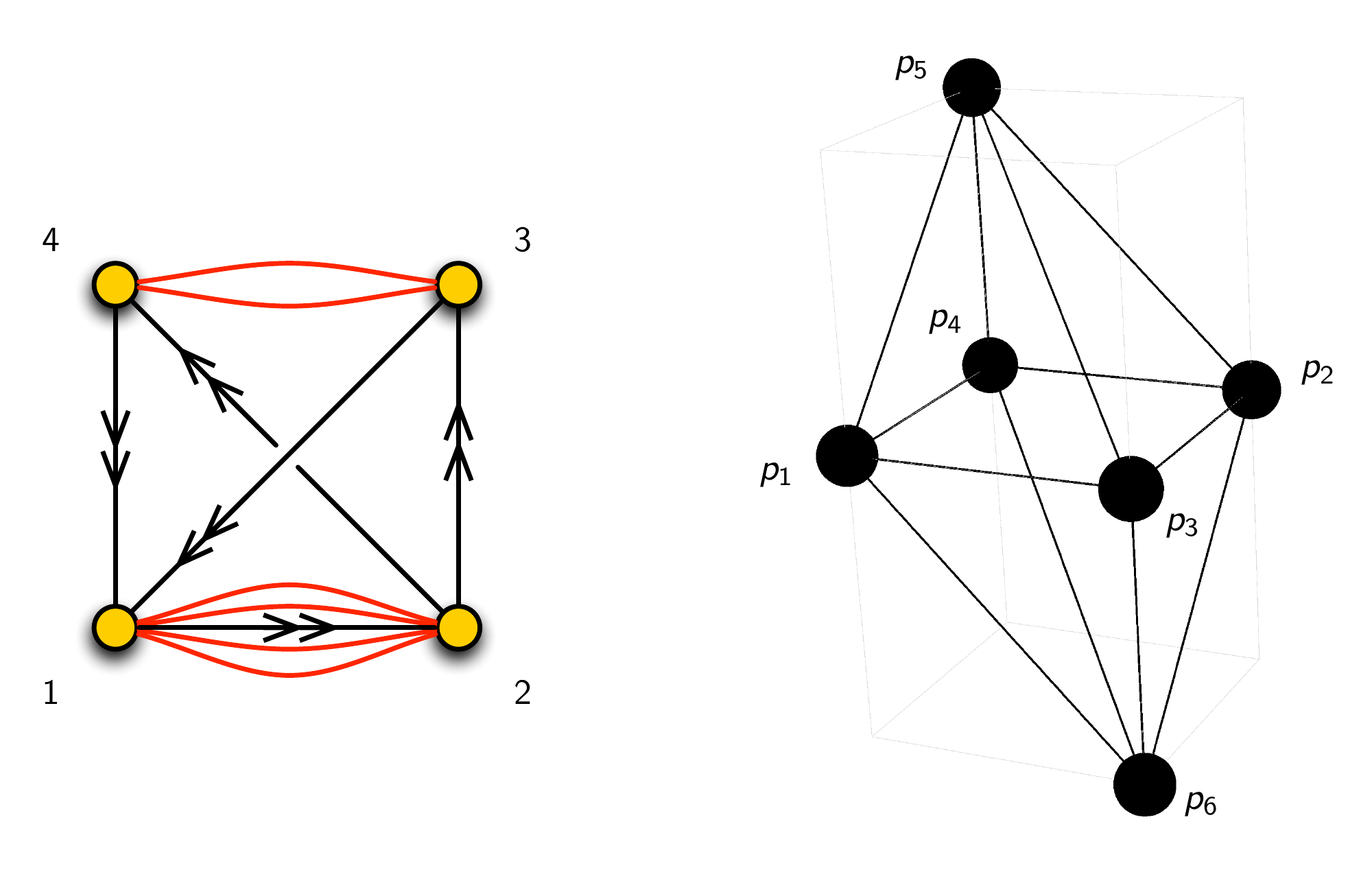}
}  
\caption{
Quiver and toric diagrams for $Q^{1,1,1}$.
\label{fq111}}
 \end{center}
 \end{figure}

The quiver diagram for the $Q^{1,1,1}$ theory is shown in \fref{fq111}. The $J$-and $E$-terms are
{\footnotesize
\beq
\begin{array}{rcrcccrcc}
& & J  \ \ \ \ \ \ \ \ \ \ \ \ \ \ \ \ \ \ \ \ \ & & & & E \ \ \ \ \ \ \ \ \ \ \ \ & & \\
\Lambda_{21}^{1} : & \ \ \ & D_{12}\cdot Z_{24}\cdot Y_{41}\cdot X_{12} - X_{12}\cdot Z_{23}\cdot D_{31}\cdot D_{12}& = & 0 & \ \ \ \ &X_{23}\cdot Y_{31} - X_{24}\cdot D_{41}&= & 0 \\
\Lambda_{21}^{2}: & \ \ \ & X_{12}\cdot X_{24}\cdot Y_{41}\cdot D_{12}-D_{12}\cdot Z_{23}\cdot Y_{31}\cdot X_{12}& = & 0 & \ \ \ \ &X_{23}\cdot D_{31} - Z_{24}\cdot D_{41}&= & 0 \\
\Lambda_{21}^{3} : & \ \ \ & D_{12}\cdot Z_{24}\cdot D_{41}\cdot X_{12} - X_{12}\cdot X_{23}\cdot D_{31}\cdot D_{12}& = & 0 & \ \ \ \ & X_{24}\cdot Y_{41} - Z_{23}\cdot Y_{31}&= & 0 \\
\Lambda_{21}^{4} : & \ \ \ & D_{12}\cdot X_{23}\cdot Y_{31}\cdot X_{12} - X_{12}\cdot X_{24}\cdot D_{41}\cdot D_{12}& = & 0 & \ \ \ \ &Z_{23}\cdot D_{31} - Z_{24}\cdot Y_{41}&= & 0 \\
\Lambda_{43} : & \ \ \ & D_{31}\cdot X_{12}\cdot X_{24} - Y_{31}\cdot X_{12}\cdot Z_{24}& = & 0 & \ \ \ \ &D_{41}\cdot D_{12}\cdot Z_{23} - Y_{41}\cdot D_{12}\cdot X_{23}&= & 0 \\
\Lambda_{34} : & \ \ \ & Y_{41}\cdot X_{12}\cdot X_{23}-D_{41}\cdot X_{12}\cdot Z_{23}& = & 0 & \ \ \ \ &D_{31}\cdot D_{12}\cdot X_{24} - Y_{31}\cdot D_{12}\cdot Z_{24}&= & 0 
\end{array} 
\label{es110a8}
\eeq}
In this case the $U(1)^3 \times U(1)_R$ global symmetry is enhanced to $SU(3)^3 \times U(1)_R$. A complete analysis is provided in appendix \ref{section_Hilbert_Q111}.

The brick matching matrix is
 \beal{es399a2}
 \small
 P_\Lambda=
 \left(
\begin{array}{c|cccccccc}
\; & p_1 & p_2 & p_3 & p_4 & p_5 & p_6 \\
\hline
 D_{12} & 1 & 0 & 0 & 0 & 0 & 0 \\
 X_{12} & 0 & 1 & 0 & 0 & 0 & 0\\
 Y_{31} & 0 & 0 & 1 & 0 & 0 & 0 \\
D_{31} & 0 & 0 & 0 & 1 & 0 & 0  \\
D_{41} & 0 & 0 & 0 & 0 & 1 & 0 \\
Y_{41} & 0 & 0 & 0 & 0 & 0 & 1 \\
X_{24} & 0 & 0 & 1 & 0 & 0 & 0 \\
Z_{24} & 0 & 0 & 0 & 1 & 0 & 0  \\
X_{23} & 0 & 0 & 0 & 0 & 1 & 0  \\
Z_{23} & 0 & 0 & 0 & 0 & 0 & 1 \\
\hline
\Lambda_{21}^{1} & 0 & 0 & 1 & 0 & 1 & 0 \\
\Lambda_{21}^{2} & 0 & 0 & 0 & 1 & 1 & 0 \\
\Lambda_{21}^{3} & 0 & 0 & 1 & 0 & 0 & 1\\
\Lambda_{21}^{4} & 0 & 0 & 0 & 1 & 0 & 1\\
\Lambda_{43} & 1 & 0 & 0 & 0 & 1 & 1\\
\Lambda_{34} & 1 & 0 & 1 & 1 & 0 & 0\\
\end{array}
\right)
 ~.~
 \eea
The phase boundary matrix is then
 \beal{es399cc1}
 \small
 H=
 \left(
\begin{array}{c|cccccccccccc}
\; & \eta_{13} & \eta_{14} & \eta_{15} &\eta_{16} & \eta_{24} & \eta_{23} & \eta_{26} & \eta_{25} & \eta_{35} & \eta_{36} & \eta_{46} & \eta_{45}
\\
\hline
D_{12} & 
 1 & 1 & 1 & 1 & 0 & 0 & 0 & 0 & 0 & 0 & 0 & 0 \\
X_{12} & 
 0 & 0 & 0 & 0 & 1 & 1 & 1 & 1 & 0 & 0 & 0 & 0 \\
Y_{31} & 
 -1 & 0 & 0 & 0 & 0 & -1 & 0 & 0 & 1 & 1 & 0 & 0 \\
D_{31} & 
 0 & -1 & 0 & 0 & -1 & 0 & 0 & 0 & 0 & 0 & 1 & 1 \\
D_{41} & 
 0 & 0 & -1 & 0 & 0 & 0 & 0 & -1 & -1 & 0 & 0 & -1 \\
Y_{41} & 
 0 & 0 & 0 & -1 & 0 & 0 & -1 & 0 & 0 & -1 & -1 & 0 \\
X_{24} &
 -1 & 0 & 0 & 0 & 0 & -1 & 0 & 0 & 1 & 1 & 0 & 0 \\ 
Z_{24} & 
 0 & -1 & 0 & 0 & -1 & 0 & 0 & 0 & 0 & 0 & 1 & 1 \\
X_{23} & 
 0 & 0 & -1 & 0 & 0 & 0 & 0 & -1 & -1 & 0 & 0 & -1 \\
Z_{23} & 
 0 & 0 & 0 & -1 & 0 & 0 & -1 & 0 & 0 & -1 & -1 & 0 \\
\hline
\Lambda_{21}^{1} & 
 -1 & 0 & -1 & 0 & 0 & -1 & 0 & -1 & 0 & 1 & 0 & -1 \\
\Lambda_{21}^{2} & 
 0 & -1 & -1 & 0 & -1 & 0 & 0 & -1 & -1 & 0 & 1 & 0 \\
\Lambda_{21}^{3} & 
 -1 & 0 & 0 & -1 & 0 & -1 & -1 & 0 & 1 & 0 & -1 & 0 \\
\Lambda_{21}^{4} & 
 0 & -1 & 0 & -1 & -1 & 0 & -1 & 0 & 0 & -1 & 0 & 1 \\
\Lambda_{43} & 
 1 & 1 & 0 & 0 & 0 & 0 & -1 & -1 & -1 & -1 & -1 & -1 \\
\Lambda_{34} & 
 0 & 0 & 1 & 1 & -1 & -1 & 0 & 0 & 1 & 1 & 1 & 1 \\
\end{array}
\right)
 ~.~
 \eea
 
\begin{figure}[ht!!]
\begin{center}
\resizebox{0.9\hsize}{!}{
\includegraphics[trim=0cm 0cm 0cm 0cm,totalheight=10 cm]{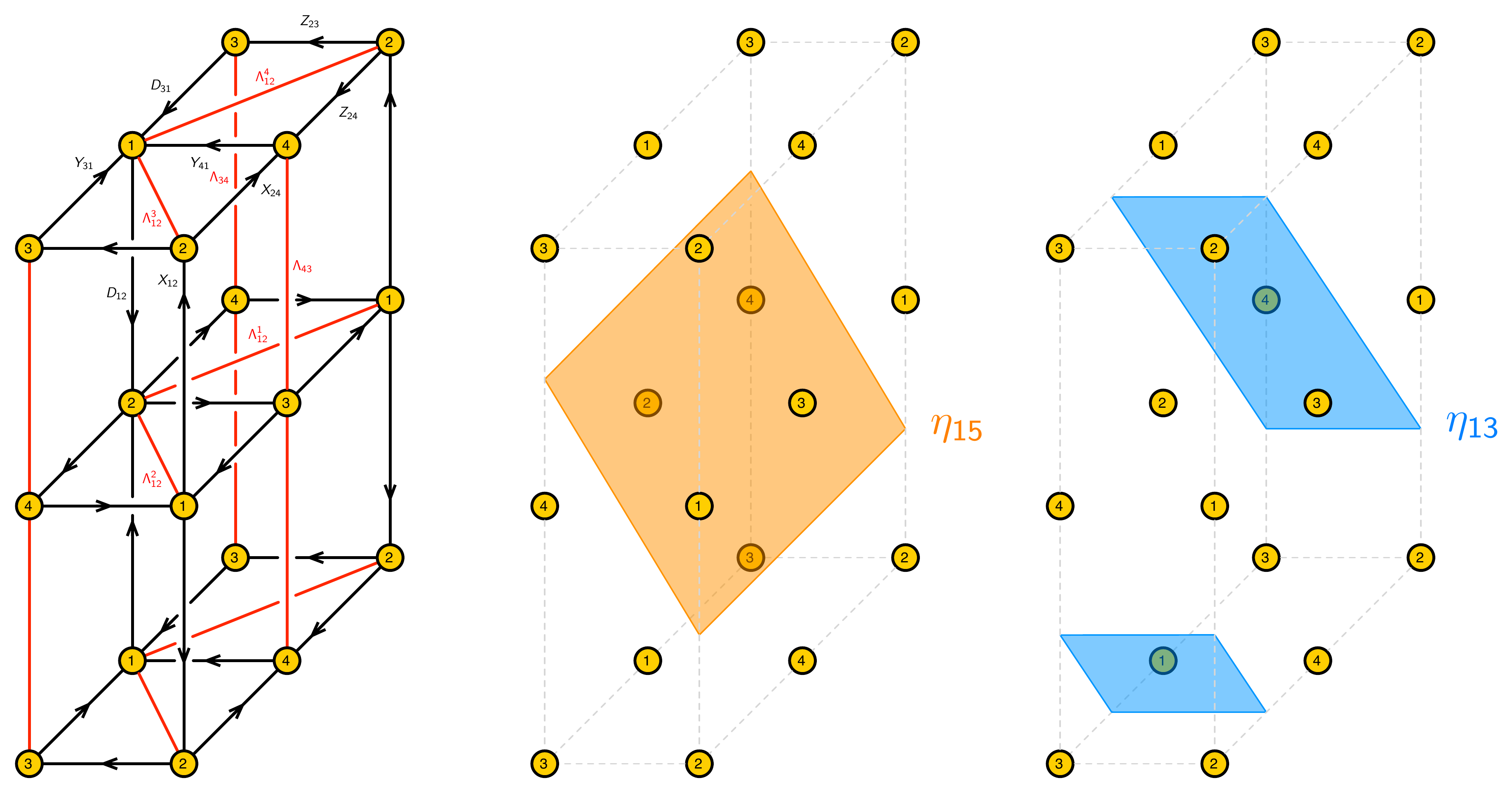}
}  
\caption{The periodic quiver for $Q^{1,1,1}$ and a pair of representative phase boundaries: $\eta_{15}$ and $\eta_{13}$.
\label{periodic_Q111}}
 \end{center}
 \end{figure}

 Using the $U(1)^3$ global symmetry charges determined in appendix \ref{section_Hilbert_Q111}, we construct the periodic quiver shown in \fref{periodic_Q111}.
 
\begin{figure}[ht!!]
\begin{center}
\resizebox{0.8\hsize}{!}{
\includegraphics[trim=0cm 0cm 0cm 0cm,totalheight=10 cm]{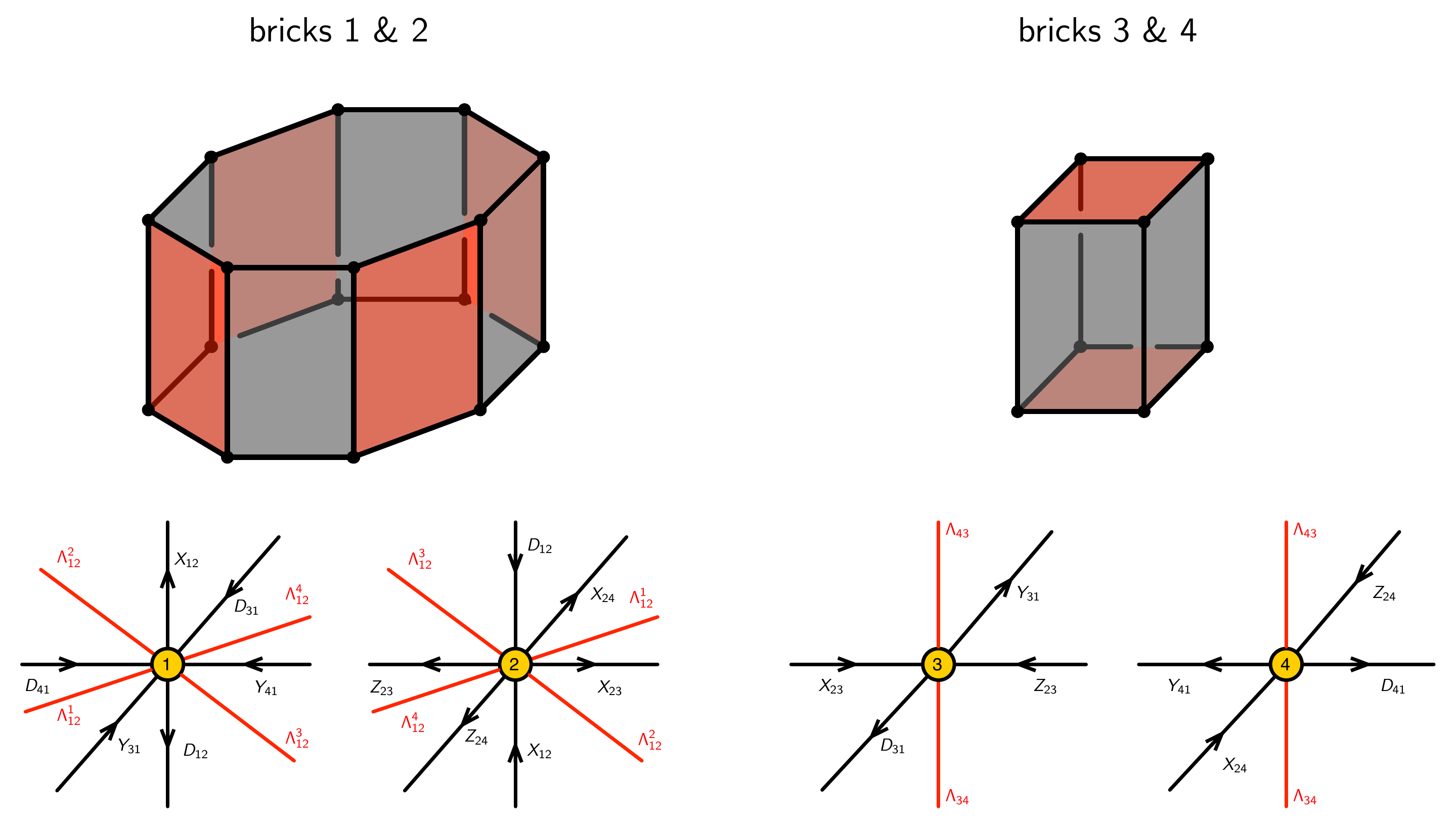}
}  
\caption{
The four bricks of the $Q^{1,1,1}$ brane brick model and the corresponding nodes in the periodic quiver.
\label{fq111bricks}}
 \end{center}
 \end{figure}

The brane brick model contains four bricks, two octagonal cylinders and two cubes, as shown in \fref{fq111bricks}.

\begin{figure}[ht!!]
\begin{center}
\resizebox{0.5\hsize}{!}{
\includegraphics[trim=0cm 0cm 0cm 0cm,totalheight=10 cm]{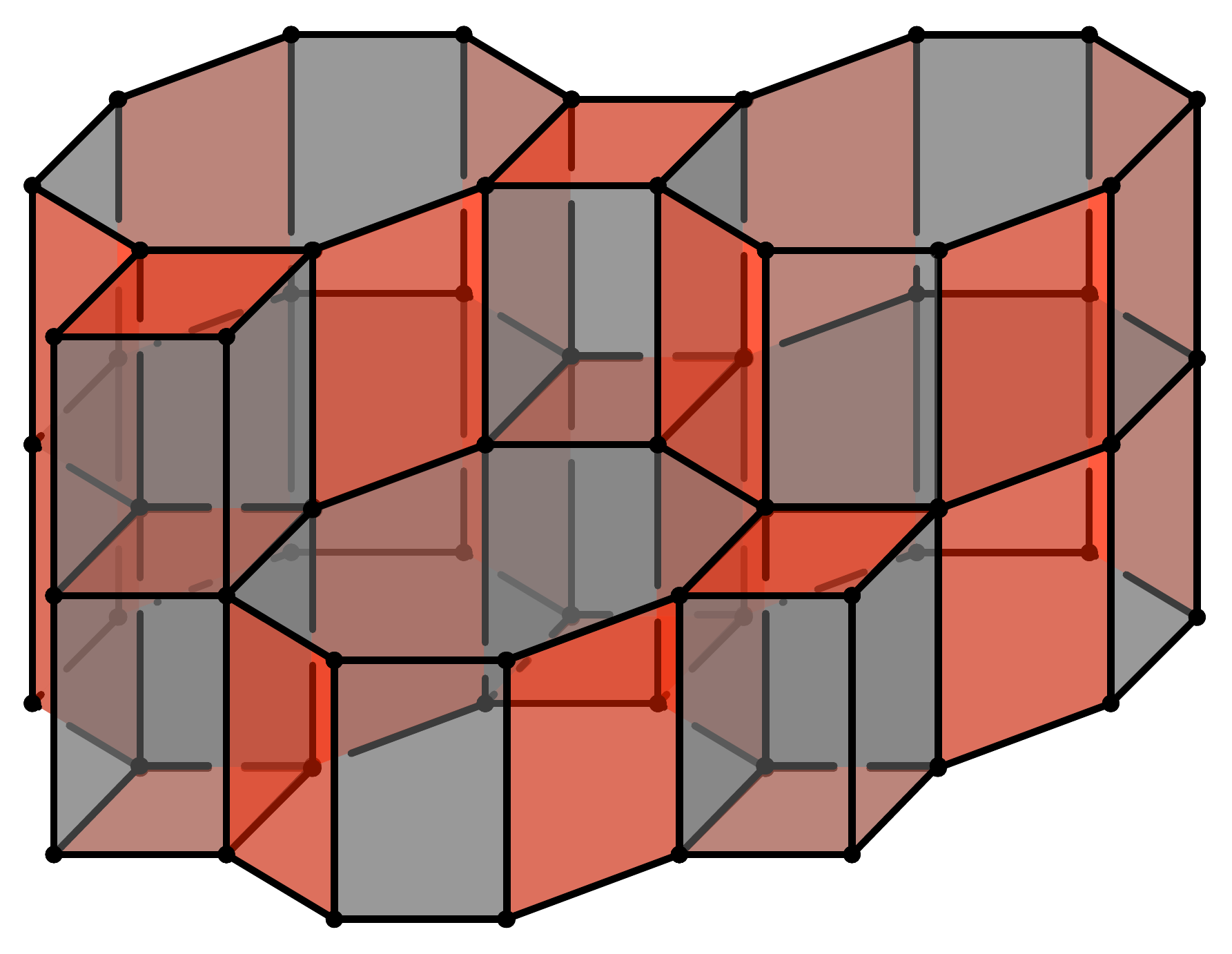}
}  
\caption{
The brane brick model for $Q^{1,1,1}$.
\label{fq111branebricks}}
 \end{center}
 \end{figure}

\begin{figure}[ht!!]
\begin{center}
\resizebox{0.9\hsize}{!}{
\includegraphics[trim=0cm 0cm 0cm 0cm,totalheight=10 cm]{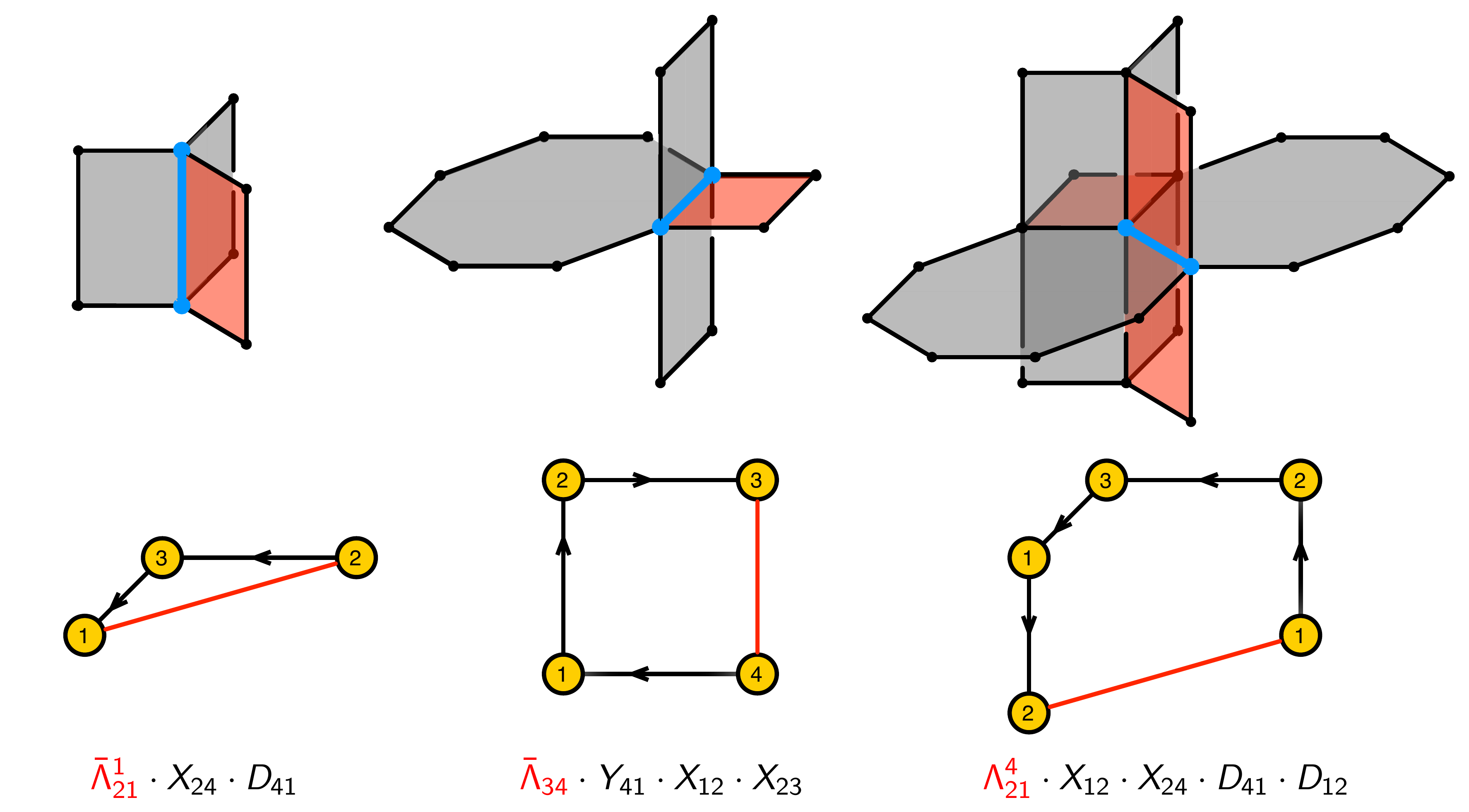}
}  
\caption{
The three types of plaquettes in the $Q^{1,1,1}$ theory and the corresponding edges in the brane brick model. In the third case, we observe two Fermi fields coinciding on the same edge.
\label{fq111plaquettes}}
 \end{center}
 \end{figure}

The brane brick model is shown in \fref{fq111branebricks}. It exhibits the interesting feature that certain edges are adjacent to more than one Fermi face. As discussed in section \sref{section_dictionary}, this results from the fact that the chiral fields in some of the plaquettes are subsets of those in larger ones. This feature can be already noticed in the explicit expressions for the $J$- and $E$-terms in \eref{es110a8}. \fref{fq111plaquettes} shows the three types of plaquettes that appear in this theory and their realization in the brane brick model.

\section{Conclusions}

\label{section_conclusions}

\begin{figure}[!ht]
\begin{center}
\includegraphics[width=10cm]{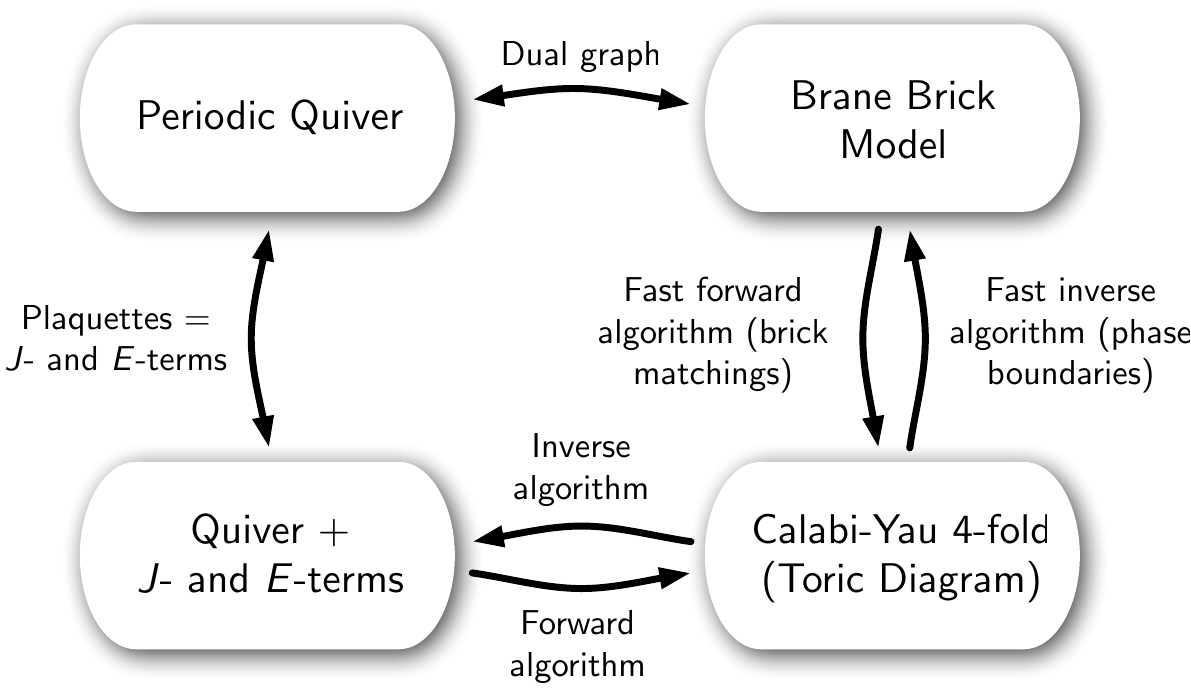}
\caption{A summary of the connections between brane brick models, gauge theories and geometry.}
\label{flowchart}
\end{center}
\end{figure}

We introduced brane brick models, a novel type of Type IIA brane configurations consisting of an NS5-brane and D4-branes. Brane brick models are T-dual to D1-branes over toric CY$_4$ singularities. They fully encode the infinite class of $2d$ (generically) $\mathcal{N}=(0,2)$ gauge theories on the worldvolume of the D1-branes and streamline the connection to the probed geometries. We introduced the fast inverse algorithm, which constructs brane brick models from geometry in terms of phase boundaries. We also developed the fast forward algorithm, which identifies the probed CY$_4$, i.e. the mesonic moduli space of the gauge theory, in terms of some newly found combinatorial objects denoted brick matchings. \fref{flowchart} schematically shows all the connections between branes, geometry and gauge theories found in this paper and \cite{Franco:2015tna}. We developed brane brick models in several additional directions, including a lifting algorithm that constructs brane brick models for CY$_4=\mathrm{CY}_3 \times \mathbb{C}$ singularities from the brane tilings corresponding to the CY$_3$.

There are several interesting directions that deserve future investigation. A few of them are:

\begin{itemize}
\item Some of the results presented in this work are well supported by substantial evidence but still require rigorous proofs.  We are at a stage that closely resembles the one after several of the initial seminal papers on brane tilings.

\item While the determination of brick matchings is rather straightforward and can be easily automated, it would be interesting to explore whether an even simpler method to find them exists, analogous to the one based on the Kasteleyn matrix and its generalizations for (almost) perfect matchings on bipartite graphs \cite{Franco:2005rj,Franco:2012mm}. This would result in a further simplification of the fast forward algorithm.

\item Both here and in \cite{Franco:2015tna}, we have discussed a single gauge theory for every singularity. It is however natural to expect that, in general, there will be more than one field theory per geometry. This assumption is well motivated by our knowledge of $2d$ $(0,2)$ trialities \cite{Gadde:2013lxa} and dualities \cite{Gadde:2015wta}, and $2d$ $(2,2)$ dualities \cite{Benini:2014mia}. Indeed, brane brick models nicely account for many of these equivalences. This will be the topic of a forthcoming publication \cite{topub2}.

\item In \cite{Franco:2007ii}, brane tilings were extended to systematically describe the $4d$ gauge theories on D3-branes probing orientifolds of toric CY$_3$ singularities. It would be interesting to investigate whether a similar extension of brane brick models to orientifolds of toric CY$_4$ singularities exists. If so, it would considerably expand the range of geometries amenable to study with these tools and can potentially lead to qualitatively different gauge theories.

\item To our knowledge, brane brick models are a completely new class of objects that have not yet been studied in the mathematical literature. This is in sharp contrast with dimer models, which were thoroughly studied from a mathematical viewpoint before its incarnation as brane tilings defining $4d$ gauge theories. There are numerous questions, e.g. the combinatorics of brick matchings, that we consider will be of great interest to the math community. It would be exciting to see a mathematical exploration of brane brick models.

\item Recently, dualities in $2d$ $(2,2)$ gauge theories have been related to cluster mutations in cluster algebras \cite{Benini:2014mia}. As we explained, in the case of D1-branes on CY$_4$ singularities, $(2,2)$ theories arise when the probed geometries are of the form CY$_3 \times \mathbb{C}$. On a parallel line of developments, there has been considerable interest in the combinatorial interpretation of cluster variables in systems related to brane tilings, which turn out to be given by certain partition functions of perfect matchings \cite{2007arXiv0710.3574M,2008arXiv0810.3638M,2011arXiv1106.0952L,Eager:2011ns,2013arXiv1308.3926L}. It would then be interesting to investigate whether the combinatorics of cluster algebras in the $2d$ $(2,2)$ are captured by brick matchings.

\end{itemize}

Based on the extensive list of applications and successes of brane tilings in the study of $4d$ $\mathcal{N}=1$ gauge theories on D3-branes over toric CY$_3$ singularities, we expect brane brick models will have a similar transformative impact in the case of $2d$ $(0,2)$ theories.

\section*{Acknowledgements}

We would like to thank M. Nisse, B. Sturmfels and A. Uranga for enjoyable and helpful discussions. We are also grateful to D. Ghim and D. Yokoyama for collaboration on related topics. S. F. and R.-K. S. gratefully acknowledge hospitality at the Simons Center for Geometry and Physics, Stony Brook University where some of the research for this paper was performed. S. F. would also like to thank the Aspen Center for Physics (NSF grant 1066293) for hospitality during part of this project. This work was initiated in the ICMS workshop ``Gauge theories: quivers, tilings and Calabi-Yaus" in May 2014. The work of S. F. is supported by the U.S. National Science Foundation grant PHY-1518967 and by a PSC-CUNY award. The work of S. L. is supported by Samsung Science and Technology Foundation under Project Number SSTF-BA1402-08 and by the IBM Einstein Fellowship of the Institute for Advanced Study.


\appendix

\section{The Fast Forward Algorithm: $\mathbb{C}^4/\mathbb{Z}_2\times\mathbb{Z}_2\times\mathbb{Z}_2$ \label{sfast}}

In this appendix, we illustrate the fast forward algorithm introduced in section \sref{section_fast_forward_algorithm} in the case of the $\mathbb{C}^4/\mathbb{Z}_2\times\mathbb{Z}_2\times\mathbb{Z}_2$ brane brick model. In particular, we show how the two prescriptions for assigning coordinates in the toric diagram to brick matchings are implemented.

\begin{figure}[ht!!]
\begin{center}
\includegraphics[trim=0cm 0cm 0cm 0cm,totalheight=6cm]{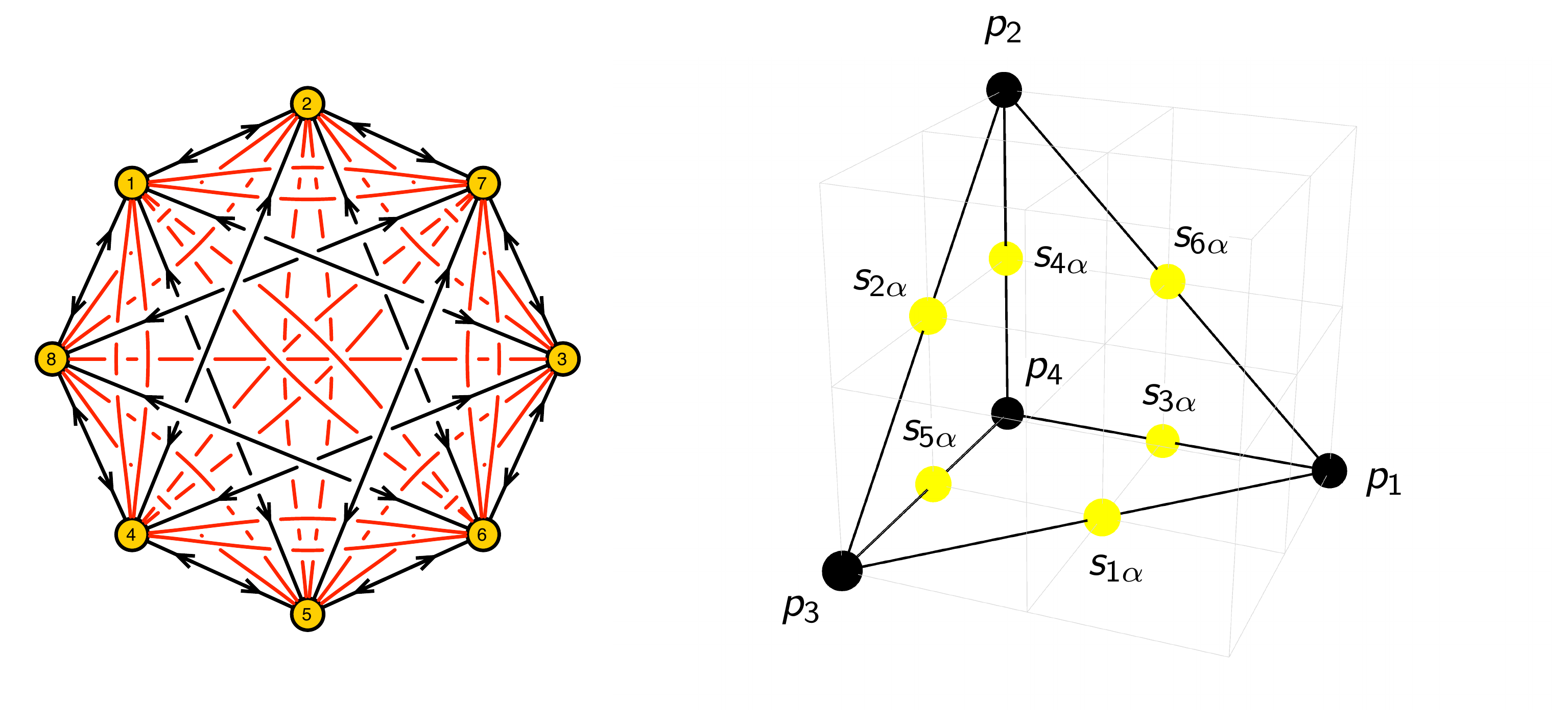}
\caption{The quiver and toric diagrams for $\mathbb{C}^4/\mathbb{Z}_2\times\mathbb{Z}_2\times\mathbb{Z}_2$. The $\alpha$ indices run over the multiple brick matchings associated to a given point in the toric diagram.
\label{fquiverc4z2z2z2}}
 \end{center}
 \end{figure}

The quiver diagram is shown in \fref{fquiverc4z2z2z2} and its $J$- and $E$-terms are as follows,

\beq
\begin{array}{rclccclcc}
& & \ \ \ \ \ \ \ \ \ \ \ \ \ \ \ \ \ \ \ \ J & & & & \ \ \ \ \ \ \ \ \ \ \ \ \ \ E & & \\
 \Lambda_{14} : & \ \ \ & Y_{43}\cdot Z_{31} - Z_{42}\cdot Y_{21}& = & 0  & \ \ \ \ & D_{18}\cdot X_{84} - X_{15}\cdot D_{54} & = & 0  \\
 \Lambda_{41} : & \ \ \ & Y_{12}\cdot Z_{24} - Z_{13}\cdot Y_{34}& = & 0  & \ \ \ \ & D_{45}\cdot X_{51} - X_{48}\cdot D_{81} & = & 0  \\
 \Lambda_{16} : & \ \ \ & X_{62}\cdot Y_{21} - Y_{65}\cdot X_{51}& = & 0  & \ \ \ \ & D_{18}\cdot Z_{86} - Z_{13}\cdot D_{36} & = & 0  \\
 \Lambda_{61} : & \ \ \ & X_{15}\cdot Y_{56} - Y_{12}\cdot X_{26}& = & 0  & \ \ \ \ & D_{63}\cdot Z_{31} - Z_{68}\cdot D_{81} & = & 0  \\
 \Lambda_{17} : & \ \ \ & Z_{75}\cdot X_{51} - X_{73}\cdot Z_{31}& = & 0  & \ \ \ \ & D_{18}\cdot Y_{87} - Z_{13}\cdot D_{36} & = & 0  \\
 \Lambda_{71} : & \ \ \ & Z_{13}\cdot X_{37} - X_{15}\cdot Z_{57}& = & 0  & \ \ \ \ & D_{72}\cdot Y_{21} - Z_{68}\cdot D_{81} & = & 0  \\
 \Lambda_{23} : & \ \ \ & Y_{34}\cdot Z_{42} - Z_{31}\cdot Y_{12}& = & 0  & \ \ \ \ & D_{27}\cdot X_{73} - X_{26}\cdot D_{63} & = & 0  \\
 \Lambda_{32} : & \ \ \ & Y_{21}\cdot Z_{13} - Z_{24}\cdot Y_{43}& = & 0  & \ \ \ \ & D_{36}\cdot X_{62} - X_{37}\cdot D_{72} & = & 0  \\
 \Lambda_{25} : & \ \ \ & X_{51}\cdot Y_{12} - Y_{56}\cdot X_{62}& = & 0  & \ \ \ \ & D_{27}\cdot Z_{75} - Z_{24}\cdot D_{45} & = & 0  \\
 \Lambda_{52} : & \ \ \ & X_{26}\cdot Y_{65} - Y_{21}\cdot X_{15}& = & 0  & \ \ \ \ & D_{54}\cdot Z_{42} - Z_{57}\cdot D_{72} & = & 0  \\
 \end{array} \nonumber
\eeq

\beq
\begin{array}{rclccclcc}
& & \ \ \ \ \ \ \ \ \ \ \ \ \ \ \ \ \ \ \ \ \ & & & & \ \ \ \ \ \ \ \ \ \ \ \ \ \ \ & & \\
 \Lambda_{28} : & \ \ \ & Z_{86}\cdot X_{62} - X_{84}\cdot Z_{42}& = & 0  & \ \ \ \ & D_{27}\cdot Y_{78} - Y_{21}\cdot D_{18} & = & 0  \\
 \Lambda_{82} : & \ \ \ & Z_{24}\cdot X_{48} - X_{26}\cdot Z_{68}& = & 0  & \ \ \ \ & D_{81}\cdot Y_{12} - Y_{87}\cdot D_{72} & = & 0  \\
 \Lambda_{35} : & \ \ \ & Z_{57}\cdot X_{73} - X_{51}\cdot Z_{13}& = & 0  & \ \ \ \ & D_{36}\cdot Y_{65} - Y_{34}\cdot D_{45} & = & 0  \\
 \Lambda_{53} : & \ \ \ & Z_{31}\cdot X_{15} - X_{37}\cdot Z_{75}& = & 0  & \ \ \ \ & D_{54}\cdot Y_{43} - Y_{56}\cdot D_{63} & = & 0  \\
 \Lambda_{38} : & \ \ \ & X_{84}\cdot Y_{43} - Y_{87}\cdot X_{73}& = & 0  & \ \ \ \ & D_{36}\cdot Z_{68} - Z_{31}\cdot D_{18} & = & 0  \\
 \Lambda_{38} : & \ \ \ & X_{37}\cdot Y_{78} - Y_{34}\cdot X_{48}& = & 0  & \ \ \ \ & D_{81}\cdot Z_{13} - Z_{86}\cdot D_{63} & = & 0  \\
 \Lambda_{46} : & \ \ \ & Z_{68}\cdot X_{84} - X_{62}\cdot Z_{24}& = & 0  & \ \ \ \ & D_{45}\cdot Y_{56} - Y_{43}\cdot D_{36} & = & 0  \\
 \Lambda_{64} : & \ \ \ & Z_{42}\cdot X_{26} - X_{48}\cdot Z_{86}& = & 0  & \ \ \ \ & D_{63}\cdot Y_{34} - Y_{65}\cdot D_{54} & = & 0  \\
 \Lambda_{47} : & \ \ \ & X_{73}\cdot Y_{34} - Y_{78}\cdot X_{84}& = & 0  & \ \ \ \ & D_{45}\cdot Z_{57} - Z_{42}\cdot D_{27} & = & 0  \\
 \Lambda_{74} : & \ \ \ & X_{48}\cdot Y_{87} - Y_{43}\cdot X_{37}& = & 0  & \ \ \ \ & D_{72}\cdot Z_{24} - Z_{75}\cdot D_{54} & = & 0  \\
 \Lambda_{58} : & \ \ \ & Y_{87}\cdot Z_{75} - Z_{86}\cdot Y_{65}& = & 0  & \ \ \ \ & D_{54}\cdot X_{48} - X_{51}\cdot D_{18} & = & 0  \\
 \Lambda_{85} : & \ \ \ & Y_{56}\cdot Z_{68} - Z_{57}\cdot Y_{78}& = & 0  & \ \ \ \ & D_{81}\cdot X_{15} - X_{84}\cdot D_{45} & = & 0  \\
 \Lambda_{67} : & \ \ \ & Y_{78}\cdot Z_{86} - Z_{75}\cdot Y_{56}& = & 0  & \ \ \ \ & D_{63}\cdot X_{37} - X_{62}\cdot D_{27} & = & 0  \\
 \Lambda_{76} : & \ \ \ & Y_{65}\cdot Z_{57} - Z_{68}\cdot Y_{87}& = & 0  & \ \ \ \ & D_{72}\cdot X_{26} - X_{73}\cdot D_{36} & = & 0  \\
\end{array}
\label{esapp10e1}
\eeq

\bigskip

The brick matchings are 

\beal{esapp10e2}
\footnotesize
\ba{ccrllllllllllllllll}
p_1 &=& \{ Z_{13},& Z_{24},& Z_{31},& Z_{42},& Z_{57},& Z_{68},& Z_{75},& Z_{86},& \Lambda_{16}^{1},& \Lambda_{16}^{2},& \Lambda_{25}^{1},& \Lambda_{25}^{2},& \Lambda_{38}^{1},& \Lambda_{38}^{2},& \Lambda_{47}^{1},& \Lambda_{47}^{2} \}
\\[.2cm]
p_2 &=& \{ Y_{12},& Y_{21},& Y_{34},& Y_{43},& Y_{56},& Y_{65},& Y_{78},& Y_{87},& \Lambda_{17}^{1},& \Lambda_{17}^{2},& \Lambda_{28}^{1},& \Lambda_{28}^{2},& \Lambda_{35}^{1},& \Lambda_{35}^{2},& \Lambda_{46}^{1},& \Lambda_{46}^{2} \},~
\\[.2cm]
p_3 &=& \{ X_{15},& X_{26},& X_{37},& X_{48},& X_{51},& X_{62},& X_{73},& X_{84},& \Lambda_{14}^{1},& \Lambda_{14}^{2},& \Lambda_{23}^{1},& \Lambda_{23}^{2},& \Lambda_{58}^{1},& \Lambda_{58}^{2},& \Lambda_{67}^{1},& \Lambda_{67}^{2} \}
\\[.2cm]
p_4 &=& \{ D_{18},& D_{27},& D_{36},& D_{45},& D_{54},& D_{63},& D_{72},& D_{81},& \Lambda_{14}^{1},& \Lambda_{14}^{2},& \Lambda_{16}^{1},& \Lambda_{16}^{2},& \Lambda_{17}^{1},& \Lambda_{17}^{2},& \Lambda_{23}^{1},& \Lambda_{23}^{2}  
\\[.05cm]
& 
& \Lambda_{25}^{1},& \Lambda_{25}^{2},& \Lambda_{28}^{1},& \Lambda_{28}^{2},& \Lambda_{35}^{1},& \Lambda_{35}^{2},& \Lambda_{38}^{1},& \Lambda_{38}^{2} 
& \Lambda_{46}^{1},& \Lambda_{46}^{2},& \Lambda_{47}^{1},& \Lambda_{47}^{2},& \Lambda_{58}^{1},& \Lambda_{58}^{2},& \Lambda_{67}^{1},& \Lambda_{67}^{2} 
\}
\\[.2cm]
s_{11} &=& \{ X_{37},& X_{48},& X_{51},& X_{62},& Z_{31},& Z_{42},& Z_{57},& Z_{68},& 
\Lambda_{14}^{2},& \Lambda_{16}^{2},& \Lambda_{23}^{2},& \Lambda_{25}^{2},& \Lambda_{38}^{1},& \Lambda_{47}^{1},& \Lambda_{67}^{1},& \Lambda_{58}^{1}
\}
\\[.2cm]
s_{12} &=& \{ X_{15},& X_{26},& X_{73},& X_{84},& Z_{13},& Z_{24},& Z_{75},& Z_{86},& 
\Lambda_{14}^{1},& \Lambda_{16}^{1},& \Lambda_{23}^{1},& \Lambda_{25}^{1},& \Lambda_{38}^{2},& \Lambda_{47}^{2},& \Lambda_{58}^{2},& \Lambda_{67}^{2}
\}
\\[.2cm]
s_{21} &= & \{  X_{26},& X_{48},& X_{51},& X_{73},& Y_{21},& Y_{43},& Y_{56},& Y_{78},& 
\Lambda_{14}^{2},& \Lambda_{17}^{2},& \Lambda_{23}^{1},& \Lambda_{28}^{1},& \Lambda_{35}^{2},& \Lambda_{46}^{1},& \Lambda_{58}^{1},& \Lambda_{67}^{2}
\}
\\[.2cm]
s_{22} &= & \{ X_{15},& X_{37},& X_{62},& X_{84},& Y_{12},& Y_{34},& Y_{65},& Y_{87},& 
\Lambda_{14}^{1},& \Lambda_{17}^{1},& \Lambda_{23}^{2},& \Lambda_{28}^{2},& \Lambda_{35}^{1},& \Lambda_{46}^{2},& \Lambda_{58}^{2},& \Lambda_{67}^{1}
\}
\\[.2cm]
s_{31} &= & \{ D_{36},& D_{45},& D_{72},& D_{81},& Z_{31},& Z_{42},& Z_{75},& Z_{86},& 
\Lambda_{14}^{2},& \Lambda_{16}^{1},& \Lambda_{16}^{2},& \Lambda_{17}^{2},& \Lambda_{23}^{2},& \Lambda_{25}^{1},& \Lambda_{25}^{2},& \Lambda_{28}^{2},& 
\\[.05cm]
&&
\Lambda_{35}^{1},& \Lambda_{38}^{1},& \Lambda_{38}^{2},& \Lambda_{46}^{1},& \Lambda_{47}^{1},& \Lambda_{47}^{2},& \Lambda_{58}^{2},& \Lambda_{67}^{2}
\}
\\[.2cm]
s_{32} &= & \{ D_{18},& D_{27},& D_{54},& D_{63},& Z_{13},& Z_{24},& Z_{57},& Z_{68},& 
\Lambda_{14}^{1},& \Lambda_{16}^{1},& \Lambda_{16}^{2},& \Lambda_{17}^{1},& \Lambda_{23}^{1},& \Lambda_{25}^{1},& \Lambda_{25}^{2},& \Lambda_{28}^{1},& 
\\[.05cm]
&&
\Lambda_{35}^{2},& \Lambda_{38}^{1},& \Lambda_{38}^{2},& \Lambda_{46}^{2},& \Lambda_{47}^{1},& \Lambda_{47}^{2},& \Lambda_{58}^{1},& \Lambda_{67}^{1}
\}
\\[.2cm]
s_{41} &= & \{ D_{27},& D_{45},& D_{63},& D_{81},& Y_{21},& Y_{43},& Y_{65},& Y_{87},& 
\Lambda_{14}^{2},&  \Lambda_{16}^{2},& \Lambda_{17}^{1},& \Lambda_{17}^{2},& \Lambda_{23}^{1},& \Lambda_{25}^{1},& \Lambda_{28}^{1},& \Lambda_{28}^{2},& 
\\[.05cm]
&&
\Lambda_{35}^{1},& \Lambda_{35}^{2},& \Lambda_{38}^{2},& \Lambda_{46}^{1},&  \Lambda_{46}^{2},& \Lambda_{47}^{1},& \Lambda_{58}^{2},&  \Lambda_{67}^{1}\}
\\[.2cm]
s_{42} &= & \{ D_{18},& D_{36},& D_{54},& D_{72},& Y_{12},& Y_{34},& Y_{56},& Y_{78},& 
\Lambda_{14}^{1},& \Lambda_{16}^{1},& \Lambda_{17}^{1},& \Lambda_{17}^{2},& \Lambda_{23}^{2},& \Lambda_{25}^{2},& \Lambda_{28}^{1},& \Lambda_{28}^{2},& 
\\[.05cm]
&&
\Lambda_{35}^{1},& \Lambda_{35}^{2},& \Lambda_{38}^{1},& \Lambda_{46}^{1},& \Lambda_{46}^{2},& \Lambda_{47}^{2},& \Lambda_{58}^{1},& \Lambda_{67}^{2}\}
\\[.2cm]
\ea
\nn
\eea

\beal{esapp10e2}
\footnotesize
\ba{ccrllllllllllllllll}
s_{51} &= & \{ D_{18},& D_{27},& D_{36},& D_{45},& X_{15},& X_{26},& X_{37},& X_{48},& 
\Lambda_{14}^{1},& \Lambda_{14}^{2},& \Lambda_{16}^{1},& \Lambda_{17}^{1},& \Lambda_{23}^{1},& \Lambda_{23}^{2},& \Lambda_{25}^{1},&  \Lambda_{28}^{1},& 
\\[.05cm]
&&
\Lambda_{35}^{1},& \Lambda_{38}^{1},& \Lambda_{46}^{1},& \Lambda_{47}^{1},& \Lambda_{58}^{1},& \Lambda_{58}^{2},& \Lambda_{67}^{1},& \Lambda_{67}^{2}\}
\\[.2cm]
s_{52} &= & \{ D_{54},& D_{63},& D_{72},& D_{81},& X_{51},& X_{62},& X_{73},& X_{84},& 
\Lambda_{14}^{1},& \Lambda_{14}^{2},& \Lambda_{16}^{2},& \Lambda_{17}^{2},& \Lambda_{23}^{1},& \Lambda_{23}^{2},& \Lambda_{25}^{2},& \Lambda_{28}^{2},& 
\\[.05cm]
&&
\Lambda_{35}^{2},& \Lambda_{38}^{2},& \Lambda_{46}^{2},& \Lambda_{47}^{2},& \Lambda_{58}^{1},& \Lambda_{58}^{2},& \Lambda_{67}^{1},& \Lambda_{67}^{2}\}
\\[.2cm]
s_{61} &= & \{ Y_{21},& Y_{34},& Y_{65},& Y_{78},& Z_{24},& Z_{31},& Z_{68},& Z_{75},& 
\Lambda_{16}^{2},& \Lambda_{17}^{2},& \Lambda_{25}^{1},& \Lambda_{28}^{1},& \Lambda_{35}^{1},& \Lambda_{38}^{1},& \Lambda_{47}^{2},& \Lambda_{46}^{2}\}
\\[.2cm]
s_{62} &= & \{ Y_{12},& Y_{43},& Y_{56},& Y_{87},& Z_{13},& Z_{42},& Z_{57},& Z_{86},& 
\Lambda_{16}^{1},& \Lambda_{17}^{1},& \Lambda_{25}^{2},& \Lambda_{28}^{2},& \Lambda_{35}^{2},& \Lambda_{38}^{2},& \Lambda_{46}^{1},& \Lambda_{47}^{1}\}.~
\ea
\nn\\
\eea

\begin{figure}[ht!]
\begin{center}
\resizebox{0.8\hsize}{!}{
\includegraphics[trim=0cm 0cm 0cm 0cm,totalheight=10 cm]{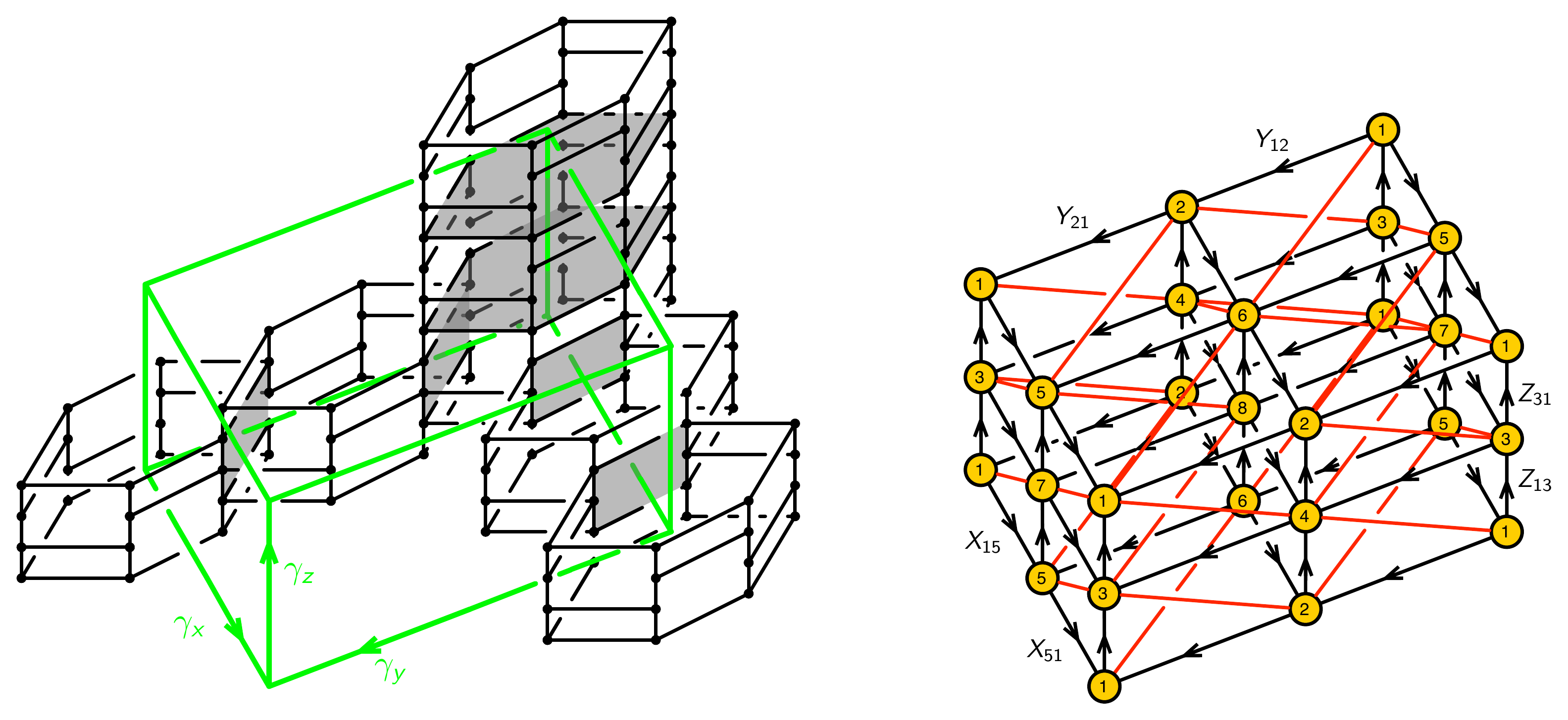}
}  
\caption{The unit cell of the brane brick model and the periodic quiver for $\mathbb{C}^4/\mathbb{Z}_{2}\times\mathbb{Z}_{2}\times\mathbb{Z}_{2}$. The faces associated to chiral fields that intersect the edges of the unit cell are shown in grey.
\label{funitcellintersections3}}
 \end{center}
 \end{figure}

\paragraph{Toric Diagram from Face Intersections.} As shown in \fref{funitcellintersections3}, the edges of the unit cell of the brane brick model intersect the faces associated to the following chiral fields
\beal{esapp10e5}
X_{15}, X_{51}, Y_{12}, Y_{21}, Z_{13}, Z_{31} ~.~
\eea
The intersection vectors $\vec{n}(X_{ij})=(\vec{X_{ij}}\cdot \hat{x},\vec{X_{ij}}\cdot \hat{y},\vec{X_{ij}}\cdot \hat{z})$ for these faces are
\beal{esapp10e6}
\ba{lll}
\vec{n}(X_{15}) &=\vec{n}(X_{51})&=(1,0,0)~,~ 
\nn\\
\vec{n}(Y_{12}) &=\vec{n}(Y_{21})&=(0,1,0)~,~ 
\nn\\
\vec{n}(Z_{13}) &=\vec{n}(Z_{31})&=(0,0,1) ~.~
\ea
\eea
Combining this information with \eref{esapp10e2}, we can compute the coordinates of the points in the toric diagram corresponding to the brick matchings, which are summarized in the following matrix
\beal{esapp10e7}
G=
\left(
\begin{array}{cccccccccccccccc}
 p_1 & p_2 & p_3 & p_4 & s_{11} & s_{12} & s_{21} & s_{22} & s_{31} & s_{32} & s_{41} & s_{42} & s_{51} & s_{52} & s_{61} & s_{62} \\
 \hline
 0 & 0 & 2 & 0 & 1 & 1 & 1 & 1 & 0 & 0 & 0 & 0 & 1 & 1 & 0 & 0 \\
 2 & 0 & 0 & 0 & 1 & 1 & 0 & 0 & 1 & 1 & 0 & 0 & 0 & 0 & 1 & 1 \\
 0 & 2 & 2 & 0 & 1 & 1 & 2 & 2 & 0 & 0 & 1 & 1 & 1 & 1 & 1 & 1 \\
\end{array}
\right)
~.~
\eea
Up to an $SL(3,\mathbb{Z})$ transformation, this matrix is in precise agreement with the toric diagram in \fref{fquiverc4z2z2z2}.

\begin{figure}[ht!]
\begin{center}
\resizebox{0.8\hsize}{!}{
\includegraphics[trim=0cm 0cm 0cm 0cm,totalheight=10 cm]{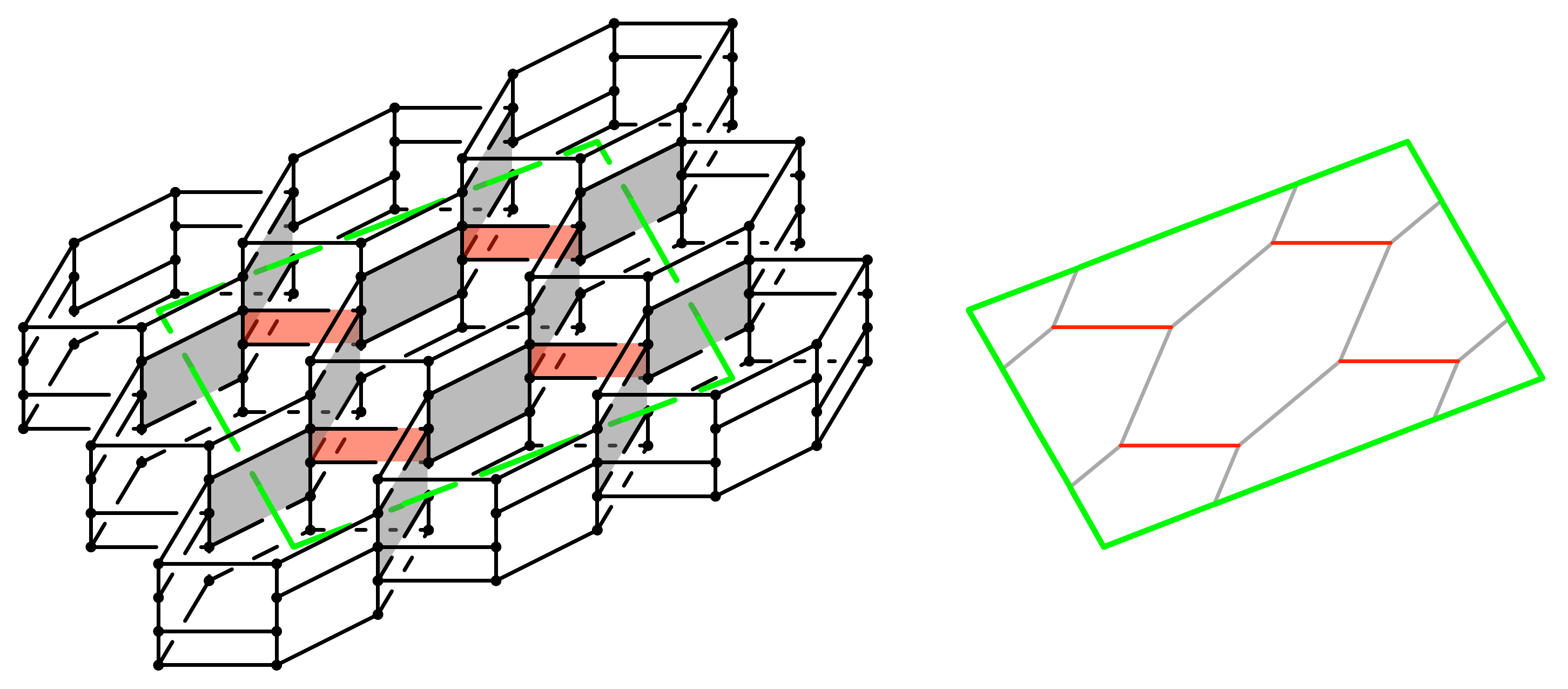}
}  
\caption{The intersection between the brane brick model for $\mathbb{C}^4/\mathbb{Z}_{2}\times\mathbb{Z}_{2}\times\mathbb{Z}_{2}$ and one of the faces of its unit cell boundary.
\label{funitcellintersections2}}
 \end{center}
 \end{figure}

\begin{figure}[ht!]
\begin{center}
\resizebox{1\hsize}{!}{
\includegraphics[trim=0cm 0cm 0cm 0cm,totalheight=10 cm]{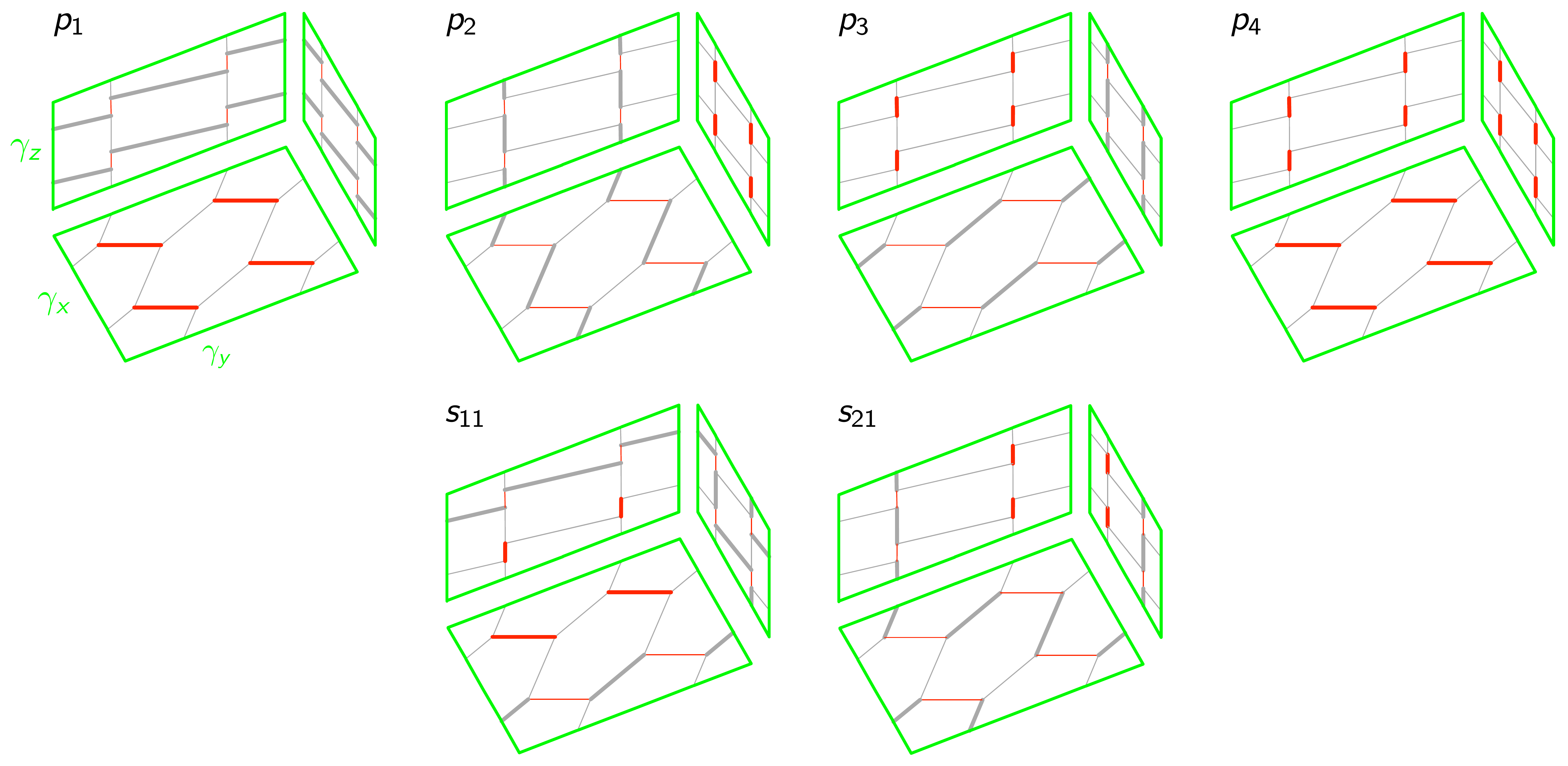}
}  
\caption{Intersections of some of the brick matchings of the $\mathbb{C}^4/\mathbb{Z}_{2}\times\mathbb{Z}_{2}\times\mathbb{Z}_{2}$ brane brick model with the faces of the unit cell. The intersections of faces contained in the brick matchings are represented by thickened lines.
\label{funitcellmatchings}}
 \end{center}
 \end{figure}

\paragraph{Toric Diagram from the Height Function.} In order to visualize the height function, it is useful to consider its values over the boundaries of the unit cell of the brane brick model. In addition, these projections contain sufficient information for computing the slope of the height function. \fref{funitcellintersections2} shows the intersection between the brane brick model and the boundary of the unit cell. Similar intersections for brick matchings are shown in \fref{funitcellmatchings}.

\begin{figure}[ht!]
\begin{center}
\resizebox{0.75\hsize}{!}{
\includegraphics[trim=0cm 0cm 0cm 0cm,totalheight=10 cm]{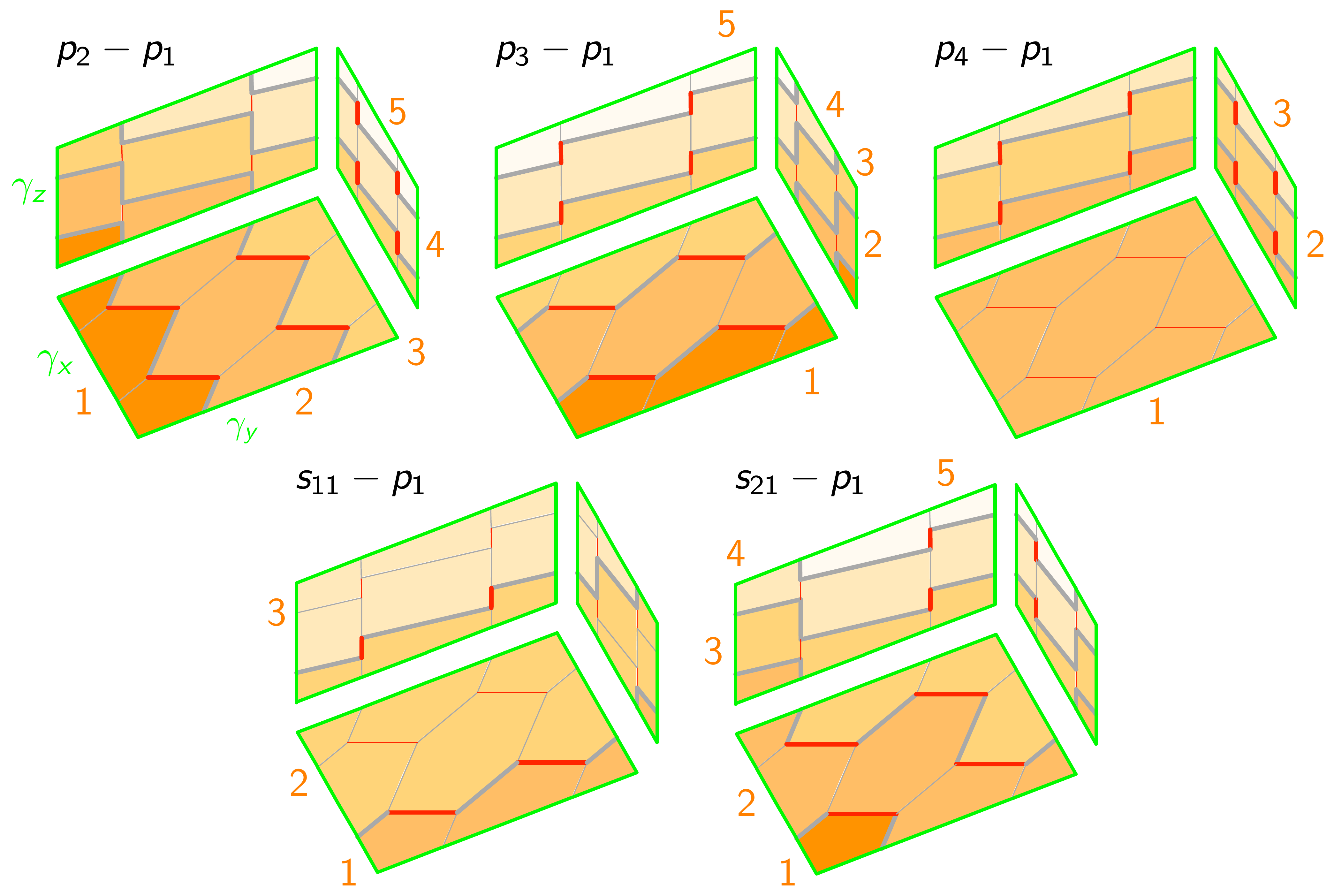}
}  
\caption{
Projections of the height function on the faces of the unit cell for some of the brick matchings of $\mathbb{C}^4/\mathbb{Z}_{2}\times\mathbb{Z}_{2}\times\mathbb{Z}_{2}$. 
\label{funitcellheights}}
 \end{center}
 \end{figure}

Let us take the reference brick matching to be $p_1$. For every brick matching $p_\mu$, we consider the intersection between the closed surfaces corresponding to $p_\mu-p_1$ and the faces in the unit cell boundary. This is shown in \fref{funitcellheights}. It is straightforward to use these projections to determine the slope of the height function $\Delta \vec{h}=(\Delta_x h_\mu,\Delta_y h_\mu,\Delta_z h_\mu)$, which contains the variations in the height function when moving by a period along the three fundamental directions of $T^3$. As we explained, the slope $\Delta \vec{h}(p_\mu)$ is precisely the coordinate of the point in the toric diagram associated to $p_\mu$. For the example at hand, the slopes are summarized in the following matrix
 \beal{esapp10e10}
 G=
\left(
\begin{array}{cccccccccccccccc}
 p_1 & p_2 & p_3 & p_4 & s_{11} & s_{12} & s_{21} & s_{22} & s_{31} & s_{32} & s_{41} & s_{42} & s_{51} & s_{52} & s_{61} & s_{62} \\
 \hline
 0 & 0 & 2 & 0 & 1 & 1 & 1 & 1 & 0 & 0 & 0 & 0 & 1 & 1 & 0 & 0 \\
 0 & 2 & 0 & 0 & 0 & 0 & 1 & 1 & 0 & 0 & 1 & 1 & 0 & 0 & 1 & 1 \\
 0 & 2 & 2 & 2 & 1 & 1 & 2 & 2 & 1 & 1 & 2 & 2 & 2 & 2 & 1 & 1 \\
\end{array}
\right)
~.~
 \eea
This is identical to \eref{esapp10e7} up to an $SL(3,\mathbb{Z})$ transformation and corresponds to the toric diagram in \fref{fquiverc4z2z2z2}. It is interesting to remark that the slope of the height function can be computed without determining the height function in the bulk of the unit cell.

\section{Hilbert Series and Plethystics \label{shilbert}}

The Hilbert series \cite{Benvenuti:2006qr} is a generating function that counts chiral gauge invariant operators of a supersymmetric gauge theory. It fully characterizes the algebraic structure of the moduli space of the theory, which can be identified using plethystics \cite{Benvenuti:2006qr,Feng:2007ur}. Hilbert series have been used extensively in a wide variety of contexts, such as for understanding the algebraic structure of various $4d$ $\mathcal{N}=1$ theories associated to brane tilings \cite{Hanany:2012hi,Hanany:2012vc}.

For the $2d$ wordvolume theories on D1-branes, the classical mesonic moduli space of the abelian theory corresponds to the Calabi-Yau 4-fold probed by the branes. The mesonic moduli space, as an affine algebraic variety, takes the form 
\beal{es10a0a}
\mathcal{M} = \master // U(1)^G  ~,~
\eea 
where the generalized master space \cite{Forcella:2008bb} is 
\beal{es10a0b}
\master = \mathbb{C}^E [X_1 ,\dots X_E] / \langle J_{ij} , E_{ij}\rangle  ~.~
\eea
Above, $G$ indicates the number of $U(1)$ gauge groups and $E$ the number of chiral fields $X_{ij}$. The quotienting ideal arises from the $J$- and $E$-terms. The mesonic moduli space can also be expressed as
\beal{es10a0b1}
\mathcal{M} = (\mathbb{C}[p_1,\dots,p_c]// Q_{EJ}) // Q_{D} ~,~
\eea
where $p_i$ are the GLSM fields of the theory, and $Q_{EJ}$ and $Q_{D}$ are matrices of complexified $U(1)$ charges arising from the $J$-, $E$- and $D$-terms. In \cite{Franco:2015tna}, the forward algorithm was developed for obtaining $Q_{EJ}$ and $Q_{D}$ and determining the toric diagram of the Calabi-Yau 4-fold, $G_t = \ker (Q_{EJ}, Q_D)$. 

Using Hilbert series, we count gauge invariant operators in terms of GLSM fields which are invariant under the complexified $U(1)$, whose charges are given by $Q_{EJ}$ and $Q_{D}$ \cite{Franco:2015tna}. The Hilbert series can be computed using the Molien integral as follows 
\beal{es10a1}
g(t_\alpha; \mathcal{M}) = 
\prod_{i=1}^{c-4} \oint_{|z_i|=1} \frac{\ud z_i}{2\pi i z_i} \prod_{\alpha=1}^c \frac{1}{1-t_{\alpha} \prod_{j=1}^{c-4} z_{j}^{(Q_t)_{j\alpha}}}
~,~
\eea
where $\alpha= 1,\dots, c$ labels GLSM fields, $t_\alpha$ is the fugacity associated to the GLSM field $p_\alpha$, and $z_i$ are the fugacities corresponding to the $c-4$ complexified $U(1)$ charges arising from $Q_{EJ}$ and $Q_D$. $Q_{t}= (Q_{EJ},Q_{D})$ is the concatenated matrix of $Q_{EJ}$ and $Q_D$.

The generators and relations of the mesonic moduli space can be identified from the Hilbert series using plethystics. The plethystic logarithm of the Hilbert series \cite{Benvenuti:2006qr,Feng:2007ur} is defined as
\beal{es10a2}
\PL\left[
g(t_\alpha; \mathcal{M})
\right]
 = 
\sum_{k=1}^{\infty}
\frac{\mu(k)}{k} \log\left[
g(t_\alpha^k; \mathcal{M})
\right]
=
\sum_{i} n_i M_i(t_\alpha)
~,~
\eea
where $\mu$ is the M\"obius function, $M_i(t_\alpha)$ are monomials made of fugacities $t_\alpha$, and $n_i$ are integer coefficients. When the plethystic logarithm of the Hilbert series contains a finite number of terms, the mesonic moduli space is identified as a \textit{complete intersection}. It is parameterized by the generators corresponding to the monomials $\{M_i(t_\alpha); n_i > 0\}$ satisfying a finite number of relations associated to the monomials $\{M_i(t_\alpha); n_i < 0\}$. When the expansion of the plethystic logarithm is not finite, the mesonic moduli space is identified as a \textit{non-complete intersection}.

\subsection{Hilbert Series for the $D_3$ Theory}
\label{section_Hilbert_D3}

Using the forward algorithm for the theory defined by the quiver in \fref{quiver_toric_D3} and the $J$- and $E$-terms in \eref{es130a1}, we obtain the following charge matrices for GLSM fields
\beal{es130aa2}
Q_{EJ}=
0
~,~
Q_D=
\left(
\begin{array}{cccccc}
 p_1 & p_2 & p_3 & p_4 & p_5 & p_6 
\\
\hline
 1 & 0 & -1 & -1 & 0 & 1 \\
 -1 & -1 & 0 & 1 & 1 & 0 \\
\end{array}
\right)
~.~
\eea
They give rise to the toric diagram
\beal{es130aa3}
G=
\left(
\begin{array}{cccccc}
 p_1 & p_2 & p_3 & p_4 & p_5 & p_6 
\\
\hline
  1 &  1 &  1  &  1  & 1 & 1 \ \\
 0 & 0 & 0 & 1 & -1 & 1 \\
 0 & 1 & 0 & 0 & 1 & 0 \\
 0 & 0 & 1 & 0 & 0 & 1 \\
 \end{array}
\right)
~,~
\eea
which is also shown in \fref{quiver_toric_D3}.
 
The symmetries of the gauge theory imply that all bifundamental chiral fields carry the same $U(1)_R$ charge, which we denote $r_1$. Similarly, all adjoint chiral fields carry the same $U(1)_R$-charge $r_2$. \tref{tchd3} summarizes the charges of the chiral fields under the full $U(1)^3 \times U(1)_R$ symmetry.

 \begin{table}[ht!!]
 \centering
 \begin{tabular}{r|cccc}
 field & $U(1)_x$ & $U(1)_y$ & $U(1)_z$ & $U(1)_R$ \\
 \hline
 $Y_{12}$ & $+1/3$ & $+1/2$ & $0$ & $r_1$ \\ 
 $D_{23}$ & $+1/3$ & $-1/2$ & $+1/2$ & $r_1$ \\
 $X_{31}$ & $+1/3$ & $0$ & $-1/2$ & $r_1$\\
 $X_{13}$ & $-1/3$ & $0$ & $-1/2$ & $r_1$\\
 $Z_{32}$ & $-1/3$ & $-1/2$ &$+1/2$  & $r_1$\\
 $Z_{21}$ & $-1/3$ & $+1/2$ & $0$ & $r_1$\\
 \hline
 $D_{11}$ & $0$ &$-1$ & $+1$ & $r_2$\\
 $X_{22}$ & $0$ & $0$& $-1$& $r_2$\\
 $Y_{33}$ & $0$ & $+1$& $0$& $r_2$\\
 \end{tabular}
 \caption{Global charges of chiral fields in the $D_3$ theory.
 }
 \label{tchd3}
 \end{table}

Assigning fugacities $t_\alpha$ for the GLSM fields $p_\alpha$, the Hilbert series for the $D_3$ theory is computed to be
\beal{es130aa3}
g(t_i;D_3) = 
\frac{
1- t_1 t_2 t_3 t_4 t_5 t_6
}{
(1- t_1 t_4) (1- t_2 t_5) (1- t_3 t_6) 
(1- t_1 t_3 t_5) (1- t_2 t_4 t_6)
}
~.~
\eea
Its plethystic logarithm is 
\beal{es130aa4}
\PL\left[g(t_i;D_3)\right]
=
t_1 t_4 +  t_2 t_5+  t_3 t_6+ t_1 t_3 t_5+  t_2 t_4 t_6 -t_1 t_2 t_3 t_4 t_5 t_6 ~.~
\eea 
Introducing fugacities $x$, $y$, $z$ for the global $U(1)_x$, $U(1)_y$ and $U(1)_z$ respectively, and fugacities $\bar{t}_i$ counting the number of fields with $U(1)_R$ charge $r_i$, the above plethystic logarithm can be rewritten as follows,
\beal{es130aa5}
\PL\left[g(t,x,y,z;D_3)\right]
=
y^{-1} z \bar{t}_1^2 \bar{t}_2^2 
+ z^{-1} \bar{t}_1^2 \bar{t}_2^2 
+ y \bar{t}_1^2 \bar{t}_2^2 + x \bar{t}_1^3 \bar{t}_2^3+  x^{-1} \bar{t}_1^3 \bar{t}_2^3 -\bar{t}_1^6 \bar{t}_2^6 ~,~
\eea
where the fugacity map is
\beal{es130aa500}
t_1 = y^{-1} z \bar{t}_1 \bar{t}_2 ~,~
t_2 =  \bar{t}_1 \bar{t}_2~,~
t_3 =  x y \bar{t}_1 \bar{t}_2 ~,~ 
t_4 = \bar{t}_1 \bar{t}_2 ~,~
t_5 = z^{-1} \bar{t}_1 \bar{t}_2 ~,~
t_6 = x^{-1}  \bar{t}_1 \bar{t}_2~.~
\nn\\
\eea
The above charge assignment agrees with the charges of chiral fields in \tref{tchd3}. 

The generators of the mesonic moduli space are
\begin{center}
\begin{tabular}{c|c|c|c|c}
 & & Generator & Generator & $U(1)^3 \times U(1)_R$
\\
PL term & Generator &in GLSM fields &  in chiral fields & charges
\\
\hline
$+ t_1 t_4$ & $A_1$ &$p_1 p_4$ & $ D_{11}=  D_{23} Z_{32}$ & $y^{-1} z \bar{t}_1^2 \bar{t}_2^2$
\\
$+ t_2 t_5$ & $A_2$ & $p_2 p_5$ & $X_{22}=  X_{13} X_{31}$ & $z^{-1} \bar{t}_1^2 \bar{t}_2^2$
\\
$+ t_3 t_6$ &$A_3$ & $p_3 p_6$ & $Y_{33}=  Y_{12} Z_{21}$ & $y \bar{t}_1^2 \bar{t}_2^2$
\\
$+ t_1 t_3 t_5$ & $B_1$& $p_1 p_3 p_5$ & $D_{23} X_{31} Y_{12}$ & $x \bar{t}_1^3 \bar{t}_2^3$
\\
$+ t_2 t_4 t_6$ & $B_2$&$p_2 p_4 p_6$ & $ Z_{21} X_{13} Z_{32}$ & $x^{-1} \bar{t}_1^3 \bar{t}_2^3$
\end{tabular}
\end{center}
The mesonic moduli space can then be expressed as an algebraic variety as follows
\beal{es130aa5b}
\mathcal{M}_{D_3} = \mathbb{C}[A_1,A_2,A_3,B_1,B_2] / \langle A_1 A_2 A_3 = B_1 B_2 \rangle ~.~
\eea

The generators of the mesonic moduli space map to closed non-trivial cycles on the $T^3$ on which the periodic quiver lives. By identifying the $U(1)_x$, $U(1)_y$ and $U(1)_z$ charges of these generators with the winding numbers on $T^3$ of the corresponding cycles, it is possible to construct the periodic quiver without ambiguity. The periodic quiver for the $D_3$ theory is presented in \fref{fd3}.

\subsection{Hilbert Series for the $Q^{1,1,1}$ Theory}
\label{section_Hilbert_Q111}

The quiver in \fref{fq111} and the $J$- and $E$-terms in \eref{es110a8} give rise to the following charge matrices for GLSM fields
 \beal{es399a3}
 Q_{EJ}=
 \left(
\begin{array}{cccccccc}
p_1 & p_2 & p_3 & p_4 & p_5 & p_6 & q_1 & q_2 \\
\hline
 0 & 0 & -1 & -1 & -1 & -1 & 1 & 1 \\
\end{array}
\right) ~~,~~
Q_D=
\left(
\begin{array}{cccccccc}
p_1 & p_2 & p_3 & p_4 & p_5 & p_6 & q_1 & q_2 \\
\hline
 -1 & -1 & 0 & 0 & 0 & 0 & 0 & 1 \\
 0 & 0 & 1 & 1 & 0 & 0 & 0 & -1 \\
 0 & 0 & 0 & 0 & 1 & 1 & 0 & -1 \\
\end{array}
\right)
~.~
\nn\\
 \eea
The corresponding toric diagram matrix is
\beal{es399a4}
G=
\left(
\begin{array}{cccccccc}
p_1 & p_2 & p_3 & p_4 & p_5 & p_6 & q_1 & q_2 \\
\hline
 1 & 1 & 1 & 1 & 1 & 1 & 2 & 2 \\
 0 & 1 & 1 & 0 & 0 & 1 & 1 & 1 \\
 0 & 1 & 0 & 1 & 1 & 0 & 1 & 1 \\
 0 & 0 & 0 & 0 & 1 & -1 & 0 & 0 \\
\end{array}
\right)
~,~
\eea
where we note that $q_1$ and $q_2$ are extra GLSM fields. Given the total charge matrix
\beal{es399a5}
Q_t =
\left(\ba{c}
Q_{EJ} \\ Q_D
\ea\right)
=
 \left(
\begin{array}{cccccccc}
p_1 & p_2 & p_3 & p_4 & p_5 & p_6 & q_1 & q_2 \\
\hline
 0 & 0 & 0 & 0 & 0 &0 & 1 & -1 \\
  1 & 1 & 0 & 0 & 0 & 0 & 0 & -1 \\
 0 & 0 & 1 & 1 & 0 & 0 & 0 & -1 \\
 0 & 0 & 0 & 0 & 1 & 1 & 0 & -1 \\
\end{array}
\right) ~,~
\eea
it can be noted that GLSM fields $\{p_1,p_2\}$, $\{p_3,p_4\}$ and $\{p_5,p_6\}$ form doublets with the same $Q_t$ charges. This indicates an enhancement of the global symmetry from $U(1)^3 \times U(1)_R$ to $SU(2)^3\times U(1)_R$, as expected from the isometries of $Q^{1,1,1}$. Exploiting the symmetries of the gauge theory and the toric diagram, the $U(1)_R$ charges of chiral fields can be assigned as in \tref{tchq111}.

 \begin{table}
 \centering
 \begin{tabular}{r|cccc}
 field & $SU(2)_x$ & $SU(2)_y$ & $SU(2)_z$ & $U(1)_R$ \\
 \hline
 $D_{12}$ & $+1$ & 0 & 0 & $r_1$ \\ 
 $X_{12}$ & $-1$ & 0 & 0 & $r_1$ \\
 $Y_{31}$ & 0 & $+1$ & 0 & $r_2$\\
 $D_{31}$ & 0 & $-1$ & 0 & $r_2$\\
 $X_{24}$ & 0 & $+1$ & 0 & $r_2$\\
 $Z_{24}$ & 0 & $-1$ & 0 & $r_2$\\
 $X_{23}$ & 0 & 0 & $+1$& $r_3$\\
 $Z_{23}$ & 0 & 0 & $-1$& $r_3$\\
 $D_{41}$ & 0 & 0 & $+1$& $r_3$\\
 $Y_{41}$ & 0 & 0 & $-1$& $r_3$\\
 \end{tabular}
 \caption{Global charges of chiral fields in the $Q^{1,1,1}$ theory.}
 \label{tchq111}
 \end{table}

We choose fugacities $t_i$ for the GLSM fields $p_i$ and fugacities $s_i$ for the extra GLSM fields $q_i$. The Hilbert series takes the form
\beal{es140a1}
&&
g(t_i,s_i; Q^{1,1,1}) = 
(
1 - s^2 t_1 t_2 t_3 t_4 t_5^2 - s^2 t_1 t_2 t_3^2 t_5 t_6 -  s^2 t_1^2 t_3 t_4 t_5 t_6 - 3 s^2 t_1 t_2 t_3 t_4 t_5 t_6 
\nn\\
&& \hspace{1cm}
-  s^2 t_2^2 t_3 t_4 t_5 t_6 - s^2 t_1 t_2 t_4^2 t_5 t_6 +  2 s^3 t_1^2 t_2 t_3^2 t_4 t_5^2 t_6 +  2 s^3 t_1 t_2^2 t_3^2 t_4 t_5^2 t_6 +  2 s^3 t_1^2 t_2 t_3 t_4^2 t_5^2 t_6 
\nn\\
&& \hspace{1cm}
+  2 s^3 t_1 t_2^2 t_3 t_4^2 t_5^2 t_6 -  s^4 t_1^2 t_2^2 t_3^2 t_4^2 t_5^3 t_6 - s^2 t_1 t_2 t_3 t_4 t_6^2 +  2 s^3 t_1^2 t_2 t_3^2 t_4 t_5 t_6^2 
+  2 s^3 t_1 t_2^2 t_3^2 t_4 t_5 t_6^2 
\nn\\
&& \hspace{1cm}
+  2 s^3 t_1^2 t_2 t_3 t_4^2 t_5 t_6^2 +  2 s^3 t_1 t_2^2 t_3 t_4^2 t_5 t_6^2 -  s^4 t_1^2 t_2^2 t_3^3 t_4 t_5^2 t_6^2 -  s1^4 s2^4 t_1^3 t_2 t_3^2 t_4^2 t_5^2 t_6^2 
\nn\\
&& \hspace{1cm}
-  3 s^4 t_1^2 t_2^2 t_3^2 t_4^2 t_5^2 t_6^2 -  s^4 t_1 t_2^3 t_3^2 t_4^2 t_5^2 t_6^2 -  s^4 t_1^2 t_2^2 t_3 t_4^3 t_5^2 t_6^2 -  s^4 t_1^2 t_2^2 t_3^2 t_4^2 t_5 t_6^3 +  s^6 t_1^3 t_2^3 t_3^3 t_4^3 t_5^3 t_6^3
)
\nn\\
&& \hspace{0.5cm}
\times 
\PE\big[
s t_1 t_3 t_5+ s t_2 t_3 t_5+    s t_1 t_4 t_5+ s t_2 t_4 t_5+ s t_1 t_3 t_6+    s t_2 t_3 t_6+ s t_1 t_4 t_6+ s t_2 t_4 t_6
 \big]~,~
\nn\\
\eea
where $s= s_1 s_2$. The plethystic logarithm is
\beal{es140a2}
&&
\PL\left[g(t_i, s_i;Q^{1,1,1})\right]
=
s t_1 t_3 t_5+ s t_2 t_3 t_5+    s t_1 t_4 t_5+ s t_2 t_4 t_5+ s t_1 t_3 t_6+    s t_2 t_3 t_6
 \nn\\
 &&\hspace{1cm}
+ s t_1 t_4 t_6+ s t_2 t_4 t_6
-s^2 t_1 t_2 t_3 t_4 t_5^2 - s^2 t_1 t_2 t_3^2 t_5 t_6 -  s^2 t_1^2 t_3 t_4 t_5 t_6 - 3 s^2 t_1 t_2 t_3 t_4 t_5 t_6 
  \nn\\
 &&\hspace{1cm}
-  s^2 t_2^2 t_3 t_4 t_5 t_6 - s^2 t_1 t_2 t_4^2 t_5 t_6 -  s^2 t_1 t_2 t_3 t_4 t_6^2
 + \dots ~.~
\eea
It is important to note that the extra GLSM fields do not play any role in determining the algebraic structure of the mesonic moduli space. This can be seen by setting $s=s_1 s_2 = 1$, which does not affect the information of the algebraic variety captured by the Hilbert series.

Using the fugacity map 
\beal{es140b1}
&
t_1 = x \bar{t}_1 ~,~
t_2 = x^{-1} \bar{t}_1 ~,~&
\nn\\
&
t_3 = y  \bar{t}_2 ~,~
t_4 = y^{-1} \bar{t}_2 ~,~&
\nn\\
&
t_5 = z \bar{t}_3 ~,~
t_6 = z^{-1} \bar{t}_3 ~,~&
\nn\\
&
s_1 = s_2 = 1 ~,~&
\eea
which is in accordance with the $SU(2)^3 \times U(1)_R$ global charges carried by chiral fields in \tref{tchq111}, the Hilbert series in \eref{es140a1} can be written as
\beal{es140b2}
g(t,x,y,z; Q^{1,1,1}) = \sum_{n=0}^{\infty} [n]_{SU(2)_x} [n]_{SU(2)_y} [n]_{SU(2)_z} \bar{t}_1^{n} \bar{t}_2^{n} \bar{t}_3^{n} ~,~
\eea
where $x$, $y$ and $z$ are the fugacities for $SU(2)_x \times SU(2)_y \times SU(2)_z$ charges and the $\bar{t}_i$ fugacities count the fields with $U(1)_R$ charge $r_i$. $[n]_{SU(2)_i} $ is the character of the irreducible representation of $SU(2)_i$ with highest weight $n$. The plethystic logarithm is
\beal{es140b3}
&&
\PL\left[
g(t,x,y,z; Q^{1,1,1})
\right]
= 
[1]_{SU(2)_x} [1]_{SU(2)_y} [1]_{SU(2)_z} \bar{t}_1 \bar{t}_2 \bar{t}_3
\nn\\
&&
\hspace{3cm}
- [2]_{SU(2)_x} \bar{t}_1^2 \bar{t}_2^2 \bar{t}_3^2
- [2]_{SU(2)_y} \bar{t}_1^2 \bar{t}_2^2 \bar{t}_3^2
- [2]_{SU(2)_z} \bar{t}_1^2 \bar{t}_2^2 \bar{t}_3^2
+ \dots ~.~
\eea

The generators of the Calabi-Yau moduli space are summarized below.

\begin{center}
\begin{tabular}{c|c|c|c|c}
 &  &Generator & Generator& $U(1)^3 \times U(1)_R$
\\
PL term & Generator &in GLSM fields &  in chiral fields & charges
\\
\hline
$+ s t_1 t_3 t_5$ & $A_{111}$ &$p_1 p_3 p_5 q_1 q_2$ & $ D_{41} D_{12} X_{24}=  D_{12} X_{23} Y_{31}$ & $x y z \bar{t}_1 \bar{t}_2 \bar{t}_3$
\\
$+ s t_2 t_3 t_5$ & $A_{211}$ & $p_2 p_3 p_5 q_1 q_2$ & $D_ {41} X_ {12} X_ {24}=  X_ {12} X_ {23} Y_{31}$ &$x^{-1} y z \bar{t}_1 \bar{t}_2 \bar{t}_3$ 
\\
$+ s t_1 t_4 t_5$ &$A_{121}$ & $p_1 p_4 p_5 q_1 q_2$ & $D_ {31} D_ {12} X_ {23}=  D_ {41} D_ {12} Z_{24}$ & $x y^{-1} z \bar{t}_1 \bar{t}_2 \bar{t}_3$ 
\\
$+ s t_2 t_4 t_5$ & $A_{221}$& $p_2 p_4 p_5 q_1 q_2$ & $D_ {31} X_ {12} X_ {23}=  D_ {41} X_ {12} Z_{24}$ & $x^{-1} y^{-1} z \bar{t}_1 \bar{t}_2 \bar{t}_3$
\\
$+ s t_1 t_3 t_6$ & $A_{112}$&$p_1 p_3 p_6 q_1 q_2$ & $D_ {12} X_ {24} Y_ {41}=  Y_ {31} D_ {12} Z_{23}$ & $x y z^{-1} \bar{t}_1 \bar{t}_2 \bar{t}_3$
\\
$+ s t_2 t_3 t_6$ & $A_{212}$&$p_2 p_3 p_6 q_1 q_2$ & $X_ {12} X_ {24} Y_ {41}=  Y_ {31} X_ {12} Z_{23}$ & $x^{-1} y z^{-1} \bar{t}_1 \bar{t}_2 \bar{t}_3$
\\
$+ s t_1 t_4 t_6$ & $A_{122}$&$p_1 p_4 p_6 q_1 q_2$ & $D_ {31} D_ {12} Z_ {23}=  Y_ {41} D_ {12} Z_{24}$ & $x y^{-1} z^{-1} \bar{t}_1 \bar{t}_2 \bar{t}_3$
\\
$+ s t_2 t_4 t_6$ & $A_{222}$&$p_2 p_4 p_6 q_1 q_2$ & $D_{31} X_{12} Z_{23}=  Y_{41} X_{12} Z_{24}$ & $x^{-1} y^{-1} z^{-1} \bar{t}_1 \bar{t}_2 \bar{t}_3$
\end{tabular}
\end{center}

The mesonic moduli space can then be expressed as 
\beal{es140a4}
\mathcal{M}_{Q^{1,1,1}} = \mathbb{C}[A_{ijk}] /  \mathcal{I}
~,~
\eea
where the ideal is given by
\beal{es140a4b}
\mathcal{I} =
&&  \langle
A_{122} A_{212} - A_{112} A_{222}~,~ A_{122} A_{221} - A_{121} A_{222}~,~ A_{122} A_{211} - A_{111} A_{222}~,~ 
\nn\\
&&
\ A_{212} A_{221} - A_{211} A_{222}~,~ A_{121} A_{212} - A_{111} A_{222}~,~ A_{112} A_{221} - A_{111} A_{222}~,~ 
\nn\\
&&
\ A_{112} A_{121} - A_{111} A_{122}~,~ A_{112} A_{211} - A_{111} A_{212}~,~ A_{121} A_{211} - A_{111} A_{221}
\rangle
~.~
\nn\\
\eea

\bibliographystyle{JHEP}
\bibliography{mybib}


\end{document}

%% file: pref.tex
\newcommand{\be}{\begin{equation}}
\newcommand{\ee}{\end{equation}}
\newcommand{\beq}{\begin{equation}}
\newcommand{\beql}[1]{\begin{equation}\label{#1}}
\newcommand{\eeq}{\end{equation}}
\newcommand{\ba}{\begin{array}}
\newcommand{\ea}{\end{array}}
\newcommand{\bea}{\begin{eqnarray}}
\newcommand{\beal}[1]{\begin{eqnarray}\label{#1}}
\newcommand{\eea}{\end{eqnarray}}
\newcommand{\ben}{\begin{enumerate}}
\newcommand{\een}{\end{enumerate}}
\newcommand{\bean}{\begin{eqnarray*}}
\newcommand{\eean}{\end{eqnarray*}}
\newcommand{\eref}[1]{(\ref{#1})}
\newcommand{\sref}[1]{\S\ref{#1}}
\newcommand{\tref}[1]{Table~\ref{#1}}
\newcommand{\nn}{\nonumber}

\newcommand{\fref}[1]{Figure \ref{#1}}
\newcommand{\btab}[1]{\begin{tabular}{#1}}
\newcommand{\etab}{\end{tabular}}

\newcommand{\master}{\mathcal{F}^\flat}

\def \Black {\pdfsetcolor{0 0 0 1}}

\newcommand{\comment}[1]{}

\newcommand{\ud}{\mathrm{d}}

\newcommand{\PE}{\mathrm{PE}}
\newcommand{\PL}{\mathrm{PL}}

\newcommand{\qed}{\nobreak \ifvmode \relax \else
      \ifdim\lastskip<1.5em \hskip-\lastskip
      \hskip1.5em plus0em minus0.5em \fi \nobreak
      \vrule height0.75em width0.5em depth0.25em\fi}

%% file: paper.bbl
\providecommand{\href}[2]{#2}\begingroup\raggedright\begin{thebibliography}{10}

\bibitem{Gadde:2013lxa}
A.~Gadde, S.~Gukov, and P.~Putrov, {\it {(0, 2) trialities}},  {\em JHEP} {\bf
  1403} (2014) 076, [\href{http://xxx.lanl.gov/abs/1310.0818}{{\tt
  arXiv:1310.0818}}].

\bibitem{Gadde:2013sca}
A.~Gadde, S.~Gukov, and P.~Putrov, {\it {Fivebranes and 4-manifolds}},
  \href{http://xxx.lanl.gov/abs/1306.4320}{{\tt arXiv:1306.4320}}.

\bibitem{GarciaCompean:1998kh}
H.~Garcia-Compean and A.~M. Uranga, {\it {Brane box realization of chiral gauge
  theories in two-dimensions}},  {\em Nucl.Phys.} {\bf B539} (1999) 329--366,
  [\href{http://xxx.lanl.gov/abs/hep-th/9806177}{{\tt hep-th/9806177}}].

\bibitem{Ahn:1999iu}
C.-h. Ahn and H.~Kim, {\it {Branes at C**4 / Lambda singularity from toric
  geometry}},  {\em JHEP} {\bf 9904} (1999) 012,
  [\href{http://xxx.lanl.gov/abs/hep-th/9903181}{{\tt hep-th/9903181}}].

\bibitem{Sarkar:2000iz}
T.~Sarkar, {\it {D-brane gauge theories from toric singularities of the form
  C**3 / Gamma and C**4 / Gamma}},  {\em Nucl.Phys.} {\bf B595} (2001)
  201--224, [\href{http://xxx.lanl.gov/abs/hep-th/0005166}{{\tt
  hep-th/0005166}}].

\bibitem{Franco:2015tna}
S.~Franco, D.~Ghim, S.~Lee, R.-K. Seong, and D.~Yokoyama, {\it {2d (0,2) Quiver
  Gauge Theories and D-Branes}},  {\em JHEP} {\bf 09} (2015) 072,
  [\href{http://xxx.lanl.gov/abs/1506.0381}{{\tt arXiv:1506.0381}}].

\bibitem{Kutasov:2013ffl}
D.~Kutasov and J.~Lin, {\it {(0,2) Dynamics From Four Dimensions}},  {\em
  Phys.Rev.} {\bf D89} (2014), no.~8 085025,
  [\href{http://xxx.lanl.gov/abs/1310.6032}{{\tt arXiv:1310.6032}}].

\bibitem{Kutasov:2014hha}
D.~Kutasov and J.~Lin, {\it {(0,2) ADE Models From Four Dimensions}},
  \href{http://xxx.lanl.gov/abs/1401.5558}{{\tt arXiv:1401.5558}}.

\bibitem{Tatar:2015sga}
R.~Tatar, {\it {Geometric Constructions of Two Dimensional (0,2) SUSY
  Theories}},  {\em Phys. Rev.} {\bf D92} (2015), no.~4 045006,
  [\href{http://xxx.lanl.gov/abs/1506.0537}{{\tt arXiv:1506.0537}}].

\bibitem{Witten:1993yc}
E.~Witten, {\it {Phases of N = 2 theories in two dimensions}},  {\em Nucl.
  Phys.} {\bf B403} (1993) 159--222,
  [\href{http://xxx.lanl.gov/abs/hep-th/9301042}{{\tt hep-th/9301042}}].

\bibitem{Mohri:1997ef}
K.~Mohri, {\it {D-branes and quotient singularities of Calabi-Yau fourfolds}},
  {\em Nucl.Phys.} {\bf B521} (1998) 161--182,
  [\href{http://xxx.lanl.gov/abs/hep-th/9707012}{{\tt hep-th/9707012}}].

\bibitem{Hanany:1997gh}
A.~Hanany and A.~Zaffaroni, {\it {Branes and six-dimensional supersymmetric
  theories}},  {\em Nucl. Phys.} {\bf B529} (1998) 180--206,
  [\href{http://xxx.lanl.gov/abs/hep-th/9712145}{{\tt hep-th/9712145}}].

\bibitem{Hanany:1997tb}
A.~Hanany and A.~Zaffaroni, {\it {On the realization of chiral four-dimensional
  gauge theories using branes}},  {\em JHEP} {\bf 9805} (1998) 001,
  [\href{http://xxx.lanl.gov/abs/hep-th/9801134}{{\tt hep-th/9801134}}].

\bibitem{Hanany:1998it}
A.~Hanany and A.~M. Uranga, {\it {Brane boxes and branes on singularities}},
  {\em JHEP} {\bf 9805} (1998) 013,
  [\href{http://xxx.lanl.gov/abs/hep-th/9805139}{{\tt hep-th/9805139}}].

\bibitem{Franco:2005rj}
S.~Franco, A.~Hanany, K.~D. Kennaway, D.~Vegh, and B.~Wecht, {\it {Brane Dimers
  and Quiver Gauge Theories}},  {\em JHEP} {\bf 01} (2006) 096,
  [\href{http://xxx.lanl.gov/abs/hep-th/0504110}{{\tt hep-th/0504110}}].

\bibitem{2010arXiv1001.4858F}
M.~{Futaki} and K.~{Ueda}, {\it {Tropical coamoeba and torus-equivariant
  homological mirror symmetry for the projective space}},  {\em ArXiv e-prints}
  (Jan., 2010) [\href{http://xxx.lanl.gov/abs/1001.4858}{{\tt
  arXiv:1001.4858}}].

\bibitem{Kennaway:2007tq}
K.~D. Kennaway, {\it {Brane Tilings}},  {\em Int.J.Mod.Phys.} {\bf A22} (2007)
  2977--3038, [\href{http://xxx.lanl.gov/abs/0706.1660}{{\tt
  arXiv:0706.1660}}].

\bibitem{Hanany:2005ss}
A.~Hanany and D.~Vegh, {\it {Quivers, tilings, branes and rhombi}},  {\em JHEP}
  {\bf 10} (2007) 029, [\href{http://xxx.lanl.gov/abs/hep-th/0511063}{{\tt
  hep-th/0511063}}].

\bibitem{Feng:2005gw}
B.~Feng, Y.-H. He, K.~D. Kennaway, and C.~Vafa, {\it {Dimer models from mirror
  symmetry and quivering amoebae}},  {\em Adv.Theor.Math.Phys.} {\bf 12} (2008)
  3, [\href{http://xxx.lanl.gov/abs/hep-th/0511287}{{\tt hep-th/0511287}}].

\bibitem{2004math......3015M}
G.~{Mikhalkin}, {\it {Amoebas of algebraic varieties and tropical geometry}},
  {\em ArXiv Mathematics e-prints} (Feb., 2004)
  [\href{http://xxx.lanl.gov/abs/math/0403015}{{\tt math/0403015}}].

\bibitem{2001math......8225M}
G.~{Mikhalkin}, {\it {Amoebas of algebraic varieties}},  {\em ArXiv Mathematics
  e-prints} (Aug., 2001) [\href{http://xxx.lanl.gov/abs/math/0108225}{{\tt
  math/0108225}}].

\bibitem{topub2}
S.~Franco, S.~Lee, and R.-K. Seong, {\it {Brane Brick Models and 2d (0,2)
  Triality}},  {\em to appear.}

\bibitem{2011arXiv1110.1033N}
M.~{Nisse} and F.~{Sottile}, {\it {Non-Archimedean coamoebae}},  {\em ArXiv
  e-prints} (Oct., 2011) [\href{http://xxx.lanl.gov/abs/1110.1033}{{\tt
  arXiv:1110.1033}}].

\bibitem{Franco:2012mm}
S.~Franco, {\it {Bipartite Field Theories: from D-Brane Probes to Scattering
  Amplitudes}},  {\em JHEP} {\bf 11} (2012) 141,
  [\href{http://xxx.lanl.gov/abs/1207.0807}{{\tt arXiv:1207.0807}}].

\bibitem{Franco:2006gc}
S.~Franco and D.~Vegh, {\it {Moduli spaces of gauge theories from dimer models:
  Proof of the correspondence}},  {\em JHEP} {\bf 0611} (2006) 054,
  [\href{http://xxx.lanl.gov/abs/hep-th/0601063}{{\tt hep-th/0601063}}].

\bibitem{GarciaEtxebarria:2006aq}
I.~Garcia-Etxebarria, F.~Saad, and A.~M. Uranga, {\it {Quiver gauge theories at
  resolved and deformed singularities using dimers}},  {\em JHEP} {\bf 06}
  (2006) 055, [\href{http://xxx.lanl.gov/abs/hep-th/0603108}{{\tt
  hep-th/0603108}}].

\bibitem{Benvenuti:2006qr}
S.~Benvenuti, B.~Feng, A.~Hanany, and Y.-H. He, {\it {Counting BPS operators in
  gauge theories: Quivers, syzygies and plethystics}},  {\em JHEP} {\bf 11}
  (2007) 050, [\href{http://xxx.lanl.gov/abs/hep-th/0608050}{{\tt
  hep-th/0608050}}].

\bibitem{2003math.....10326K}
R.~{Kenyon}, {\it {An introduction to the dimer model}},  {\em ArXiv
  Mathematics e-prints} (Oct., 2003)
  [\href{http://xxx.lanl.gov/abs/math/0310326}{{\tt math/0310326}}].

\bibitem{2003math.ph..11005K}
R.~{Kenyon}, A.~{Okounkov}, and S.~{Sheffield}, {\it {Dimers and Amoebae}},
  {\em ArXiv Mathematical Physics e-prints} (Nov., 2003)
  [\href{http://xxx.lanl.gov/abs/math-ph/0311005}{{\tt math-ph/0311005}}].

\bibitem{Franco:2005sm}
S.~Franco {\em et.~al.}, {\it {Gauge theories from toric geometry and brane
  tilings}},  {\em JHEP} {\bf 01} (2006) 128,
  [\href{http://xxx.lanl.gov/abs/hep-th/0505211}{{\tt hep-th/0505211}}].

\bibitem{Klebanov:1998hh}
I.~R. Klebanov and E.~Witten, {\it {Superconformal field theory on three-branes
  at a Calabi-Yau singularity}},  {\em Nucl.Phys.} {\bf B536} (1998) 199--218,
  [\href{http://xxx.lanl.gov/abs/hep-th/9807080}{{\tt hep-th/9807080}}].

\bibitem{Morrison:1998cs}
D.~R. Morrison and M.~R. Plesser, {\it {Nonspherical horizons. 1.}},  {\em
  Adv.Theor.Math.Phys.} {\bf 3} (1999) 1--81,
  [\href{http://xxx.lanl.gov/abs/hep-th/9810201}{{\tt hep-th/9810201}}].

\bibitem{Hanany:2006uc}
A.~Hanany and C.~Romelsberger, {\it {Counting BPS operators in the chiral ring
  of N = 2 supersymmetric gauge theories or N = 2 braine surgery}},  {\em Adv.
  Theor. Math. Phys.} {\bf 11} (2007) 1091--1112,
  [\href{http://xxx.lanl.gov/abs/hep-th/0611346}{{\tt hep-th/0611346}}].

\bibitem{Feng:2007ur}
B.~Feng, A.~Hanany, and Y.-H. He, {\it {Counting Gauge Invariants: the
  Plethystic Program}},  {\em JHEP} {\bf 03} (2007) 090,
  [\href{http://xxx.lanl.gov/abs/hep-th/0701063}{{\tt hep-th/0701063}}].

\bibitem{Butti:2007jv}
A.~Butti, D.~Forcella, A.~Hanany, D.~Vegh, and A.~Zaffaroni, {\it {Counting
  Chiral Operators in Quiver Gauge Theories}},  {\em JHEP} {\bf 0711} (2007)
  092, [\href{http://xxx.lanl.gov/abs/0705.2771}{{\tt arXiv:0705.2771}}].

\bibitem{Hanany:2007zz}
A.~Hanany, {\it {Counting BPS operators in the chiral ring: The plethystic
  story}},  {\em AIP Conf.Proc.} {\bf 939} (2007) 165--175.

\bibitem{Gadde:2015wta}
A.~Gadde, S.~S. Razamat, and B.~Willett, {\it {On the reduction of $4d$
  $\mathcal{N}=1$ theories on $\mathbb {S}^2$}},
  \href{http://xxx.lanl.gov/abs/1506.0879}{{\tt arXiv:1506.0879}}.

\bibitem{Benini:2014mia}
F.~Benini, D.~S. Park, and P.~Zhao, {\it {Cluster algebras from dualities of 2d
  N=(2,2) quiver gauge theories}},
  \href{http://xxx.lanl.gov/abs/1406.2699}{{\tt arXiv:1406.2699}}.

\bibitem{Franco:2007ii}
S.~Franco, A.~Hanany, D.~Krefl, J.~Park, A.~M. Uranga, {\em et.~al.}, {\it
  {Dimers and orientifolds}},  {\em JHEP} {\bf 0709} (2007) 075,
  [\href{http://xxx.lanl.gov/abs/0707.0298}{{\tt arXiv:0707.0298}}].

\bibitem{2007arXiv0710.3574M}
G.~{Musiker}, {\it {A graph theoretic expansion formula for cluster algebras of
  classical type}},  {\em ArXiv e-prints} (Oct., 2007)
  [\href{http://xxx.lanl.gov/abs/0710.3574}{{\tt arXiv:0710.3574}}].

\bibitem{2008arXiv0810.3638M}
G.~{Musiker} and R.~{Schiffler}, {\it {Cluster expansion formulas and perfect
  matchings}},  {\em ArXiv e-prints} (Oct., 2008)
  [\href{http://xxx.lanl.gov/abs/0810.3638}{{\tt arXiv:0810.3638}}].

\bibitem{2011arXiv1106.0952L}
K.~{Lee} and R.~{Schiffler}, {\it {A Combinatorial Formula for Rank 2 Cluster
  Variables}},  {\em ArXiv e-prints} (June, 2011)
  [\href{http://xxx.lanl.gov/abs/1106.0952}{{\tt arXiv:1106.0952}}].

\bibitem{Eager:2011ns}
R.~Eager and S.~Franco, {\it {Colored BPS Pyramid Partition Functions, Quivers
  and Cluster Transformations}},  {\em JHEP} {\bf 09} (2012) 038,
  [\href{http://xxx.lanl.gov/abs/1112.1132}{{\tt arXiv:1112.1132}}].

\bibitem{2013arXiv1308.3926L}
M.~{Leoni}, G.~{Musiker}, S.~{Neel}, and P.~{Turner}, {\it {Aztec Castles and
  the dP3 Quiver}},  {\em ArXiv e-prints} (Aug., 2013)
  [\href{http://xxx.lanl.gov/abs/1308.3926}{{\tt arXiv:1308.3926}}].

\bibitem{Hanany:2012hi}
A.~Hanany and R.-K. Seong, {\it {Brane Tilings and Reflexive Polygons}},  {\em
  Fortsch.Phys.} {\bf 60} (2012) 695--803,
  [\href{http://xxx.lanl.gov/abs/1201.2614}{{\tt arXiv:1201.2614}}].

\bibitem{Hanany:2012vc}
A.~Hanany and R.-K. Seong, {\it {Brane Tilings and Specular Duality}},  {\em
  JHEP} {\bf 1208} (2012) 107, [\href{http://xxx.lanl.gov/abs/1206.2386}{{\tt
  arXiv:1206.2386}}].

\bibitem{Forcella:2008bb}
D.~Forcella, A.~Hanany, Y.-H. He, and A.~Zaffaroni, {\it {The Master Space of
  N=1 Gauge Theories}},  {\em JHEP} {\bf 0808} (2008) 012,
  [\href{http://xxx.lanl.gov/abs/0801.1585}{{\tt arXiv:0801.1585}}].

\end{thebibliography}\endgroup
